\documentclass[hyper]{JHEP3} 

\usepackage{graphicx}
\usepackage{epsfig}
\usepackage{cite}
\usepackage{axodraw}
\usepackage{amsmath}

\newcommand{\beq}{\begin{equation}}
\newcommand{\eeq}{\end{equation}}
\makeatletter
\newcommand{\fmslash}[2][0mu]{%
  \mathchoice
    {\fmsl@sh\displaystyle{#1}{#2}}%
    {\fmsl@sh\textstyle{#1}{#2}}%
    {\fmsl@sh\scriptstyle{#1}{#2}}%
    {\fmsl@sh\scriptscriptstyle{#1}{#2}}}
\newcommand{\fmsl@sh}[3]{%
  \m@th\ooalign{$\hfil#1\mkern#2/\hfil$\crcr$#1#3$}}
\makeatother

\newcommand{\mptvec}{\vec{Q}_{tot}} 

\newcommand{\tma}{\tilde{M}_c^{(a)}}
\newcommand{\tmb}{\tilde{M}_c^{(b)}}
\newcommand{\tmasq}{\big(\tilde{M}_c^{(a)}\big)^2}
\newcommand{\tmbsq}{\big(\tilde{M}_c^{(b)}\big)^2}
\newcommand{\ma}{M_c^{(a)}}
\newcommand{\mb}{M_c^{(b)}}
\newcommand{\masq}{\big(M_c^{(a)}\big)^2}
\newcommand{\mbsq}{\big(M_c^{(b)}\big)^2}
\newcommand{\tmaz}{\tilde{M}_c^{(a)}}
\newcommand{\tmbz}{\tilde{M}_c^{(b)}}
\newcommand{\maz}{M_c^{(a)}}
\newcommand{\mbz}{M_c^{(b)}}
\newcommand{\mazsq}{\big(M_c^{(a)}\big)^2}
\newcommand{\mbzsq}{\big(M_c^{(b)} \big)^2}

\title{Dark Matter Particle Spectroscopy at the LHC:\\ Generalizing $M_{T2}$ to Asymmetric Event Topologies}

\author{Partha Konar \\
        Physics Department, University of Florida,
        Gainesville, FL 32611, USA \\
        E-mail: \email{konar@phys.ufl.edu}
        }

\author{Kyoungchul Kong\\
        Theoretical Physics Department, SLAC, 
        Menlo Park, CA 94025, USA \\
        E-mail: \email{kckong@slac.stanford.edu}
        }

\author{Konstantin T.~Matchev \\ 
        Physics Department, University of Florida,
        Gainesville, FL 32611, USA \\
        E-mail: \email{matchev@phys.ufl.edu}
        }

\author{Myeonghun Park\\
        Physics Department, University of Florida,
        Gainesville, FL 32611, USA\\
        E-mail: \email{ishaed@phys.ufl.edu}
        }

\received{\today}               
\accepted{\today}               

\preprint{
          January 12, 2010
          } 

\abstract{We consider SUSY-like missing energy events at hadron colliders
and critically examine the common assumption that the missing energy is 
the result of two {\em identical} missing particles. In order to 
experimentally test this hypothesis, we generalize the subsystem 
$M_{T2}$ variable to the case of asymmetric event topologies, where
the two SUSY decay chains terminate in different ``children'' particles. 
In this more general approach, the endpoint 
$M_{T2(max)}$ of the $M_{T2}$ distribution now gives the mass 
$\tilde M_p(\tilde M_c^{(a)},\tilde M_c^{(b)})$
of the parent particles as a function of {\em two} input children masses 
$\tilde M_c^{(a)}$ and $\tilde M_c^{(b)}$. 
We propose two methods for an independent determination of
the {\em individual} children masses $M_c^{(a)}$ and $M_c^{(b)}$. 
First, in the presence of upstream transverse momentum $P_{UTM}$ the corresponding 
function $\tilde M_{p}(\tilde M_c^{(a)},\tilde M_c^{(b)},P_{UTM})$ is independent of $P_{UTM}$ 
at precisely the right values of the children masses. Second, the previously 
discussed $M_{T2}$ ``kink'' is now generalized to a ``ridge''
on the 2-dimensional surface $\tilde M_p(\tilde M_c^{(a)},\tilde M_c^{(b)})$.
As we show in several examples, quite often there is a special point along that
ridge which marks the true values of the children masses. 
Our results allow collider experiments to probe a multi-component dark matter
sector directly and without any theoretical prejudice.
}


\begin{document} 

\section{Introduction}
\label{sec:intro}

A general expectation in high energy physics today is that physics
beyond the standard model (BSM) should emerge at the TeV scale
in order to stabilize the hierarchy between the Planck and 
electroweak scales. Further evidence in support of this 
belief is provided by the dark matter problem of astro-particle 
physics, which can be quite naturally solved by postulating the 
existence of a new, weakly-interacting dark matter particle
with a mass in the TeV range. Such dark matter particles are 
naturally present in the most popular BSM scenarios such as
supersymmetry \cite{Jungman:1995df}, 
extra dimensions \cite{Servant:2002aq,Burnell:2005hm,Kong:2005hn}, 
little Higgs theory \cite{Cheng:2003ju,Birkedal:2006fz} etc.
They will be produced in the upcoming high-energy collisions at 
the Large Hadron Collider (LHC) at CERN, which offers an exciting
opportunity to study dark matter in a high-energy lab. 
Since the dark matter particles are weakly interacting, 
they do not leave any deposits inside the detector and can only manifest 
themselves in the form of missing energy. Recently,
there has been a lot of theoretical effort directed at 
testing the dark matter hypothesis at the LHC
\cite{Birkedal:2004xn,Feng:2005gj,Baltz:2006fm,Chung:2007cn,Berger:2007yu,%
Baer:2008uu,Arrenberg:2008wy,Berger:2008cq,Baer:2009bu,Roszkowski:2009ye}
and the future International Linear Collider (ILC) 
\cite{Birkedal:2004xn,Birkedal:2005jq,Baltz:2006fm,Chung:2007cn,%
Berger:2007yu,Belanger:2008yc,Bernal:2008zk,Konar:2009ae}. 
Unfortunately, most of these studies have been performed in 
some very model-dependent as well as very complex 
setup\footnote{Some notable exceptions are the studies in Refs.~
\cite{Birkedal:2004xn,Feng:2005gj}.}. 
In the literature, a typical collider study of dark matter 
most often starts with the assumption of a specific model 
with a dark matter candidate
(usually supersymmetry with its myriad of parameters)
and then investigates the model's predictions for  
the expected rates at the LHC in one or several missing energy channels.
Rarely, if ever, has the question been posed in reverse: 
what does the observation of a missing energy signal at the LHC 
tell us about the dark matter particle and its properties 
in a generic and model-independent way \cite{Chang:2009dh}.

\subsection{Probing the dark matter sector at colliders}

Naturally, the most pertinent question after the discovery of any BSM
missing energy signal at the LHC is simply whether the new signal 
is indeed due to the production of new massive particles, or whether it is just
an enhancement in the production of SM neutrinos \cite{Chang:2009dh}. 
In principle, there are two handles that can be used in addressing this 
question. In order to prove dark matter production, one can 
measure the {\em mass} of the missing particle and show that it is 
different (heavier) from the SM neutrino masses.
Alternatively, one can try to measure the {\em spin} of the missing particle 
and show that it is different from $1/2$ (the spin of the neutrino).
While there is a large body of recent work on spin measurements
in missing energy events
\cite{Barr:2004ze,Smillie:2005ar,Athanasiou:2006ef,Athanasiou:2006hv,%
Goto:2004cpa,Battaglia:2005zf,Battaglia:2005ma,%
Datta:2005zs,Datta:2005vx,Barr:2005dz,Meade:2006dw,Alves:2006df,%
Wang:2006hk,SA:2006jm,Smillie:2006cd,Kong:2006pi,Kilic:2007zk,Alves:2007xt,%
Csaki:2007xm,Datta:2007xy,Buckley:2007th,Buckley:2008pp,Kane:2008kw,Burns:2008cp,
Gedalia:2009ym,Boudjema:2009fz,Ehrenfeld:2009rt},
once again very few of those methods are model-independent
\cite{Burns:2008cp,Boudjema:2009fz}.
Furthermore, in all considered examples in the literature 
the spin measurement appears to be very difficult. Therefore,
in this paper we shall concentrate on the question of measuring the  
mass(es) of the particles responsible for the missing energy.
In doing so, we are motivated by two reasons.
First, previous experience indicates that the mass question will be 
answered long before any spin measurements, and second, many of the 
spin determination methods require prior knowledge of the mass spectrum anyway.

The difficulty in measuring the mass of the dark matter particle 
at a hadron collider like the Tevatron or the LHC is widely appreciated
and has generated a lot of recent activity
\cite{Hinchliffe:1996iu, Lester:1999tx,
Bachacou:1999zb, Hinchliffe:1999zc, Allanach:2000kt, Barr:2003rg,
Nojiri:2003tu, Kawagoe:2004rz, Gjelsten:2004ki, Gjelsten:2005aw,
Birkedal:2005cm, Miller:2005zp, 
Gjelsten:2006tg,
Matsumoto:2006ws, Cheng:2007xv, Lester:2007fq, Cho:2007qv,
Gripaios:2007is, Barr:2007hy, Cho:2007dh, Ross:2007rm, Nojiri:2007pq,
Huang:2008ae, Nojiri:2008hy, Tovey:2008ui, Nojiri:2008ir,
Cheng:2008mg, Cho:2008cu, Serna:2008zk, Bisset:2008hm, Barr:2008ba,
Kersting:2008qn, Nojiri:2008vq, Cho:2008tj,Barr:2008hv,
Cheng:2008hk, Burns:2008va, Burns:2009zi, Konar:2008ei,
Cheng:2009fw,Matchev:2009iw,Han:2009ss,Barr:2009wu,
Webber:2009vm,Kim:2009nq,Kang:2009sk,MPtalk,Barr:2009jv,Matchev:2009fh,
Polesello:2009rn,Matchev:2009ad,Konar:2009wn,Autermann:2009js,Cho:2009ve}.
The main problem can be understood as follows. In a typical BSM dark 
matter scenario, the cosmological longevity of the dark matter particle
is ensured by some new symmetry\footnote{Some popular examples
are: $R$-parity in supersymmetry, KK parity in Universal Extra Dimensions,
$T$-parity in Little Higgs models, $Z$-parity in warped extra dimensions, 
$U$-parity in extended gauge theories, etc.} 
under which the SM particles are singlets.
At the same time, there are additional particles in the spectrum which 
are charged under the new symmetry. If the lightest one among those
is electrically and color neutral, it is a potential dark matter candidate, 
whose lifetime is protected by the new symmetry.
With any such setup, it is clear that {\em single} production
of dark matter particles at colliders is forbidden by the symmetry.
Therefore, each event has {\em at least two} missing particles, whose
energies and momenta are unknown. As a rule, it is typically impossible 
to fully reconstruct the kinematics of such events and observe the mass 
of the missing particle directly as an invariant mass peak\footnote{For 
studies attempting full event reconstruction in long cascade chains, 
see Refs.~\cite{Cheng:2007xv,Cheng:2008mg,Cheng:2009fw,Webber:2009vm,Autermann:2009js}.}.
Consequently, one has to resort to various indirect methods
of extracting the mass of the dark matter particle.

Unfortunately, all existing studies in the literature 
have explicitly or implicitly made the following two 
assumptions: 
\begin{itemize}
\item {\em Single dark matter component.}
A common assumption throughout the collider phenomenology literature 
is that colliders are probing only one dark matter species at a time, i.e.
that the missing energy signal at colliders is due to the 
production of one and only one type of dark matter particles.
Of course, there is no astrophysical evidence that the dark matter
is made up of a single particle species: it may very well be that the
dark matter world has a rich structure, just like ours \cite{Profumo:2009tb}.
Consequently, if there exist several types of dark matter particles, 
each contributing some fraction to the total relic density,
a priori there is no reason why they cannot {\em all} be produced 
in high energy collisions. Theoretical models with multiple dark matter 
candidates have also been proposed
\cite{Boehm:2003ha,Ma:2006uv,Hur:2007ur,Cao:2007fy,Lee:2008pc,%
Feng:2008ya,SungCheon:2008ts,Fairbairn:2008fb,Zurek:2008qg}.
\item {\em Identical missing particles in each event.}
A separate assumption, common to most previous studies, is that
the two missing particles in each event are {\em identical}.
This assumption could in principle be violated as well, even if
the single dark matter component hypothesis is true. The point is
that one of the missing particles in the event may {\em not} 
be a dark matter particle, but simply some heavier cousin which 
decays invisibly. An invisibly decaying heavy neutralino 
($\tilde\chi^0_i\to \nu\bar{\nu}\tilde\chi^0_1$ with $i>1$)
and an invisibly decaying sneutrino ($\tilde\nu\to \nu \tilde\chi^0_1$)
are two such examples from supersymmetry.
As far as the event kinematics is concerned, the mass of the heavier cousin
{\em is} a relevant parameter and approximating it with the 
mass of the dark matter particle will simply give nonsensical
results. Another relevant example is provided by models in which
the SUSY cascade may terminate in any one of several light neutral
particles \cite{Arvanitaki:2009hb}.
\end{itemize}

Given our utter ignorance about the structure of the dark matter sector, 
in this paper we set out to develop the necessary formalism for
carrying out missing energy studies at hadron colliders in a very general 
and model-independent way, without relying on any assumptions 
about the nature of the missing particles. 
In particular, we shall {\em not} assume that the 
two missing particles in each event are the same.
We shall also allow for the simultaneous production of {\em several}
dark matter species, or alternatively, for the production of a dark matter
candidate in association with a heavier, invisibly decaying 
particle. Under these very general circumstances, we shall try to develop a method 
for measuring the individual masses of {\em all} relevant particles - 
the various missing particles which are responsible for the missing energy, 
as well as their parents which were originally produced in the event. 

\subsection{Generalizing $M_{T2}$ to asymmetric event topologies}

In general, by now there is a wide variety of techniques 
available for mass measurements in SUSY-like missing energy events.
Such events are characterized by the pair production of 
two new particles, each of which undergoes a sequence of 
cascade decays ending up in a particle which is invisible in the detector. 
Each technique has its own advantages and disadvantages\footnote{For 
a comparative review of the three main techniques, see \cite{Burns:2008va}.}.
For our purposes, we chose to revamp the method of the Cambridge 
$M_{T2}$ variable \cite{Lester:1999tx} and adapt it to the more general case
of an asymmetric event topology shown in Fig.~\ref{fig:metevent}.
\FIGURE[ht]{
{
\unitlength=1.3 pt
\SetScale{1.3}
\SetWidth{1.0}      
\normalsize    
{} \qquad\allowbreak
\begin{picture}(285,200)(15,0)
\CBoxc(136,100)(138,116){Blue}{White}
\SetColor{Gray}
\Line( 13,185)(260,185) 
\Line( 13, 15)(260, 15)
%
\Line( 10,190)(55,120)
\Line( 10, 10)(55, 80)
\Text( 5,195)[c]{\Black{$p(\bar{p})$}}
\Text( 5,  5)[c]{\Black{$p(\bar{p})$}}
%
\DashLine(130,170)(130,180){1}
\DashLine(130, 20)(130, 30){1}
\Line( 50,165)(260,165)
\Line( 50, 35)(260, 35)
%
%
\Line(85,110)(105,140)
\Line(95,110)(115,140)
\Line(105,110)(125,140)
\Line(130,110)(150,140)
\Line(85, 90)(105, 60)
\Line(130,90)(150, 60)
\DashLine(210,110)(220,125){2}
\DashLine(230,110)(240,125){2}
\DashLine(210, 90)(220, 75){2}
\DashLine(230, 90)(240, 75){2}
\DashLine(120,125)(135,125){1}
\DashLine(100, 75)(135, 75){1}
\DashLine(219,117)(231,117){1}
\DashLine(219, 83)(231, 83){1}
%
\SetColor{Red}
\Line(50,110)(175,110)
\Line(50, 90)(175, 90)
\Text( 70,117)[l]{\Red{$M_p$}}
\Text( 70, 83)[l]{\Red{$M_p$}}
\Text(140,117)[l]{\Red{$M_c^{(a)}$}}
\Text(140, 83)[l]{\Red{$M_c^{(b)}$}}
\DashLine(190,110)(245,110){2}
\DashLine(190, 90)(245, 90){2}
\CBoxc(258,100)(15,190){Black}{Yellow}
\rText(258,100)[c][l]{\Blue{$\vec{P}_{UTM}$}}
\CBoxc(125,145)(60,15){Black}{Yellow}
\Text(125,145)[c]{\Blue{$\vec{p}_T^{~(a)},m_{(a)}$}}
\CBoxc(125, 55)(60,15){Black}{Yellow}
\Text(125, 55)[c]{\Blue{$\vec{p}_T^{~(b)},m_{(b)}$}}
\CBoxc(180,110)(42,15){Red}{Yellow}
\Text( 180,110)[c]{\Red{$\vec{q}_T^{~(a)},\tilde M_c^{(a)}$}}
\CBoxc(180, 90)(42,15){Red}{Yellow}
\Text(180, 90)[c]{\Red{$\vec{q}_T^{~(b)},\tilde M_c^{(b)}$}}
\COval(50,100)(80,12)(0){Black}{Green}
\CBoxc(285,100)(15,150){Green}{Yellow}
\rText(285, 100)[c][l]{{$(\vec{q}^{~(a)}_T + \vec{q}^{~(b)}_T) 
= -(\vec{p}_T^{~(a)}+\vec{p}_T^{~(b)}+\vec{P}_{UTM}) $}}
\end{picture}
}
\caption{\it The generic event topology under consideration in this paper.  We
consider the inclusive pair-production of two ``parent'' particles
with identical masses $M_p$. The parents may be accompanied
by ``upstream'' objects, e.g.~jets from initial state radiation,
visible decay products of even heavier particles, etc. 
The transverse momentum of all upstream objects is measured and
denoted by $\vec{P}_{UTM}$. In turn, each parent particle initiates a
decay chain (shown in red) which produces a certain number $n^{(\lambda)}$ of SM
particles (shown in gray) and an intermediate ``child'' particle of
mass $M_c^{(\lambda)}$, where $\lambda=a$ ($\lambda=b$) for the branch above 
(below).  In general, the child particle does
not have to be the dark matter candidate, and may decay further
as shown by the dashed lines.  The $M_{T2}$ variable is defined for the subsystem
inside the blue box and is defined in terms of two
arbitrary children ``test'' masses $\tilde M_c^{(a)}$ and $\tilde M_c^{(b)}$. 
The $n^{(\lambda)}$ SM particles from each branch form a composite particle of transverse momentum
$\vec{p}_T^{~(\lambda)}$ and invariant mass $m_{(\lambda)}$,
correspondingly. The trial transverse momenta $\vec{q}_T^{~(\lambda)}$
of the children obey the transverse momentum conservation
relation shown inside the green box. In general, the number $n^{(\lambda)}$,
as well as the type of SM decay products in each branch do not have to be the same.  
}
\label{fig:metevent} 
}
Consider the inclusive production of two identical\footnote{In principle,
the assumption of identical parents can also be relaxed, by a suitable generalization
of the $M_{T2}$ variable, in which the mass ratio of the two parents is treated as an
additional input parameter \cite{Barr:2009jv}.} parents
of mass $M_p$ as shown in Fig.~\ref{fig:metevent}. The parent particles 
may be accompanied by any number of ``upstream'' objects,
such as jets from initial state radiation \cite{Gripaios:2007is,Barr:2007hy,Barr:2008ba}, 
or visible decay products of even heavier (grandparent) particles \cite{Burns:2008va}. 
The exact origin and nature of the upstream objects will be of no
particular importance to us, and the only information about 
them that we shall use will be their total transverse momentum $\vec{P}_{UTM}$. 
In turn, each parent particle initiates a decay chain (shown in red) which
produces a certain number $n^{(\lambda)}$ of Standard Model (SM) particles (shown in gray) 
and an intermediate ``child'' particle of mass $M_c^{(\lambda)}$.
Throughout this paper we shall use the index $\lambda$ 
to classify various objects as belonging to the upper ($\lambda=a$) 
or lower ($\lambda=b$) branch in Fig.~\ref{fig:metevent}.
The child particle may or may not be a dark
matter candidate: in general, it may decay further 
as shown by the dashed lines in Fig.~\ref{fig:metevent}.  
We shall apply the ``subsystem'' $M_{T2}$ concept \cite{Serna:2008zk,Burns:2008va}
to the subsystem within the blue rectangular frame. The SM 
particles from each branch within the subsystem 
form a composite particle of known\footnote{We assume that 
there are no neutrinos among the SM decay products in each branch.} transverse momentum
$\vec{p}_T^{~(\lambda)}$ and invariant mass $m_{(\lambda)}$.
Since the children masses $M_c^{(a)}$ and $M_c^{(b)}$ are a priori unknown, 
the subsystem $M_{T2}$ will be defined in terms of two ``test'' masses
$\tilde M_c^{(a)}$ and $\tilde M_c^{(b)}$.
In Fig.~\ref{fig:metevent}, $\vec{q}_T^{~(\lambda)}$
are the trial transverse momenta of the two children.
The individual momenta $\vec{q}_T^{~(\lambda)}$ are also a priori unknown,
but they are constrained by transverse momentum conservation:
\begin{equation}
\vec{q}^{~(a)}_T + \vec{q}^{~(b)}_T \equiv \vec{Q}_{tot} 
= -(\vec{p}_T^{~(a)}+\vec{p}_T^{~(b)}+\vec{P}_{UTM}) .
\label{momcon}
\end{equation}

Given this very general setup, in Section~\ref{sec:mt2dlsp} 
we shall consider a generalization\footnote{The possibility 
of applying the $M_{T2}$ variable to an event topology with different 
children was previously mentioned in Refs.~\cite{MPtalk,Barr:2009jv}.} 
of the usual $M_{T2}$ variable which can apply to the 
asymmetric event topology of Fig.~\ref{fig:metevent}.
There will be two different aspects of the asymmetry:
\begin{itemize}
\item First and foremost, we shall avoid the common assumption that the
two children have the same mass. This will be important for two reasons. 
On the one hand, it will allow us to study events in which there are indeed
two different types of missing particles. We shall give several such examples 
in the subsequent sections. More importantly, the endpoint of the 
asymmetric $M_{T2}$ variable will allow us to measure the two 
children masses {\em separately}.
Therefore, even when the events contain identical missing particles,
as is usually assumed throughout the literature,
one would be able to establish this fact experimentally from the data,
instead of relying on an ad hoc theoretical assumption.
\item As can be seen from Fig.~\ref{fig:metevent}, in general, 
the number as well as the types of SM
decay products in each branch may be different as well.
Once we allow for the children to be different, 
and given the fact that we start from identical parents, 
the two branches of the subsystem will naturally involve 
different sets of SM particles. 
\end{itemize}
In what follows, when referring to the more general $M_{T2}$ variable
defined in Section~\ref{sec:mt2dlsp}, we shall interchangeably use 
the terms ``asymmetric'' or ``generalized'' $M_{T2}$. 
In contrast, we shall use the term ``symmetric'' when referring to 
the more conventional $M_{T2}$ definition with identical children.

The traditional $M_{T2}$ approach assumes that the children have a common 
test mass $\tilde M_c\equiv \tilde M_c^{(a)}=\tilde M_c^{(b)}$
and then proceeds to find one functional relation between 
the true child mass $M_c$ and the true parent mass $M_p$ as follows 
\cite{Lester:1999tx}. Construct several $M_{T2}$ distributions 
for different input values of the test children mass $\tilde M_c$
and then read off their upper kinematic endpoints
$M_{T2(max)}(\tilde M_c)$. These endpoint measurements are then 
interpreted as an output parent mass $\tilde M_p$, which is 
a function of the input test mass $\tilde M_c$:
\begin{equation}
\tilde M_p(\tilde M_c) \equiv M_{T2(max)}(\tilde M_c)\, .
\label{eq:Mptilde}
\end{equation}
The importance of this functional relation is that it is automatically 
satisfied for the {\em true} values $M_p$ and $M_c$ of the parent and child masses:
\begin{equation}
M_p = M_{T2(max)}(M_c).
\label{eq:MpatMctrue}
\end{equation}
In other words, if we could somehow guess the correct value $M_c$
of the child mass, the function (\ref{eq:Mptilde}) will provide the 
correct value $M_p$ of the parent mass.
However, since the true child mass $M_c$ is a priori unknown, 
the individual masses $M_p$ and $M_c$ still remain undetermined 
and must be extracted by some other means.

At this point, it may seem that by considering the asymmetric
$M_{T2}$ variable with non-identical children particles, we have regressed 
to some extent. Indeed, we are introducing an additional degree of freedom in
eq.~(\ref{eq:Mptilde}), which now reads
\begin{equation}
\tilde M_p(\tilde M_c^{(a)},\tilde M_c^{(b)}) \equiv M_{T2(max)}(\tilde M_c^{(a)},\tilde M_c^{(b)})\, .
\label{eq:Mptildedlsp}
\end{equation}
The standard $M_{T2}$ endpoint method will still allow us to find 
the parent mass $\tilde M_p$, but now it is a function of 
{\em two} input parameters $\tilde M_c^{(a)}$ and $\tilde M_c^{(b)}$ which are completely unknown.
Of course, if one knew the correct values of the two children masses
$M_c^{(a)}$ and $M_c^{(b)}$ entering eq.~(\ref{eq:Mptildedlsp}), the true
parent mass $M_p$ will be given in a manner analogous to 
eq.~(\ref{eq:MpatMctrue}):
\begin{equation}
M_p = M_{T2(max)}(M_c^{(a)},M_c^{(b)}).
\label{eq:MpatMctruedlsp}
\end{equation}

Our main result in this paper is that in spite of the apparent 
remaining arbitrariness in eq.~(\ref{eq:Mptildedlsp}), one can nevertheless
uniquely determine {\em all three} masses $M_p$, $M_c^{(a)}$ and $M_c^{(b)}$,
just by studying the behavior of the measured function 
$\tilde M_p(\tilde M_c^{(a)},\tilde M_c^{(b)})$. More importantly,
this determination can actually be done in two different ways! 
Our first method is simply a generalization of the 
observation made in Refs.~\cite{Cho:2007qv,Gripaios:2007is,Barr:2007hy,Cho:2007dh,Burns:2008va}
that under certain circumstances (varying $m_{(\lambda)}$ or nonvanishing
upstream momentum $P_{UTM}$), the function (\ref{eq:Mptilde})
develops a ``kink'' precisely at the correct value $M_c$ of the 
child mass:
\begin{equation}
\left ( \frac{\partial \tilde M_{p} (\tilde{M}_c)}
             {\partial \tilde{M}_c} \right )_{\tilde{M}_c + \epsilon} -
\left ( \frac{\partial \tilde M_{p} (\tilde{M}_c)}
             {\partial \tilde{M}_c} \right )_{\tilde{M}_c - \epsilon} 
\left\{ 
\begin{array}{ll}
  = 0,  &  ~~ \textrm{if}~~ \tilde M_c\ne M_c,   \\
\ne 0,  &  ~~ \textrm{if}~~ \tilde M_c = M_c .
\end{array}
\right.
\label{eq:kink}
\end{equation}
In other words, the function (\ref{eq:Mptilde}) is continuous, but not differentiable at the
point $\tilde M_c=M_c$. In the asymmetric $M_{T2}$ case, we find that the 
function (\ref{eq:Mptildedlsp}) is similarly non-differentiable 
at a {\em set} of points $\{(\tilde M_c^{(a)},\tilde M_c^{(b)})\}$, 
so that the kink of eq.~(\ref{eq:kink}) is generalized to a ``ridge'' 
on the 2-dimensional hypersurface defined by (\ref{eq:Mptildedlsp}) in the 
three-dimensional parameter space of $\{\tilde M_c^{(a)},\tilde M_c^{(b)},\tilde M_p\}$.
\footnote{Ref.~\cite{Barr:2009jv} studied the orthogonal scenario of 
different parents ($M_p^{(a)}\ne M_p^{(b)}$) and identical children
($M_c^{(a)}= M_c^{(b)}$) and found a similar non-differentiable feature, 
called a ``crease'', on the corresponding two-dimensional hypersurface 
within the three-dimensional parameter space
$\{ \tilde M_c, \tilde M_p^{(a)}, \tilde M_p^{(b)}\}$.} 
Interestingly enough, the ridge often (albeit not always) exhibits 
a special point which marks the exact location of the true values 
$(M_c^{(a)},M_c^{(b)})$.

Our second method for determining the two children masses
$\tilde M_c^{(a)}$ and $\tilde M_c^{(b)}$ is even more general
and is applicable under any circumstances. The main starting 
point is that just like the endpoint of the symmetric $M_{T2}$,
the endpoint of the asymmetric $M_{T2}$ also depends on the value 
of the upstream transverse momentum $P_{UTM}$, so that eq.~(\ref{eq:Mptildedlsp})
is more properly written as
\begin{equation}
\tilde M_p(\tilde M_c^{(a)},\tilde M_c^{(b)},P_{UTM}) =
M_{T2(max)}(\tilde M_c^{(a)},\tilde M_c^{(b)}, P_{UTM})\, .
\label{eq:MptildePT}
\end{equation}
Now we can explore the $P_{UTM}$ dependence in (\ref{eq:MptildePT}) 
and note that it is absent for precisely the right values of 
$\tilde M_c^{(a)}$ and $\tilde M_c^{(b)}$:
\begin{equation}
\frac{\partial M_{T2(max)} (\tilde{M}_c^{(a)}, \tilde{M}_c^{(b)}, P_{UTM})}
     {\partial P_{UTM}} \Big |_{  \tilde{M}_c^{(a)} = M_c^{(a)}, \tilde{M}_c^{(b)} = M_c^{(b)}}= 0 \, .
\label{eq:MpISRinv}
\end{equation}
While this property has been known, it was rarely used in the case of 
the symmetric $M_{T2}$, since it offers redundant information:
once the correct child mass $M_c$ is found through the $M_{T2}$ kink
(\ref{eq:kink}), the parent mass $M_p$ is given by (\ref{eq:Mptilde})
and there are no remaining unknowns, thus there is no need to
further investigate the $P_{UTM}$ dependence. 
In the case of the asymmetric $M_{T2}$, however, we start with 
one additional unknown parameter, which cannot always be determined from the
``ridge'' information alone. Therefore, in order to pin down the complete spectrum, 
we are forced to make use of (\ref{eq:MpISRinv}).
The nice feature of the $P_{UTM}$ method is that it always allows
us to determine {\em both} children masses $M_c^{(a)}$ and $M_c^{(b)}$,
without relying on the ``ridge'' information at all.
In this sense, our two methods are complementary and 
each can be used to cross-check the results obtained by the other.

The paper is organized as follows\footnote{Readers 
who are unfamiliar with the $M_{T2}$ concept
may benefit from consulting Refs.~\cite{Barr:2003rg,Cho:2007dh,Burns:2008va,Barr:2009jv} first.}. 
In Sec.~\ref{sec:mt2} we begin with a review of the 
conventional symmetric $M_{T2}$ variable and its properties.
Then in Sec.~\ref{sec:mt2dlsp} we introduce the asymmetric 
$M_{T2}$ variable and highlight its properties which are relevant
for our mass measurements. We also discuss some experimental subtleties 
in the construction of the asymmetric $M_{T2}$ distribution,
which are not present in the case of the symmetric $M_{T2}$. 
Sections \ref{sec:110}, \ref{sec:220off} and \ref{sec:220on}
present some simple examples of asymmetric event topologies.
Finally, Sec.~\ref{sec:conclusions} summarizes our main results
and outlines some possible directions for future work.
Appendix \ref{app:infpt} revisits the examples of Section \ref{sec:110}
in the case of $P_{UTM}\to \infty$, which can be handled 
by purely analytical means \cite{Barr:2009jv}.

\section{The conventional symmetric $M_{T2}$ 
}
\label{sec:mt2}

\subsection{Definition}

We begin our discussion by revisiting the conventional definition of the
symmetric $M_{T2}$ variable with identical daughters, 
following the general notation introduced
in Fig.~\ref{fig:metevent}.  Let us consider the inclusive production of two
parent particles with common mass $M_p$. Each parent initiates a
decay chain producing a certain number $n^{(\lambda)}$ of SM particles.
In this section we assume that
the two chains terminate in children particles of the same mass:
$M_c^{(a)}=M_c^{(b)}=M_c$. (From Section~\ref{sec:mt2dlsp} on we 
shall remove this assumption.)
In most applications of $M_{T2}$ in the literature,
the children particles are identified with the very last particles in the 
decay chains, i.e. the dark matter candidates. However, the symmetric $M_{T2}$ 
can also be usefully applied to a subsystem of the original event topology, 
where the children are some other pair of (identical) particles appearing 
further up the decay chain \cite{Serna:2008zk,Burns:2008va}.
The $M_{T2}$ variable is defined in terms of the measured 
invariant mass $m_{(\lambda)}$ and transverse momentum $\vec{p}^{~(\lambda)}_T$ 
of the visible particles on each side (see Fig.~\ref{fig:metevent}). 
With the assumption of identical children, the transverse 
mass of each parent is
\begin{equation}
M_{T}^{(\lambda)}\big( \vec{p}^{~(\lambda)}_T;~ \vec{q}^{~(\lambda)}_T;~m_{(\lambda)};~\tilde{M}_c \big ) 
= \sqrt{m_{(\lambda)}^2 + \tilde{M}_c^2 + 2\left(e^{(\lambda)} \tilde{e}^{(\lambda)} 
                 -  \vec{p}^{~(\lambda)}_T \cdot \vec{q}^{~(\lambda)}_T  \right)} \, ,
\label{eq:MTp}
\end{equation}
where $\tilde{M}_c$ is the common test mass for the children, 
which is an input to the $M_{T2}$ calculation, while
$\vec{q}^{~(\lambda)}_T$ is the unknown transverse momentum 
of the child particle in the $\lambda$-th chain. In eq.~(\ref{eq:MTp})
we have also introduced shorthand notation for the transverse energy 
of the composite particle made from the visible SM particles in the $\lambda$-th chain
\begin{equation}
e^{(\lambda)} = \sqrt{m_{(\lambda)}^2 + \vec{p}^{~(\lambda)}_T
\cdot \vec{p}^{~(\lambda)}_T}
\label{elambda}
\end{equation}
and for the transverse energy of the corresponding child particle
in the $\lambda$-th chain
\begin{equation}
\tilde{e}^{(\lambda)} =
\sqrt{\tilde{M}_c^2 + \vec{q}^{~(\lambda)}_T\cdot
\vec{q}^{~(\lambda)}_T}\, .
\label{etilde}
\end{equation} 

Then the event-by-event symmetric $M_{T2}$ variable
is defined through a minimization procedure over all possible partitions of
the two children momenta $\vec{q}^{~(\lambda)}_T$ \cite{Lester:1999tx}
{\setlength\arraycolsep{2pt}
\begin{eqnarray}
&&M_{T2}{\scriptstyle\big( \vec{p}^{~(a)}_T, \vec{p}^{~(b)}_T;
~m_{(a)},m_{(b)};~\tilde{M}_c,~P_{UTM} \big )}= \nonumber \\
&& \min_{{\scriptscriptstyle\vec{q}^{~(a)}_T + \vec{q}^{~(b)}_T = \vec{Q}_{tot}}}
{\Bigg[\max\left\{
    M_{T}^{(a)} {\scriptstyle \big( \vec{p}^{~(a)}_T;~ \vec{q}^{~(a)}_T;~m_{(a)};~\tilde{M}_c \big )},~
    M_{T}^{(b)}  {\scriptstyle\big( \vec{p}^{~(b)}_T;~ \vec{q}^{~(b)}_T;~m_{(b)};~\tilde{M}_c\big )}
\right\} \Bigg]} \, ,
\label{eq:mt2} 
\end{eqnarray} 
consistent with the momentum conservation constraint (\ref{momcon}) in the transverse plane.

\subsection{Computation}
\label{sec:mt2comp}

The standard definition (\ref{eq:mt2}) of the $M_{T2}$ variable is
sufficient to compute the value of $M_{T2}$ numerically, given 
a set of input values for its arguments. The right-hand side of
eq.~(\ref{eq:mt2}) represents a simple minimization problem in two variables,
which can be easily handled by a computer. In fact, there are
publicly available computer codes for computing 
$M_{T2}$~\cite{Lester_code,Davis_code}. 
The public codes have even been optimized for speed 
\cite{Cheng:2008hk} and give results consistent with each 
other (as well as with our own code)\footnote{Unfortunately, the assumption of identical children
is hardwired in the public codes and they cannot be used 
to calculate the asymmetric $M_{T2}$ introduced below in Section~\ref{sec:mt2dlsp} without 
additional hacking. We shall return to this point in Section~\ref{sec:mt2dlsp}.}.
Nevertheless, it is useful to have an analytical formula for calculating the
event-by-event $M_{T2}$ for several reasons. First, an analytical formula
is extremely valuable when it comes to understanding the properties 
and behavior of complex mathematical functions like (\ref{eq:mt2}). Second, computing $M_{T2}$ 
from a formula will be faster than any numerical scanning algorithm. The
computing speed becomes an issue especially when one considers variations 
of $M_{T2}$ like $M_{T2gen}$, where in addition one needs to scan over all 
possible partitions of the visible objects into two decay chains
\cite{Lester:2007fq}. Therefore, in this paper we shall pay special 
attention to the availability of analytical formulas and we shall quote 
such formulas whenever they are available.

In the symmetric case with identical children,
an analytical formula for the event-by-event $M_{T2}$
exists only in the special case $P_{UTM}=0$.
It was derived in \cite{Lester:2007fq}
and we provide it here for completeness. (In the next
section we shall present its generalization for the asymmetric
case of different children.)
The symmetric $M_{T2}$
is known to have two types of solutions: ``balanced'' and 
``unbalanced'' \cite{Barr:2003rg, Lester:2007fq}. 
The balanced solution is achieved when the minimization 
procedure in eq.~(\ref{eq:mt2}) selects a momentum configuration
for $\vec{q}^{~(\lambda)}_T$ in which the transverse masses of the 
two parents are the same: $M_T^{(a)}=M_T^{(b)}$. In that case, 
typically neither $M_T^{(a)}$ nor $M_T^{(b)}$ is at its
global (unconstrained) minimum. 
In what follows, we shall use a superscript $B$ to refer to such 
balanced-type solutions. The formula for the balanced solution $M_{T2}^{B}$
of the symmetric $M_{T2}$ variable is given by
\cite{Lester:2007fq,Cho:2007dh} 
\begin{equation}  
\Big[M_{T2}^{B}{\scriptstyle\big(\vec{p}^{~(a)}_T,\vec{p}^{~(b)}_T;~m_{(a)},m_{(b)};~\tilde{M}_c\big)}\Big]^{2}
= \tilde{M}_c^2 + A_T + 
          \sqrt{ \left ( 1 + \frac{4 \tilde{M}_c^2}{2A_T-m_{(a)}^2-m_{(b)}^2} \right ) 
             \left ( A_T^2 - m_{(a)}^2 ~m_{(b)}^2  \right ) } \, ,
\label{eq:mt2oldB}
\end{equation}
where $A_T$ is a convenient shorthand notation introduced in \cite{Cho:2007dh} 
\begin{equation}
A_T = e^{(a)} e^{(b)} + \vec{p}^{~(a)}_T\cdot\vec{p}^{~(b)}_T
\label{ATdef}
\end{equation}
and $e^{(\lambda)}$ was already defined in eq.~(\ref{elambda}).

On the other hand, unbalanced solutions arise when one of the two 
parent transverse masses ($M_T^{(a)}$ or $M_T^{(b)}$, as the case may be) 
is at its global (unconstrained) minimum.
Denoting the two unbalanced solutions with a superscript $U\lambda$, we have \cite{Barr:2003rg}
\begin{eqnarray}
M_{T2}^{Ua}{\scriptstyle\big(\vec{p}^{~(a)}_T,\vec{p}^{~(b)}_T;~m_{(a)},m_{(b)};~\tilde{M}_c\big)}
&=& m_{(a)} + \tilde{M}_c \, , \label{mt2U1} \\ [2mm]
M_{T2}^{Ub}{\scriptstyle\big(\vec{p}^{~(a)}_T,\vec{p}^{~(b)}_T;~m_{(a)},m_{(b)};~\tilde{M}_c\big)}
&=& m_{(b)} + \tilde{M}_c \, . \label{mt2U2}
\end{eqnarray}

Given the three possible options for $M_{T2}$, eqs.~(\ref{eq:mt2oldB}), 
(\ref{mt2U1}) and (\ref{mt2U2}), it remains to specify which one actually
takes place for a given set of values for 
$\vec{p}^{~(a)}_T$, $\vec{p}^{~(b)}_T$, $m_{(a)}$, $m_{(b)}$, $\tilde{M}_c$
and $P_{UTM}=0$ in the event\footnote{Recall that (\ref{eq:mt2oldB}) 
only applies for $P_{UTM}=0$.}.
The balanced solution (\ref{eq:mt2oldB}) applies when
the following two conditions are simultaneously satisfied: 
\begin{eqnarray}
{M^{(b)}_T}{\scriptstyle\big(\vec{p}^{~(b)}_T;~\vec{q}^{~(b)}_T
    =-\vec{q}^{~(a)}_{T(0)} +~\mptvec;~m_{(b)};~\tilde{M}_c\big)} &\ge& 
{M^{(a)}_T} {\scriptstyle\big(\vec{p}^{~(a)}_T;~\vec{q}^{~(a)}_T
=\vec{q}^{~(a)}_{T(0)};~m_{(a)},~\tilde{M}_c\big)}  = m_{(a)} + \tilde{M}_c ,
~~~~~~~
\label{eq:ubA}
\\ [2mm]
{M^{(a)}_T}{\scriptstyle\big(\vec{p}^{~(a)}_T;~\vec{q}^{~(a)}_T
    =-\vec{q}^{~(b)}_{T(0)} +~\mptvec;~m_{(a)};~\tilde{M}_c\big)} &\ge& 
{M^{(b)}_T} {\scriptstyle\big(\vec{p}^{~(b)}_T;~\vec{q}^{~(b)}_T
=\vec{q}^{~(b)}_{T(0)};~m_{(b)};~\tilde{M}_c\big)}  = m_{(b)}+ \tilde{M}_c , 
\label{eq:ubB}
\end{eqnarray} 
where 
\begin{equation}
\vec{q}^{~(\lambda)}_{T(0)}= \frac{\tilde{M}_c}{m_{(\lambda)}} \, \vec{p}^{~(\lambda)}_T,
\quad (\lambda=a,b),
\label{q0}
\end{equation}
gives the global (unconstrained) minimum of the corresponding 
parent transverse mass $M^{(\lambda)}_{T}$.
The unbalanced solution $M_{T2}^{Ua}$ applies when the condition
(\ref{eq:ubA}) is false and condition (\ref{eq:ubB}) is true, 
while the unbalanced solution $M_{T2}^{Ub}$ applies when the condition
(\ref{eq:ubA}) is true and condition (\ref{eq:ubB}) is false.
It is easy to see that conditions (\ref{eq:ubA}) and (\ref{eq:ubB})
cannot be simultaneously violated, so these three cases exhaust all
possibilities.

\subsection{Properties}
\label{subsec:mt2prop}

Given its definition (\ref{eq:mt2}), one can readily form and study the 
differential $M_{T2}$ distribution. Although its shape in general 
does carry some information about the underlying process,
it has become customary to focus on the upper endpoint $M_{T2(max)}$, 
which is simply the maximum value of $M_{T2}$ found in the event sample:
\begin{eqnarray}
M_{T2(max)}(\tilde M_c, P_{UTM})=\max_{all~
events} \Big [ M_{T2}{\scriptstyle\big( \vec{p}^{~(a)}_T,
\vec{p}^{~(b)}_T;~m_{(a)},m_{(b)};~\tilde{M}_c\big )} \Big ]  \, .
\label{mt2max}
\end{eqnarray}
Notice that in the process of maximizing over all events, the dependence 
on $\vec{p}^{~(a)}_T$, $\vec{p}^{~(b)}_T$, $m_{(a)}$ and $m_{(b)}$
disappears, and $M_{T2(max)}$ depends only on two input parameters:
$\tilde M_c$ and $P_{UTM}$, the latter entering through $\vec{Q}_{tot}$
in the momentum conservation constraint (\ref{momcon}).
The measured function (\ref{mt2max}) is the starting point of any
$M_{T2}$-based mass determination analysis. We shall now review
its three basic properties which make it suitable for such studies \cite{Konar:2009wn}.

\subsubsection{Property I: Knowledge of $M_p$ as a function of $M_c$}

This property was already identified in the original
papers and served as the main motivation for introducing 
the $M_{T2}$ variable in the first place \cite{Lester:1999tx,Barr:2003rg}. 
Mathematically it can be expressed as 
\begin{eqnarray}
\tilde M_p (\tilde{M}_c, P_{UTM}) \equiv M_{T2(max)}(\tilde M_c, P_{UTM}).
\label{Mptilde}
\end{eqnarray}
This is the same as eq.~(\ref{eq:Mptilde}), but now we have been careful to
include the explicit dependence on $P_{UTM}$, which will be important in our
subsequent discussion. As indicated in eq.~(\ref{Mptilde}), the function 
$\tilde M_p (\tilde{M}_c, P_{UTM})$ can be experimentally measured 
from the $M_{T2}$ endpoint (\ref{mt2max}). 
The crucial point now is that the relation (\ref{Mptilde}) is satisfied 
by the true values $M_p$ and $M_c$ of the parent and child mass,
correspondingly:
\begin{eqnarray}
M_p = M_{T2(max)} (M_c, P_{UTM}) \, .
\label{eq:mt2maxrel}
\end{eqnarray}
Notice that eq.~(\ref{eq:mt2maxrel}) holds for {\em any} value of $P_{UTM}$,
so in practical applications of this method one could choose the most 
populated $P_{UTM}$ bin to reduce the statistical error.
On the other hand, since a priori we do not know the true mass $M_c$ of the
missing particle, eq.~(\ref{eq:mt2maxrel}) gives only one relation 
between the masses of the mother and the child. This is illustrated in 
Fig.~\ref{fig:110ww}(a), where we consider the simple example of 
direct slepton pair production\footnote{The 
corresponding event topology is shown in Fig.~\ref{fig:dlsp}(a) 
below with $\maz=\mbz=M_c$.}, 
where each slepton ($\tilde \ell$) decays to 
the lightest neutralino ($\tilde\chi^0_1$)
by emitting a single lepton $\ell$:
$\tilde\ell\to \ell+\tilde\chi^0_1$. Here the slepton
is the parent and the neutralino is the child. 
Their masses were chosen to be $M_p=300$ GeV and $M_c=100$ GeV,
correspondingly, as indicated with the black dotted lines in Fig.~\ref{fig:110ww}(a).
In this example, the upstream transverse momentum $P_{UTM}$ 
is provided by jets from initial state radiation. 
In Fig.~\ref{fig:110ww}(a) we plot the function
(\ref{Mptilde}) versus $\tilde M_c$, for several fixed values of $P_{UTM}$.
The green solid line represents the case of no upstream momentum $P_{UTM}=0$.
In agreement with eq.~(\ref{eq:mt2maxrel}), this line
passes through the point $(M_c,M_p)$ corresponding to the 
true values of the mass parameters. Notice that the property 
(\ref{eq:mt2maxrel}) continues to hold for other values of 
$P_{UTM}$. Fig.~\ref{fig:110ww}(a) shows three
more cases: $P_{UTM}=500$ GeV (dotdashed black line),
$P_{UTM}=1$ TeV (dashed red line) and
$P_{UTM}=2$ TeV (dotted blue line).
All those curves still pass through the point $(M_c,M_p)$
with the correct values of the masses, illustrating the
robustness of the property (\ref{eq:mt2maxrel}) with
respect to variations in $P_{UTM}$.

%
\FIGURE[t]{
\centerline{
\epsfig{file=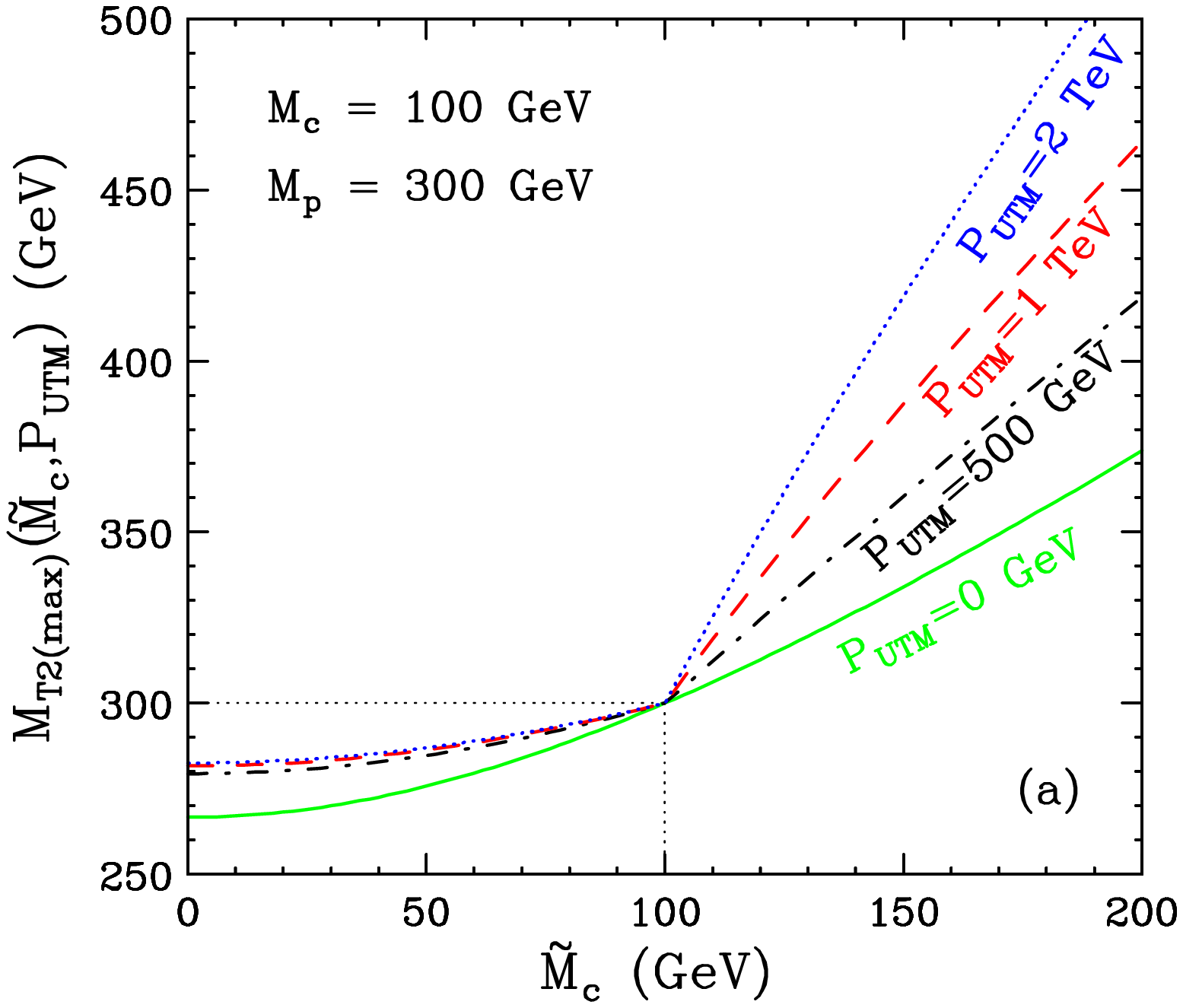,width=7.8cm} 
\hspace*{0.2cm}
\epsfig{file=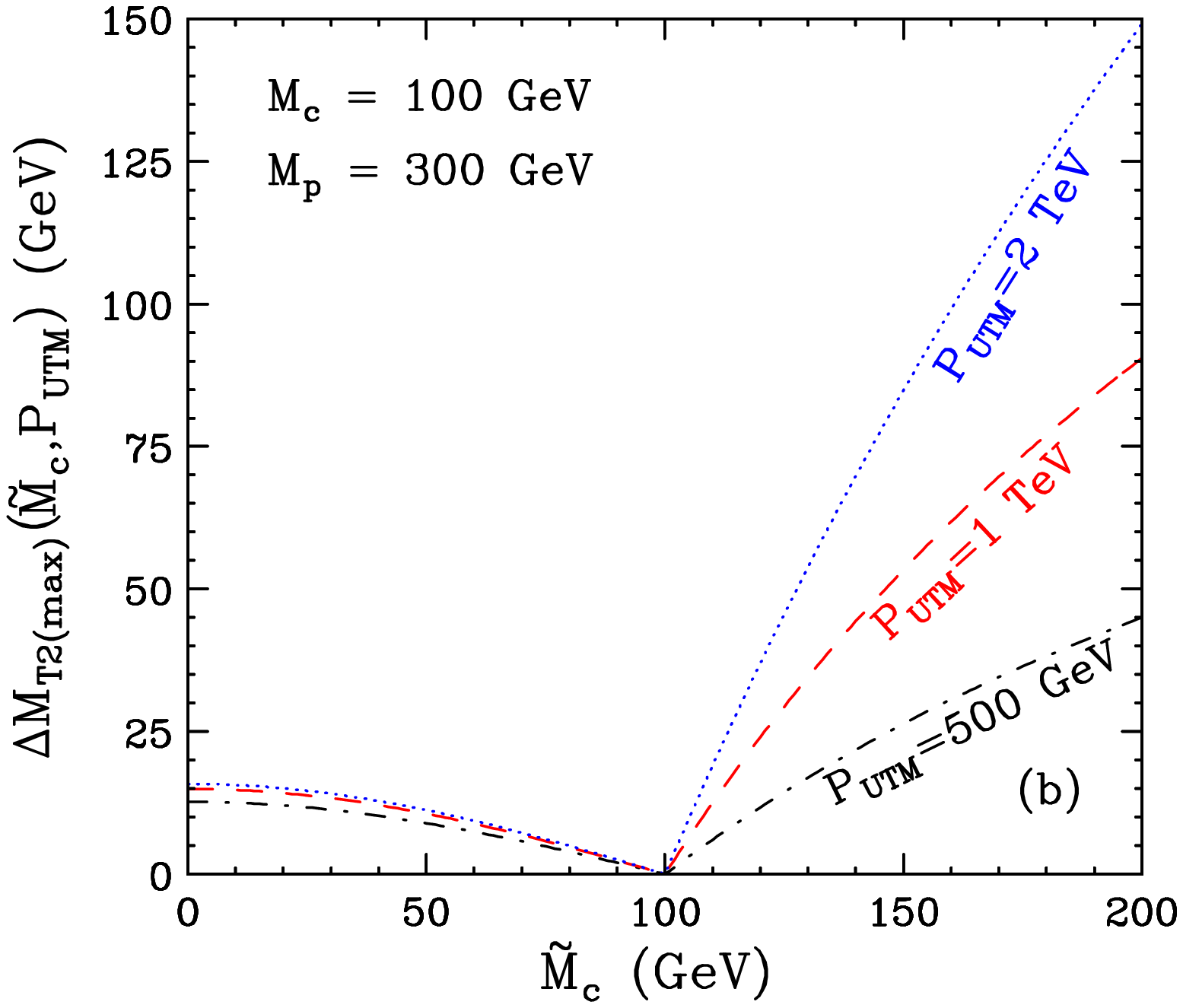,width=7.8cm}}
\caption{\it Plots of
(a) the $M_{T2}$ endpoint $M_{T2(max)}(\tilde{M}_c, P_{UTM})$ 
defined in eq.~(\ref{mt2max}), and 
(b) the function 
$\Delta M_{T2(max)}(\tilde{M}_c, P_{UTM})$ 
defined in (\ref{deltamt2def})
as a function of the test child mass $\tilde M_c$, 
for several fixed values of $P_{UTM}$:
$P_{UTM}=0$ GeV (solid, green),
$P_{UTM}=500$ GeV (dot-dashed, black),
$P_{UTM}=1$ TeV (dashed, red), and
$P_{UTM}=2$ TeV (dotted, blue).
The process under consideration is pair production
of sleptons of mass $M_p=300$ GeV, which decay 
directly to the lightest neutralino $\tilde\chi^0_1$ 
of mass $M_c=100$ GeV.}
\label{fig:110ww}
}

\subsubsection{Property II: Kink in $M_{T2(max)}$ at the true $M_c$}

The second important property of the $M_{T2}$ variable was identified 
rather recently 
\cite{Cho:2007dh,Gripaios:2007is,Cho:2007qv,Barr:2007hy,Burns:2008va}.
Interestingly, the $M_{T2}$ endpoint $M_{T2(max)}$, when considered 
as a function of the unknown input test mass $\tilde M_c$, 
often develops a kink (\ref{eq:kink})
at precisely the correct value $\tilde M_c=M_c$ of the child mass. 
The appearance of the kink is a rather general 
phenomenon and occurs under various circumstances.
It was originally noticed in event topologies with 
composite visible particles, whose invariant mass $m_{(\lambda)}$
is a variable parameter \cite{Cho:2007dh,Cho:2007qv}. 
Later it was realised that a kink also occurs in the
presence of non-zero upstream momentum $P_{UTM}$ 
\cite{Gripaios:2007is,Barr:2007hy,Burns:2008va}, 
as in the example of Fig.~\ref{fig:110ww}(a), where 
$P_{UTM}$ arises due to initial state radiation.
As can be seen in the figure, the kink is absent for
$P_{UTM}=0$, but as soon as there is some non-vanishing 
$P_{UTM}$, the kink becomes readily apparent.
As expected, the kink location (marked by the 
vertical dotted line) is at the true child mass
($M_c=100$ GeV), where the corresponding value of $M_{T2(max)}$ 
(marked by the horizontal dotted line) is at the 
true parent mass ($M_p=300$ GeV).
Fig.~\ref{fig:110ww}(a) also demonstrates that 
with the increase in $P_{UTM}$, the kink becomes more pronounced, 
thus the most favorable situations for the observation 
of the kink are cases with large $P_{UTM}$, e.g.
when the upstream momentum is due to the decays of
heavier (grandparent) particles \cite{Burns:2008va}.

In Sec.~\ref{sec:mt2dlspprop} we shall see how 
the kink feature (\ref{eq:kink}) of the symmetric $M_{T2}$ endpoint 
$\tilde M_p(\tilde M_c)$ defined by eq.~(\ref{eq:Mptilde})
is generalized to a ``ridge'' feature on the asymmetric $M_{T2}$ endpoint 
$\tilde M_p(\tilde M_c^{(a)},\tilde M_c^{(b)})$
defined in (\ref{eq:Mptildedlsp}).

\subsubsection{Property III: $P_{UTM}$ invariance of $M_{T2(max)}$ at the true $M_c$}
\label{sec:PUTM}

This property is the one which has been least emphasized in the literature.
Notice that the $M_{T2}$ endpoint function (\ref{Mptilde})
in general depends on the value of $P_{UTM}$. However, the first property
(\ref{eq:mt2maxrel}) implies that the $P_{UTM}$ dependence disappears at the
correct value $M_c$ of the child mass:
\begin{equation}
\frac{\partial M_{T2(max)} (\tilde{M}_c, P_{UTM})}
     {\partial P_{UTM}} \Big |_{\tilde{M}_c = M_c}= 0 \, .
\label{eq:isr_inv}
\end{equation}
In order to quantify this feature, 
let us define the function
\begin{equation}
\Delta M_{T2(max)}(\tilde{M}_c,P_{UTM}) \equiv 
M_{T2(max)}(\tilde{M}_c, P_{UTM}) - M_{T2(max)}(\tilde{M}_c, 0 ),
\label{deltamt2def}
\end{equation}
which measures the shift of the $M_{T2}$ endpoint
due to variations in $P_{UTM}$. The function 
$\Delta M_{T2(max)}(\tilde{M}_c,P_{UTM})$
can be measured experimentally: the first term 
on the right-hand side of (\ref{deltamt2def})
is simply the $M_{T2}$ endpoint observed in a subsample
of events with a given (preferably the most common) 
value of $P_{UTM}$, 
while the second term on the right-hand side of (\ref{deltamt2def})
contains the endpoint $M_{T2_\perp}^{(max)}$
of the 1-dimensional $M_{T2_\perp}$ variable 
introduced in \cite{Konar:2009wn}:
\begin{equation}
M_{T2(max)}(\tilde{M}_c, 0 ) = M_{T2_\perp}^{(max)}(\tilde{M}_c).
\end{equation}
Given the definition 
(\ref{deltamt2def}), the third property (\ref{eq:isr_inv}) 
can be rewritten as
\begin{equation}
\Delta M_{T2(max)}(\tilde M_c,P_{UTM}) \ge 0,
\label{deltamt2gt0}
\end{equation}
where the equality holds only for $\tilde M_c=M_c$:
\begin{equation}
\Delta M_{T2(max)}(M_c,P_{UTM}) = 0, ~\forall\, P_{UTM}.
\label{deltamt2eq0}
\end{equation}
Eqs.~(\ref{deltamt2gt0}) and (\ref{deltamt2eq0}) 
provide an alternative way to determine the true child mass $M_c$:
simply find the value of $\tilde M_c$ which 
minimizes the function $\Delta M_{T2(max)}(\tilde M_c,P_{UTM})$.
This procedure is illustrated in Fig.~\ref{fig:110ww}(b), where
we revisit the slepton pair production example of Fig.~\ref{fig:110ww}(a)
and plot the function $\Delta M_{T2(max)}(\tilde M_c,P_{UTM})$
defined in (\ref{deltamt2def}) versus the test mass
$\tilde{M}_c$, for the same set of (fixed) values of $P_{UTM}$.
Clearly, the zero of the function (\ref{deltamt2def}) occurs at
the true child mass $\tilde{M}_c= M_c=\,100$ GeV, in 
agreement with eq.~(\ref{deltamt2eq0}). In our studies 
of the asymmetric $M_{T2}$ case in the next sections, 
we shall find that the third property (\ref{deltamt2eq0})
is extremely important, since it will {\em always} allow us
the complete determination of the mass spectrum, 
including {\em both} children masses $M_c^{(a)}$ and $M_c^{(b)}$.

\section{The generalized asymmetric ${M_{T2}}$}
\label{sec:mt2dlsp}

After this short review of the basic properties of the 
conventional symmetric $M_{T2}$ variable (\ref{eq:mt2}), 
we now turn our attention to the less trivial case of 
$\tilde{M}^{(a)}_c\ne \tilde{M}^{(b)}_c$. Following the 
logic of Sec.~\ref{sec:mt2}, in Sec.~\ref{sec:mt2dlspdef}
we first introduce the asymmetric $M_{T2}$ variable 
and then in Secs.~\ref{sec:mt2dlspcomp} and
\ref{sec:mt2dlspprop} we discuss its computation and
mathematical properties, correspondingly.

\subsection{Definition}
\label{sec:mt2dlspdef}

The generalization of the usual definition (\ref{eq:mt2})
to the asymmetric case of $\tilde{M}^{(a)}_c\ne \tilde{M}^{(b)}_c$
is straightforward \cite{Barr:2009jv}. We continue to follow the conventions and 
notation of Fig.~\ref{fig:metevent}, but now we simply
avoid the assumption that the children masses are equal, 
and we let each one be an independent input parameter $\tilde M_c^{(\lambda)}$.
Without loss of generality, in what follows we assume 
$M_c^{(b)} \ge M_c^{(a)}$. The transverse mass of each 
parent (\ref{eq:MTp}) is now a function of the corresponding
child mass $\tilde M_c^{(\lambda)}$:
\begin{equation}
M_{T}^{(\lambda)}\big( \vec{p}^{~(\lambda)}_T;~ \vec{q}^{~(\lambda)}_T;~m_{(\lambda)};~\tilde{M}_c^{(\lambda)} \big ) 
= \sqrt{m_{(\lambda)}^2 + \left(\tilde{M}_c^{(\lambda)}\right)^2 + 2\left(e^{(\lambda)} \tilde{e}^{(\lambda)} 
                 -  \vec{p}^{~(\lambda)}_T \cdot \vec{q}^{~(\lambda)}_T  \right)} \, ,
\label{eq:MTpdlsp}
\end{equation}
where the transverse energy $e^{(\lambda)}$ 
of the composite SM particle on the $\lambda$-th side of the event 
was already defined in (\ref{elambda}), while the 
transverse energy $\tilde{e}^{(\lambda)}$ 
of the child is now generalized from (\ref{etilde}) to
\begin{equation}
\tilde{e}^{(\lambda)} =
\sqrt{\left(\tilde{M}_c^{(\lambda)}\right)^2 + \vec{q}^{~(\lambda)}_T\cdot
\vec{q}^{~(\lambda)}_T}\, .
\label{etildedlsp}
\end{equation} 
The event-by-event asymmetric $M_{T2}$ variable is defined 
in analogy to (\ref{eq:mt2}) and is given by \cite{Barr:2009jv}
\begin{eqnarray}
&&M_{T2}{\scriptstyle\big(\vec{p}^{~(a)}_T, \vec{p}^{~(b)}_T; m_{(a)}, m_{(b)};
        \tilde{M}_c^{(a)}, \tilde{M}_c^{(b)}, P_{UTM}\big)} =\nonumber\\
&&\min_{{\scriptscriptstyle\vec{q}^{~(a)}_T + \vec{q}^{~(b)}_T = \mptvec}}
{\Bigg[\max\left\{
    M_{T}^{(a)} {\scriptstyle \big( \vec{p}^{~(a)}_T;~ \vec{q}^{~(a)}_T;
                              ~m_{(a)};~\tilde{M}_c^{(a)} \big )},
    M_{T}^{(b)}  {\scriptstyle\big( \vec{p}^{~(b)}_T;~ \vec{q}^{~(b)}_T;
                              ~m_{(b)};~\tilde{M}_c^{(b)}\big )}
\right\} \Bigg]} \, ,
\label{eq:DLSPmt2} 
\end{eqnarray} 
which is now a function of two input test children masses $\tilde{M}_c^{(a)}$ and $\tilde{M}_c^{(b)}$.
In the special case of 
$\tilde{M}_c^{(a)}=\tilde{M}_c^{(b)}\equiv\tilde{M}_c$,
the asymmetric $M_{T2}$ variable defined in (\ref{eq:DLSPmt2})
reduces to the conventional symmetric $M_{T2}$ variable (\ref{eq:mt2}).

\subsection{Computation}
\label{sec:mt2dlspcomp}

In this subsection we generalize the discussion in Section~\ref{sec:mt2comp}
and present an analytical formula for computing 
the event-by-event asymmetric $M_{T2}$ variable (\ref{eq:DLSPmt2}).
Just like the formula (\ref{eq:mt2oldB}) for the symmetric case, 
our formula will hold only in the special case of $P_{UTM}=0$. 
As before, the asymmetric $M_{T2}$ variable has two
types of solutions -- balanced and unbalanced.
The balanced solution occurs when the following two conditions are
simultaneously satisfied (compare to the analogous conditions 
(\ref{eq:ubA}) and (\ref{eq:ubB}) for the symmetric case)
\begin{eqnarray}
&&{M^{(b)}_T}{\scriptstyle\big(\vec{p}^{~(b)}_T;\vec{q}^{~(b)}_T
    =-\vec{q}^{~(a)}_{T(0)} +~\mptvec;m_{(b)};\tilde{M}_c^{(b)}\big)} \ge 
{M^{(a)}_T} {\scriptstyle\big(\vec{p}^{~(a)}_T;\vec{q}^{~(a)}_T
    =\vec{q}^{~(a)}_{T(0)};m_{(a)},\tilde{M}_c^{(a)}\big)} 
 = m_{(a)} + \tilde{M}_c^{(a)} \, ,
\label{eq:DubA}
\\ [2mm]
&&{M^{(a)}_T}{\scriptstyle\big(\vec{p}^{~(a)}_T;\vec{q}^{~(a)}_T
    =-\vec{q}^{~(b)}_{T(0)} +~\mptvec;m_{(a)};\tilde{M}_c^{(a)}\big)} \geq 
{M^{(b)}_T} {\scriptstyle\big(\vec{p}^{~(b)}_T;\vec{q}^{~(b)}_T=
    \vec{q}^{~(b)}_{T(0)};m_{(b)};\tilde{M}_c^{(b)}\big)}  
= m_{(b)}+ \tilde{M}_c^{(b)} \, ,~~~~~
\label{eq:DubB}
\end{eqnarray} 
where, in analogy to (\ref{q0}), 
\begin{equation}
\vec{q}^{~(\lambda)}_{T(0)}= \frac{\tilde{M}_{c}^{(\lambda)}}{m_{(\lambda)}}
\, \vec{p}^{~(\lambda)}_T,
\quad (\lambda=a,b),
\label{q0dlsp}
\end{equation}
is the test child momentum at the  
global unconstrained minimum of $M^{(\lambda)}_{T}$. 
The balanced solution for $M_{T2}$ is now given by
\begin{eqnarray}  
&&\Big[M_{T2}^{B}{\scriptstyle\big(\vec{p}^{~(a)}_T,\vec{p}^{~(b)}_T;
      ~m_{(a)},m_{(b)};~\tilde{M}_c^{(a)},\tilde{M}_c^{(b)}\big)}\Big]^{2}=\tilde{M}_+^2 + A_T 
            + \left( \frac{m_{(b)}^2-m_{(a)}^2}{2A_T-m_{(a)}^2-m_{(b)}^2}\right) 
       \tilde{M}_-^2 \nonumber\\
&& \qquad\quad\pm\, \sqrt{1 + \frac{4\tilde{M}_+^2}{2A_T-m_{(a)}^2-m_{(b)}^2} 
                     + \left(\frac{2\tilde{M}_-^2}{2A_T-m_{(a)}^2-m_{(b)}^2}\right)^2}
              \times \sqrt{A_T^2 - m_{(a)}^2 m_{(b)}^2} \, , ~~~~~
\label{mt2b}
\end{eqnarray}
where $A_T$ was defined in (\ref{ATdef}).
For convenience, in (\ref{mt2b}) we have introduced two alternative mass parameters
\begin{eqnarray}
\tilde{M}_+^2 &\equiv& \frac{1}{2}\Big\{\tmbsq + \tmasq\Big\}  \, , \label{eq:m+} \\
\tilde{M}_-^2 &\equiv& \frac{1}{2}\Big\{\tmbsq - \tmasq\Big\}  \, , \label{eq:m-}
\end{eqnarray}
in place of the original trial masses $\tma$ and $\tmb$.
The new parameters $\tilde{M}_{+}$ and $\tilde{M}_{-}$ 
are simply a different parametrization of the two degrees of freedom
corresponding to the unknown child masses $\tilde{M}_c^{(a)}$ and 
$\tilde{M}_c^{(b)}$ entering the definition of the asymmetric $M_{T2}$. 
The parameters $\tilde{M}_{+}$ and $\tilde{M}_{-}$ 
allow us to write formula (\ref{mt2b}) in a more compact form.
More importantly, they also allow to make easy contact with the 
known results from Section~\ref{sec:mt2} by taking the symmetric limit 
$\tilde{M}_c^{(a)}=\tilde{M}_c^{(b)}\equiv\tilde{M}_c$ as
\begin{equation}
\tilde{M}_+ \to \tilde{M}_c,  \quad
\tilde{M}_- \to 0.
\label{symlimit}
\end{equation}
It is easy to see that in the symmetric limit (\ref{symlimit})
our balanced solution (\ref{mt2b}) for the asymmetric $M_{T2}$ 
reduces to the known result (\ref{eq:mt2oldB}) for the symmetric $M_{T2}$.

An interesting feature of the asymmetric balanced solution 
is the appearance of a $\pm$ sign on the second line of (\ref{mt2b}).
In principle, this sign ambiguity is present in the 
symmetric case as well, but there the minus sign 
always turns out to be unphysical and the sign issue does not
arise \cite{Lester:2007fq}. However, in the asymmetric case, 
both signs can be physical sometimes and one must make the proper
sign choice in eq.~(\ref{mt2b}) as follows.
For the given set of test masses $(\tilde M_c^{(a)},\tilde M_c^{(b)})$, 
calculate the transverse center-of-mass energy
\begin{eqnarray}
\sqrt{\hat{s}}_{\scriptscriptstyle T}^{\,\pm} = 
e^{(a)} + e^{(b)} + && \frac{2 (e^{(b)} - e^{(a)}) \tilde{M}_-^2}{2 A_T - m^2_{(a)} -m^2_{(b)}} \pm 
\frac{(e^{(b)} + e^{(a)}) A_T- (e^{(b)}m^2_{(a)} + e^{(a)}m^2_{(b)})}{\sqrt{ A_T^2 - m^2_{(a)} m^2_{(b)}}} \nonumber \\
\times &&\sqrt{1 + \frac{4 \tilde{M}_+^2}{2 A_T - m^2_{(a)} -m^2_{(b)}} +  
               \left ( \frac{2 \tilde{M}_-^2}{2A_T - m^2_{(a)} -m^2_{(b)}}  \right )^2 } \, ,
\end{eqnarray}
corresponding to each sign choice in eq.~(\ref{mt2b}), and compare the 
result to the minimum allowed value of $\sqrt{\hat s}_{\scriptscriptstyle T}$
\begin{eqnarray}
\sqrt{\hat{s}}_{{\scriptscriptstyle T}(min)} = e^{(a)}+e^{(b)} +
\sqrt{Q_{tot}^2 + \left(\tilde{M}_c^{(a)} + \tilde{M}_c^{(b)}\right)^2}  \, .
\end{eqnarray}
The minus sign in eq.~(\ref{mt2b}) takes precedence and applies whenever it is physical, i.e.
whenever $\sqrt{\hat{s}}_{\scriptscriptstyle T}^{-} \,> \sqrt{\hat{s}}_{{\scriptscriptstyle T}(min)}$.
In the remaining cases when $\sqrt{\hat{s}}_{\scriptscriptstyle T}^{-} \,< \sqrt{\hat{s}}_{{\scriptscriptstyle T}(min)}$
and the minus sign is unphysical, the plus sign in eq.~(\ref{mt2b}) applies.

If one of the conditions (\ref{eq:DubA}), (\ref{eq:DubB}) is not
satisfied, the asymmetric $M_{T2}$ is given by an unbalanced solution,
in analogy to (\ref{mt2U1}) and (\ref{mt2U2}):
\begin{eqnarray}
M_{T2}^{Ua}{\scriptstyle\big(\vec{p}^{~(a)}_T,\vec{p}^{~(b)}_T;
           ~m_{(a)},m_{(b)};~\tilde{M}_c^{(a)},\tilde{M}_c^{(b)}\big)} 
&=& m_{(a)} + \tma 
\, ,  \label{mt2u1} 
\\[2mm]
M_{T2}^{Ub}{\scriptstyle\big(\vec{p}^{~(a)}_T,\vec{p}^{~(b)}_T;
           ~m_{(a)},m_{(b)};~\tilde{M}_c^{(a)},\tilde{M}_c^{(b)}\big)} 
&=& m_{(b)} + \tmb 
\, .
\label{mt2u2}  
\end{eqnarray}
The unbalanced solution $M_{T2}^{Ua}$ of eq.~(\ref{mt2u1}) 
applies when the condition
(\ref{eq:DubA}) is false and condition (\ref{eq:DubB}) is true, 
while the unbalanced solution $M_{T2}^{Ub}$ of eq.~(\ref{mt2u2}) 
applies when the condition
(\ref{eq:DubA}) is true and condition (\ref{eq:DubB}) is false.

Eqs.~(\ref{mt2b}), (\ref{mt2u1}) and (\ref{mt2u2}) 
represent one of our main results. They generalize the analytical
results of Refs.~\cite{Lester:2007fq,Cho:2007dh} and allow 
the direct computation of the asymmetric $M_{T2}$ variable
without the need for scanning and numerical minimizations.
This is an important benefit, since the existing public codes 
for $M_{T2}$ \cite{Lester_code,Davis_code} only apply in the
symmetric case $M_c^{(a)}=M_c^{(b)}$.

\subsection{Properties} 
\label{sec:mt2dlspprop}

All three properties of the symmetric $M_{T2}$ discussed in 
Section~\ref{subsec:mt2prop} readily generalize to the asymmetric case.

\subsubsection{Property I: Knowledge of $M_p$ as a function of $M_c^{(a)}$ and $M_c^{(b)}$}
\label{sec:mt2dlspprop1}

In the asymmetric case, the endpoint $M_{T2(max)}$
of the $M_{T2}$ distribution still gives the mass of the parent, 
only this time it is a function of two input test masses
for the children:
\begin{equation}
\tilde M_p(\tilde M_c^{(a)},\tilde M_c^{(b)},P_{UTM}) =
M_{T2(max)}(\tilde M_c^{(a)},\tilde M_c^{(b)}, P_{UTM})\, .
\label{mt2maxdlsp}
\end{equation}
The important property is that this
relation is satisfied by the true
values of the children and parent masses:
\begin{equation}
M_p = M_{T2(max)}(M_c^{(a)},M_c^{(b)}, P_{UTM})\, .
\label{Mptruedlsp}
\end{equation}
Thus the true parent mass $M_p$ will be known once we determine 
the two children masses $M_c^{(a)}$ and $M_c^{(b)}$.

\subsubsection{Property II: Ridge in $M_{T2(max)}$ through the true $M_c^{(a)}$ and $M_c^{(b)}$}
\label{sec:kinkdlsp}

In the symmetric $M_{T2}$ case, the endpoint function 
(\ref{Mptilde}) is not continuously differentiable and 
has a ``kink'' at the true child mass $\tilde M_c=M_c$.
In the asymmetric $M_{T2}$ case, the endpoint function 
(\ref{mt2maxdlsp}) is similarly non-differentiable at a 
set of points 
\begin{equation}
\left\{\left(\tilde M_c^{(a)}(\theta), \tilde M_c^{(b)} (\theta)\right)\right\}
\label{ridge}
\end{equation}
parametrized by a single continuous parameter $\theta$.
The gradient of the endpoint function (\ref{mt2maxdlsp})
suffers a discontinuity as we cross the curve defined by (\ref{ridge}).
Since (\ref{mt2maxdlsp}) represents a hypersurface 
in the three-dimensional parameter space of 
$\{\tilde M_c^{(a)},\tilde M_c^{(b)},\tilde M_p\}$,
the gradient discontinuity will appear as a ``ridge''
(sometimes also referred to as a ``crease'' \cite{Barr:2009jv})
on our three-dimensional plots below. 
The important property of the ridge is that it passes through
the correct values for the children masses, even when they 
are different:
\begin{eqnarray}
M_c^{(a)}&=&\tilde M_c^{(a)}(\theta_0), \label{thetaa} \\ [2mm]
M_c^{(b)}&=&\tilde M_c^{(b)}(\theta_0), \label{thetab} 
\end{eqnarray}
for some $\theta_0$. Thus the ridge information provides
a relation among the two children masses and leaves us with
just a single unknown degree of freedom --- 
the parameter $\theta$ in eq.~(\ref{ridge}).

Interestingly, the shape of the ridge provides a quick test whether the two
missing particles are identical  or not\footnote{To be more precise, 
the ridge shape tests whether the two missing particles have the same 
mass or not.}. If the shape of the ridge in the 
$(\tilde M_c^{(a)},\tilde M_c^{(b)})$ plane
is symmetric with respect to the interchange 
$\tilde M_c^{(a)} \leftrightarrow \tilde M_c^{(b)}$,
i.e. under a mirror reflection with respect to the 
$45^\circ$ line $\tilde M_c^{(a)} = \tilde M_c^{(b)}$,
then the two missing particles are the same.
Conversely, when the shape of the ridge is {\em not} symmetric under 
$\tilde M_c^{(a)} \leftrightarrow \tilde M_c^{(b)}$, the
missing particles are in general expected to have different masses.

\subsubsection{Property III: $P_{UTM}$ invariance of $M_{T2(max)}$ at the true $M_c^{(a)}$ and $M_c^{(b)}$}
\label{sec:PUTMdlsp}

The third $M_{T2}$ property, which was discussed in Section~\ref{sec:PUTM},
is readily generalized to the asymmetric case as well. Note that
eq.~(\ref{Mptruedlsp}) implies that the $P_{UTM}$ dependence 
of the asymmetric $M_{T2}$ endpoint (\ref{mt2maxdlsp})
disappears at the true values of the children masses:
\begin{equation}
\frac{\partial M_{T2(max)} (\tilde{M}_c^{(a)}, \tilde{M}_c^{(b)},P_{UTM})}
     {\partial P_{UTM}} \Big |_{  \tilde{M}_c^{(a)} = M_c^{(a)}, \tilde{M}_c^{(b)} = M_c^{(b)}}= 0 \, .
\label{eq:isr_inv2}
\end{equation}
This equation is the asymmetric analogue of eq.~(\ref{eq:isr_inv}).
Proceeding as in Sec.~\ref{sec:PUTM}, let us define the function
\begin{equation}
\Delta M_{T2(max)}(\tilde{M}_c^{(a)},\tilde{M}_c^{(b)},P_{UTM}) \equiv 
  M_{T2(max)}(\tilde{M}_c^{(a)},\tilde{M}_c^{(b)}, P_{UTM}) 
- M_{T2(max)}(\tilde{M}_c^{(a)},\tilde{M}_c^{(b)}, 0 ),
\label{deltamt2defdlsp}
\end{equation}
which quantifies the shift of the asymmetric $M_{T2}$ endpoint (\ref{mt2maxdlsp})
in the presence of $P_{UTM}$. By definition,
\begin{equation}
\Delta M_{T2(max)}(\tilde M_c^{(a)},\tilde{M}_c^{(b)},P_{UTM}) \ge 0,
\label{deltamt2gt0dlsp}
\end{equation}
with equality being achieved only for the correct values of the 
children masses:
\begin{equation}
\Delta M_{T2(max)}(M_c^{(a)},M_c^{(b)},P_{UTM}) = 0, ~\forall\, P_{UTM}.
\label{deltamt2eq0dlsp}
\end{equation}
The last equation reveals the power of the $P_{UTM}$ invariance method.
Unlike the kink method discussed in Sec.~\ref{sec:kinkdlsp}, which was only
able to find a relation between the two children masses 
$M_c^{(a)}$ and $M_c^{(b)}$, the $P_{UTM}$ invariance 
implied by eq.~(\ref{deltamt2eq0dlsp}) allows us to determine 
{\em each} individual children mass, without any theoretical assumptions, and 
even in the case when the two children masses happen to be different
($M_c^{(a)}\ne M_c^{(b)}$).

\subsection{Examples}
\label{sec:examples}

In the next two sections we shall illustrate the three properties discussed 
so far in Section~\ref{sec:mt2dlspprop} with some concrete examples.
Instead of the most general event topology depicted Fig.~\ref{fig:metevent},
here we limit ourselves to the three simple examples shown in Fig.~\ref{fig:dlsp}.
%
\FIGURE[ht]{
{
\unitlength= 1.0 pt
\SetScale{1.0}
\SetWidth{1.0}      
\normalsize    
\begin{picture}(125,100)(55,0)
\SetColor{Gray}
\Line(52,50)(52,90)
\Line(48,50)(48,90)
\Line(52,90)(56,90)
\Line(48,90)(44,90)
\Line(56,90)(50,97)
\Line(44,90)(50,97)
\SetWidth{1}      
\Line( 90,65)(110,95)
\Line( 90,35)(110, 5)
\SetColor{Red}
\Text(  77, 74)[c]{\Red{$M_p$}}
\Text(  77, 44)[c]{\Red{$M_p$}}
\Text( 115, 74)[c]{\Red{$M_{c}^{(a)}$}}
\Text( 115, 44)[c]{\Red{$M_{c}^{(b)}$}}
\Text(112,104)[c]{\Black{$m_{(a)}=0$}}
\Text(112, -2)[c]{\Black{$m_{(b)}=0$}}
\SetWidth{1.2}      
\Line(50,65)(90,65)
\Line(50,35)(90,35)
\DashLine(90,65)(120,65){2}
\DashLine(90,35)(120,35){2}
\Text( 74,109)[r]{\Black{$\vec{P}_{UTM}$}}
\COval(50,50)(25,10)(0){Blue}{Green}
\SetWidth{1.0}      
\COval(90,65)(2,2)(0){Blue}{Green}
\COval(90,35)(2,2)(0){Blue}{Green}
\Text(55, -5)[r]{\bf \Black{(a)}}
\end{picture}\
{} \qquad\allowbreak
\begin{picture}(120,100)(70,0)
\SetColor{Gray}
\SetWidth{1}      
\Line( 100,65)(80,95)
\Line( 100,65)(120,95)
\Line( 100,35)(120, 5)
\Line( 100,35)(80, 5)
\SetColor{Red}
\Text(  77, 74)[c]{\Red{$M_p$}}
\Text(  77, 44)[c]{\Red{$M_p$}}
\Text( 140, 74)[c]{\Red{$M_{c}^{(a)}$}}
\Text( 140, 44)[c]{\Red{$M_{c}^{(b)}$}}
\SetWidth{1.2}      
\Line(50,65)(100,65)
\Line(50,35)(100,35)
\DashLine(100,65)(140,65){2}
\DashLine(100,35)(140,35){2}
%
\Text(112,104)[r]{\Black{$m_{(a)}$}}
\Text(112, -2)[r]{\Black{$m_{(b)}$}}
\COval(50,50)(25,10)(0){Blue}{Green}
\SetWidth{1.0}      
\COval(100,65)(2,2)(0){Blue}{Green}
\COval(100,35)(2,2)(0){Blue}{Green}
\Text(55, -5)[r]{\bf \Black{(b)}}
\end{picture}\
{} \qquad\allowbreak
\begin{picture}(120,100)(60,0)
\SetColor{Gray}
\SetWidth{1}      
\Line( 120,65)(140,95)
\Line( 90,65)(110,95)
\Line( 90,35)(110, 5)
\Line( 120,35)(140, 5)
\SetColor{Red}
\Text(  77, 74)[c]{\Red{$M_p$}}
\Text(  77, 44)[c]{\Red{$M_p$}}
\Text( 110, 74)[c]{\Red{$M_{i}^{(a)}$}}
\Text( 110, 44)[c]{\Red{$M_{i}^{(b)}$}}
\Text( 145, 74)[c]{\Red{$M_{c}^{(a)}$}}
\Text( 145, 44)[c]{\Red{$M_{c}^{(b)}$}}
\SetWidth{1.2}      
\Line(50,65)(90,65)
\Line(50,35)(90,35)
\Line(90,65)(120,65)
\Line(90,35)(120,35)
\DashLine(120,65)(150,65){2}
\DashLine(120,35)(150,35){2}
\Text(135,104)[r]{\Black{$m_{(a)}$}}
\Text(135,-2)[r]{\Black{$m_{(b)}$}}
\COval(50,50)(25,10)(0){Blue}{Green}
\SetWidth{1.0}      
\COval(90,65)(2,2)(0){Blue}{Green}
\COval(90,35)(2,2)(0){Blue}{Green}
\COval(120,65)(2,2)(0){Blue}{Green}
\COval(120,35)(2,2)(0){Blue}{Green}
\Text(55, -5)[r]{\bf \Black{(c)}}
\end{picture}
}
\caption{\it The three different event-topologies under 
consideration in this paper. In each case, two parents with mass $M_p$ 
are produced onshell and decay into two daughters of (generally different)
masses $M_c^{(a)}$ and $M_c^{(b)}$. Case (a), which is the subject of
Section~\ref{sec:110}, has a single massless visible SM particle in each leg 
and some arbitrary upstream transverse momentum $\vec{P}_{UTM}$.  
In the remaining two cases (b) and (c), which are discussed in Section~\ref{sec:220},
there are two massless visible particles in each leg, which form a composite
visible particle with varying invariant mass $m_{(\lambda)}$.  
The intermediate particle of mass $M_i^{(\lambda)}$ 
is (b) heavy and off-shell ($M_i^{(\lambda)}>M_p$), or
(c) on-shell ($M_p > M_i^{(\lambda)}>M_c^{(\lambda)}$). For simplicity,
we do not consider any upstream momentum in cases (b) and (c).}
\label{fig:dlsp} 
}

The simplest possible case is when $n^{(\lambda)}=1$, i.e.~when 
each cascade decay contains a single SM particle, as in 
Fig.~\ref{fig:dlsp}(a). In this example, $m_{(\lambda)}$ is constant.
For simplicity, we shall take $m_{(\lambda)}\approx 0$, which is the case
for a lepton or a light flavor jet. If the SM particle is a $Z$-boson 
or a top quark, its mass cannot be neglected, and one must 
keep the proper value of $m_{(\lambda)}$. This, however, is only a technical
detail, which does not affect our main conclusions below. 
In spite of its simplicity, the topology of Fig.~\ref{fig:dlsp}(a) 
is actually the most challenging case, due to the limited number 
of available measurements \cite{Burns:2008va}. In order to be able 
to determine all individual masses in that case, one must consider
events with upstream momentum $\vec{P}_{UTM}$, as 
illustrated in Fig.~\ref{fig:dlsp}(a). 
This is not a particularly restrictive assumption, since 
there is always a certain amount of $P_{UTM}$ in the event
(at the very least, from initial state radiation).
In Section~\ref{sec:110} the topology of Fig.~\ref{fig:dlsp}(a)
will be extensively studied - first for the asymmetric case of
$M_c^{(a)}\ne M_c^{(b)}$ in Sec.~\ref{subsec:110DLSP}, and then for the
symmetric case of $M_c^{(a)}=M_c^{(b)}$ in Sec.~\ref{subsec:110ELSP}.

Another simple situation arises when there are {\em two}
massless visible SM particles in each leg,
as illustrated in Figs.~\ref{fig:dlsp}(b) and \ref{fig:dlsp}(c).
In either case, the invariant mass $m_{(\lambda)}$ is not constant any more, 
but varies within a certain range $m_{(\lambda)}^{min}\le m_{(\lambda)}\le m_{(\lambda)}^{max}$,
where $m_{(\lambda)}^{min}=0$, while the value of $m_{(\lambda)}^{max}$
depends on the mass $M_i^{(\lambda)}$ of the corresponding intermediate particle. 
In Fig.~\ref{fig:dlsp}(b) we assume $M_i^{(\lambda)}>M_p$, so that the intermediate
particle is off-shell and 
\begin{equation}
m_{(\lambda)}^{max}=M_p-M_c^{(\lambda)}\, .
\label{eq:mmaxoff}
\end{equation}
The ``off-shell'' case of Fig.~\ref{fig:dlsp}(b) 
will be discussed in Sec.~\ref{sec:220off}.

In contrast, in Fig.~\ref{fig:dlsp}(c) we take $M_p > M_i^{(\lambda)}>M_c^{(\lambda)}$,
in which case the intermediate particle is on-shell and the range for $m_{(\lambda)}$ is
now limited from above by 
\begin{equation}
m_{(\lambda)}^{max}=
M_p \,
\sqrt{\left[1-\left(\frac{M_i^{(\lambda)}}{M_p             }\right)^2 \right]
      \left[1-\left(\frac{M_c^{(\lambda)}}{ M_i^{(\lambda)}}\right)^2 \right]}\, .
\label{eq:mmaxon}
\end{equation}
We shall discuss the ``on-shell'' case of Fig.~\ref{fig:dlsp}(c) 
in Sec.~\ref{sec:220on}.

In the event topologies of Figs.~\ref{fig:dlsp}(b) 
and \ref{fig:dlsp}(c), the mass $m_{(\lambda)}$ is varying and the 
ridge of eq.~(\ref{ridge}) will appear even if there were no 
upstream transverse momentum in the event. Therefore, 
in our discussion of Figs.~\ref{fig:dlsp}(b) and \ref{fig:dlsp}(c)
in Sec.~\ref{sec:220} below we shall assume $P_{UTM}=0$
for simplicity. The presence of non-zero $P_{UTM}$ will only
additionally enhance the ridge feature.


\subsection{Combinatorial issues}
\label{sec:matching}

Before going on to the actual examples in the next two sections, we 
need to discuss one minor complication, which is unique to the asymmetric 
$M_{T2}$ variable and was not present in the case of the symmetric $M_{T2}$
variable. The question is, how does one associate the visible decay products
observed in the detector with a particular decay chain $\lambda=a$
or $\lambda=b$. This is the usual combinatorics problem, which now has two
different aspects:
\begin{itemize}
\item The first issue is also present in the symmetric case, where
one has to decide how to partition the SM particles observed in the detector
into two disjoint sets, one for each cascade. In the traditional 
approach, where the children particles are assumed to be identical,
the two sets are indistinguishable and it does not matter which one is 
first and which one is second. This particular aspect of the combinatorial
problem will also be present in the asymmetric case. 
\item In the asymmetric case, however, there is an additional aspect to
the combinatorial problem: now the two cascades are distinguishable 
(by the masses of the child particles), so even if we correctly divide the 
visible objects into the proper subsets, we still do not know which subset
goes together with $M_c^{(a)}$ and thus gets a label $\lambda=a$, 
and which goes together with $M_c^{(b)}$ and gets labelled by $\lambda=b$.
This leads to an additional combinatorial factor of 2 which is absent
in the symmetric case with identical children. 
\end{itemize}

The severity of these two combinatorial problems depends on the event topology, 
as well as the type of signature objects. For example, there are cases where
the first combinatorial problem is easily resolved, or even absent altogether.
Consider the event topology of Fig.~\ref{fig:dlsp}(a) with a lepton as the 
SM particle on each side. In this case, the partition is unique, and 
the upstream objects are jets, which can be easily identified \cite{Matchev:2009fh}.
Now consider the event topologies of Figs.~\ref{fig:dlsp}(b) and 
\ref{fig:dlsp}(c), with two opposite sign, same flavor leptons on each side.
Such events result from inclusive pair production of heavier neutralinos 
in supersymmetry. By selecting events with different lepton flavors:
$e^+e^-\mu^+\mu^-$, we can overcome the first combinatorial problem above 
and uniquely associate the $e^+e^-$ pair with one cascade and the
$\mu^+\mu^-$ pair with the other. However, the second combinatorial problem 
remains, as we still have to decide which of the two lepton pairs to 
associate with $\lambda=a$ and which to associate with $\lambda=b$.
Recall that the labels $\lambda=a$ and $\lambda=b$ are already attached
to the child particles, which are distinguishable in the asymmetric case.
In this paper we use the convention that $\lambda=a$ is attached to the 
lighter child particle:
\begin{equation}
\tilde M_c^{(a)}\le\tilde M_c^{(b)}\, ,
\label{abconv}
\end{equation}
which also ensures that the $\tilde{M}_-$ parameter defined in (\ref{eq:m-}) is real.

We can put this discussion in more formal terms as follows.
The correct association of the visible particles with the 
corresponding children will yield 
\begin{equation}
M_{T2}{\scriptstyle\big(\vec{p}^{~(a)}_T,\vec{p}^{~(b)}_T;m_{(a)},m_{(b)};
        \tilde{M}_c^{(a)},\tilde{M}_c^{(b)}\big)},
\label{mt2correct}
\end{equation}
while the other, wrong association will give simply
\begin{equation}
M_{T2}{\scriptstyle\big(\vec{p}^{~(a)}_T,\vec{p}^{~(b)}_T;m_{(a)},m_{(b)};
        \tilde{M}_c^{(b)},\tilde{M}_c^{(a)}\big)}.
\label{mt2wrong}
\end{equation}
Both of these two $M_{T2}$ values can be computed from the data,
but a priori we do not know which one corresponds to the correct association.
The solution to this problem is however already known \cite{Lester:2007fq,Burns:2008va}:
one can conservatively use the smaller of the two
\begin{equation}
M_{T2}^{(<)}\equiv \min 
\left\{ M_{T2}{\scriptstyle\big(\vec{p}^{~(a)}_T,\vec{p}^{~(b)}_T;m_{(a)},m_{(b)};
        \tilde{M}_c^{(a)},\tilde{M}_c^{(b)}\big)},
M_{T2}{\scriptstyle\big(\vec{p}^{~(a)}_T,\vec{p}^{~(b)}_T;m_{(a)},m_{(b)};
        \tilde{M}_c^{(b)},\tilde{M}_c^{(a)}\big)}
\right\}
\label{mt2<}
\end{equation}
in order to preserve the location of the upper $M_{T2}$ endpoint.
\FIGURE[ht]{
\centerline{
\epsfig{file=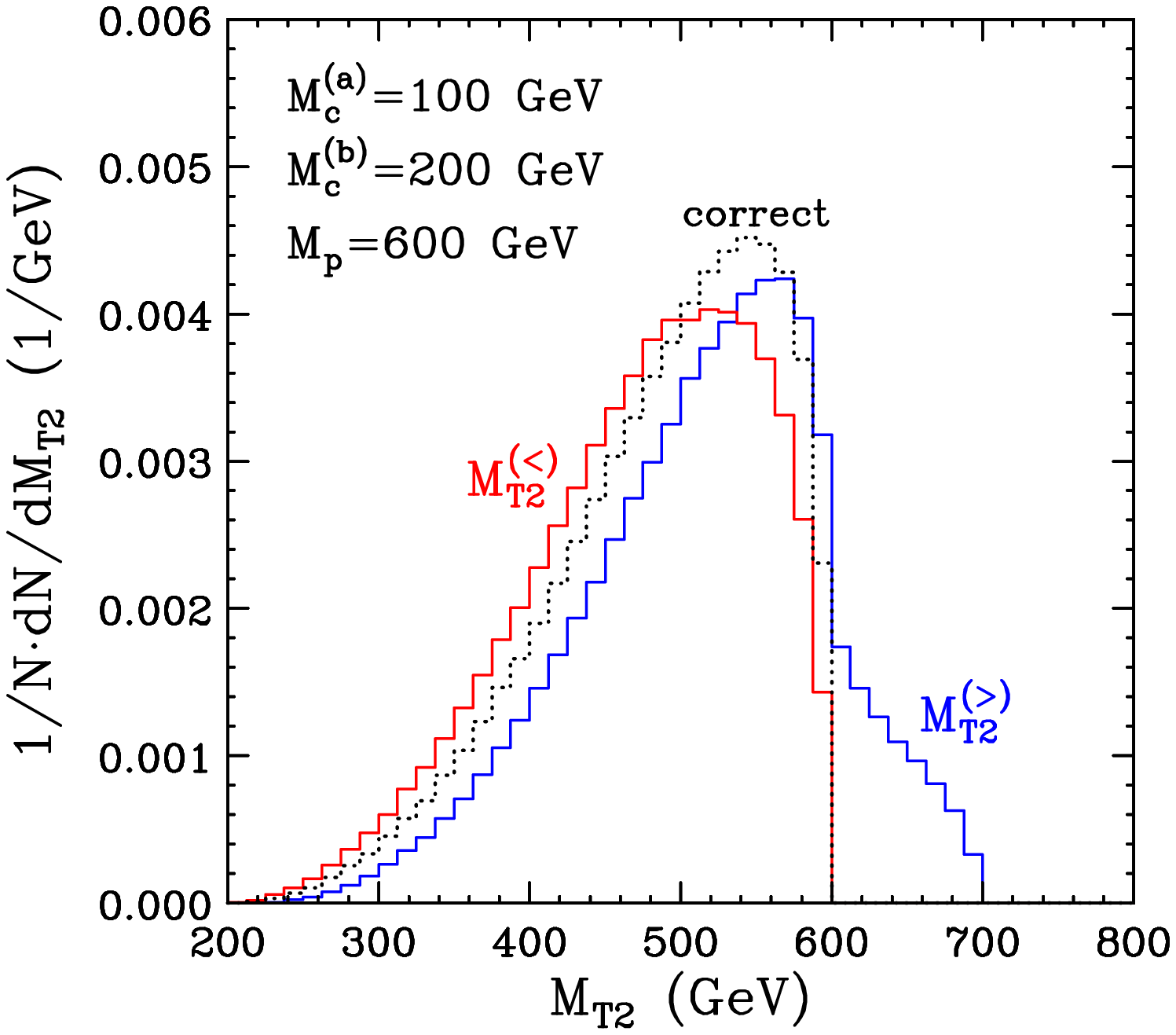,width=9cm} }
\caption{\it Unit-normalized $M_{T2}$
distributions for the event topology of Fig.~\ref{fig:dlsp}(b).
The mass spectrum is chosen as $M_c^{(a)}=100$ GeV,
$M_c^{(b)}=200$ GeV and $M_p=600$ GeV.
The test children masses are taken to be the true masses:
$\tilde M_c^{(a)}=M_c^{(a)}$ and 
$\tilde M_c^{(b)}=M_c^{(b)}$.  
The dotted black distribution is the true $M_{T2}$ distribution,
ignoring the combinatorial problem.
The red histogram shows the distribution
of the $M_{T2}^{(<)}$ variable defined in (\ref{mt2<})
while the blue histogram shows the distribution
of the $M_{T2}^{(>)}$ variable defined in (\ref{mt2>}).}
\label{fig:offshellpairing}
}
This is illustrated in Fig.~\ref{fig:offshellpairing}, where
we show results for the event topology of Fig.~\ref{fig:dlsp}(b)
with a mass spectrum as follows: $M_c^{(a)}=100$ GeV,
$M_c^{(b)}=200$ GeV and $M_p=600$ GeV.
The test children masses are taken to be the true masses:
$\tilde M_c^{(a)}=M_c^{(a)}$ and 
$\tilde M_c^{(b)}=M_c^{(b)}$.  
The dotted black distribution is the unit-normalized 
true $M_{T2}$ distribution, where one ignores the combinatorial problem
and uses the Monte Carlo information to make the correct association.
The red histogram shows the unit-normalized distribution
of the $M_{T2}^{(<)}$ variable defined in (\ref{mt2<}).
We see that the definition (\ref{mt2<}) preserves the corresponding endpoint:
\begin{equation}
M_{T2(max)}^{(<)} = M_{T2(max)}.
\label{mt2<max}
\end{equation}

Of course, we can also consider the alternative combination
\begin{equation}
M_{T2}^{(>)}\equiv \max 
\left\{ M_{T2}{\scriptstyle\big(\vec{p}^{~(a)}_T,\vec{p}^{~(b)}_T;m_{(a)},m_{(b)};
        \tilde{M}_c^{(a)},\tilde{M}_c^{(b)}\big)},
M_{T2}{\scriptstyle\big(\vec{p}^{~(a)}_T,\vec{p}^{~(b)}_T;m_{(a)},m_{(b)};
        \tilde{M}_c^{(b)},\tilde{M}_c^{(a)}\big)}
\right\},
\label{mt2>}
\end{equation}
whose unit-normalized distribution is shown in Fig.~\ref{fig:offshellpairing}
with the blue histogram. One can see that some of the wrong combination
entries in the $M_{T2}^{(>)}$ histogram violate the original endpoint 
$M_{T2(max)}$, yet there is still a well defined $M_{T2}^{(>)}$ endpoint
\begin{equation}
M_{T2(max)}^{(>)} \ge M_{T2(max)}.
\label{mt2>max}
\end{equation}

Strictly speaking, in our analysis in the next sections, we only need to study
the $M_{T2}^{(<)}$ endpoint (\ref{mt2<max}), which contains the relevant 
information about the physical $M_{T2}$ endpoint. At the same time, 
with our convention 
(\ref{abconv}) for the children masses, we only need to concentrate on
the upper half $\tilde M_c^{(b)}\ge\tilde M_c^{(a)}$ of the 
$(\tilde M_c^{(a)}, \tilde M_c^{(b)})$ plane.
However, for completeness we shall also present results for the 
$M_{T2}^{(>)}$ endpoint (\ref{mt2>max}), and we shall use
the lower ($\tilde M_c^{(b)}<\tilde M_c^{(a)}$) half of the 
$(\tilde M_c^{(a)}, \tilde M_c^{(b)})$ plane to show those. 
Thus the $M_{T2}$ endpoint shown in our plots below should be
interpreted as follows
\begin{equation}
M_{T2(max)}=\left\{
\begin{array}{l}
M_{T2(max)}^{(<)},~ {\rm if} \ \tilde M_c^{(a)}\le \tilde M_c^{(b)},
\\
M_{T2(max)}^{(>)},~ {\rm if} \ \tilde M_c^{(a)}> \tilde M_c^{(b)}.
\end{array}
\right.
\label{maxint}
\end{equation}

\section{The simplest event topology: one SM particle on each side}
\label{sec:110}

In this section, we consider the simplest topology with a
single visible particle on each side of the event.
We already introduced this example in 
Section~\ref{sec:examples}, along with its event topology
in Fig.~\ref{fig:dlsp}(a). In Section~\ref{subsec:110DLSP}
below we first discuss an asymmetric case with different children.
Later in Section~\ref{subsec:110ELSP} we consider a symmetric
situation with identical children masses.
The mass spectra for these two study points are listed in
Table~\ref{tab:mass}.

\TABLE[!h]{
\centerline{
  \begin{tabular}{|c|c|| c | c | c |}
    \hline
    Spectrum & Case                 & $M_c^{(a)}$ & $M_c^{(b)}$ & $M_p$  \\ \hline \hline
    I        & Different children   & 250         & 500         & 600    \\ \hline
    II       & Identical children   & 100         & 100         & 300    \\
    \hline
  \end{tabular}
}
\caption{Mass spectra for the two examples studied in 
Sections~\ref{subsec:110DLSP} and \ref{subsec:110ELSP}. 
All masses are given in GeV.}
\label{tab:mass}
}

\subsection{Asymmetric case}
\label{subsec:110DLSP}

Before we present our numerical results, it will be useful to derive
an analytical expression for the asymmetric $M_{T2}$ endpoint (\ref{mt2maxdlsp})
in terms of the corresponding physical spectrum of Table~\ref{tab:mass}
and the two test children masses $\tilde M_c^{(a)}$ and $\tilde M_c^{(b)}$. 
Our result will generalize the corresponding formula 
derived in \cite{Cho:2007dh} for the symmetric case of 
$\tilde M_c^{(a)}=\tilde M_c^{(b)}\equiv \tilde M_c$ and no upstream momentum 
($P_{UTM}=0$). For the event topology of Fig.~\ref{fig:dlsp}(a) 
the $M_{T2}$ endpoint is always obtained from the balanced solution 
and is given by \cite{Cho:2007dh}
\begin{equation}
M_{T2(max)}(\tilde M_c,P_{UTM}=0) = \mu_{ppc} + \sqrt{\mu_{ppc}^2+\tilde M_c^2}\, .
\label{eq:mt2maxlsp}
\end{equation}
Here we made use of the convenient shorthand notation introduced in \cite{Burns:2008va}
for the relevant combination of physical masses
\begin{equation}
\mu_{npc}
\equiv \frac{M_n}{2} \left\{ 1- \left(\frac{M_c}{M_p}\right)^2 \right\} \, .
\label{eq:munpc}
\end{equation}
The $\mu$ parameter defined in (\ref{eq:munpc})
is simply the transverse momentum of the (massless) visible particle
in those events which give the maximum value of $M_{T2}$ \cite{Matchev:2009fh}.
Squaring (\ref{eq:mt2maxlsp}), we can equivalently rewrite it as
\begin{equation}
M^2_{T2(max)}(\tilde M_c,P_{UTM}=0) = 2\,\mu^2_{ppc}+\tilde M_c^2 +  
\sqrt{4\,\mu^2_{ppc}(\mu_{ppc}^2+\tilde M_c^2)}\, .
\label{eq:mt2maxlsp2}
\end{equation}

Now let us derive the analogous expressions for the asymmetric case
$M_c^{(a)}\ne M_c^{(b)}$. Just like the symmetric case, 
the asymmetric endpoint $M_{T2(max)}$ also 
comes from a balanced solution and is given by
\begin{eqnarray}
M^2_{T2(max)} (\tmaz,\tmbz,P_{UTM}=0) 
= 2\bar\mu^2_{ppc} +\tilde{M}_+^2 
+ \sqrt{ 4\,\bar\mu^2_{ppc}(\bar\mu_{ppc}^2+\tilde M_+^2) + \tilde M_-^4} \, ,
\label{mt2max110}
\end{eqnarray}
where the parameters $\tilde M_+^2$ and $\tilde M_-^2$ 
were already defined in (\ref{eq:m+}) and (\ref{eq:m-}), while
$\bar\mu_{ppc}$ is now the geometric average of the
corresponding individual $\mu_{ppc}$ parameters
\begin{equation}
\bar\mu^2_{ppc} \equiv \mu_{ppc_a}\, \mu_{ppc_b} \equiv 
\frac{(M_p^2 - \mazsq)(M_p^2 - \mbzsq)}{4 M_p^2} \, .
\end{equation}
It is easy to check that in the symmetric limit
\begin{equation}
\tilde M_c^{(b)}\to \tilde M_c^{(a)} ~ \Longrightarrow ~
\bar\mu_{ppc} \to \mu_{ppc}, \quad 
\tilde{M}_+ \to \tilde{M}_c,  \quad
\tilde{M}_- \to 0,
\end{equation}
eq.~(\ref{mt2max110}) reduces to 
its symmetric counterpart (\ref{eq:mt2maxlsp2}),
as it should.

%
\FIGURE[ht]{
\centerline{
\epsfig{file=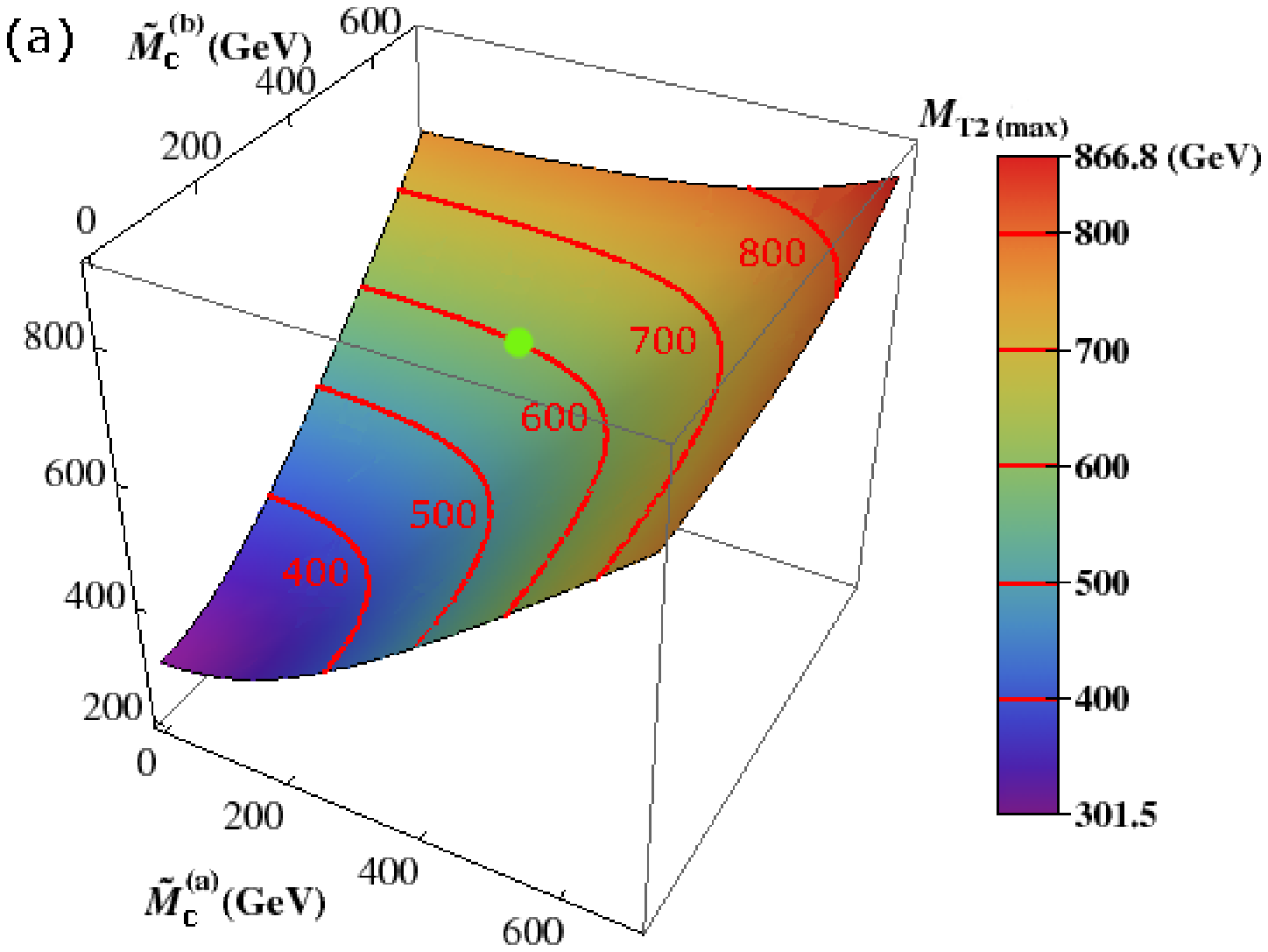,      height=6.3cm}
\epsfig{file=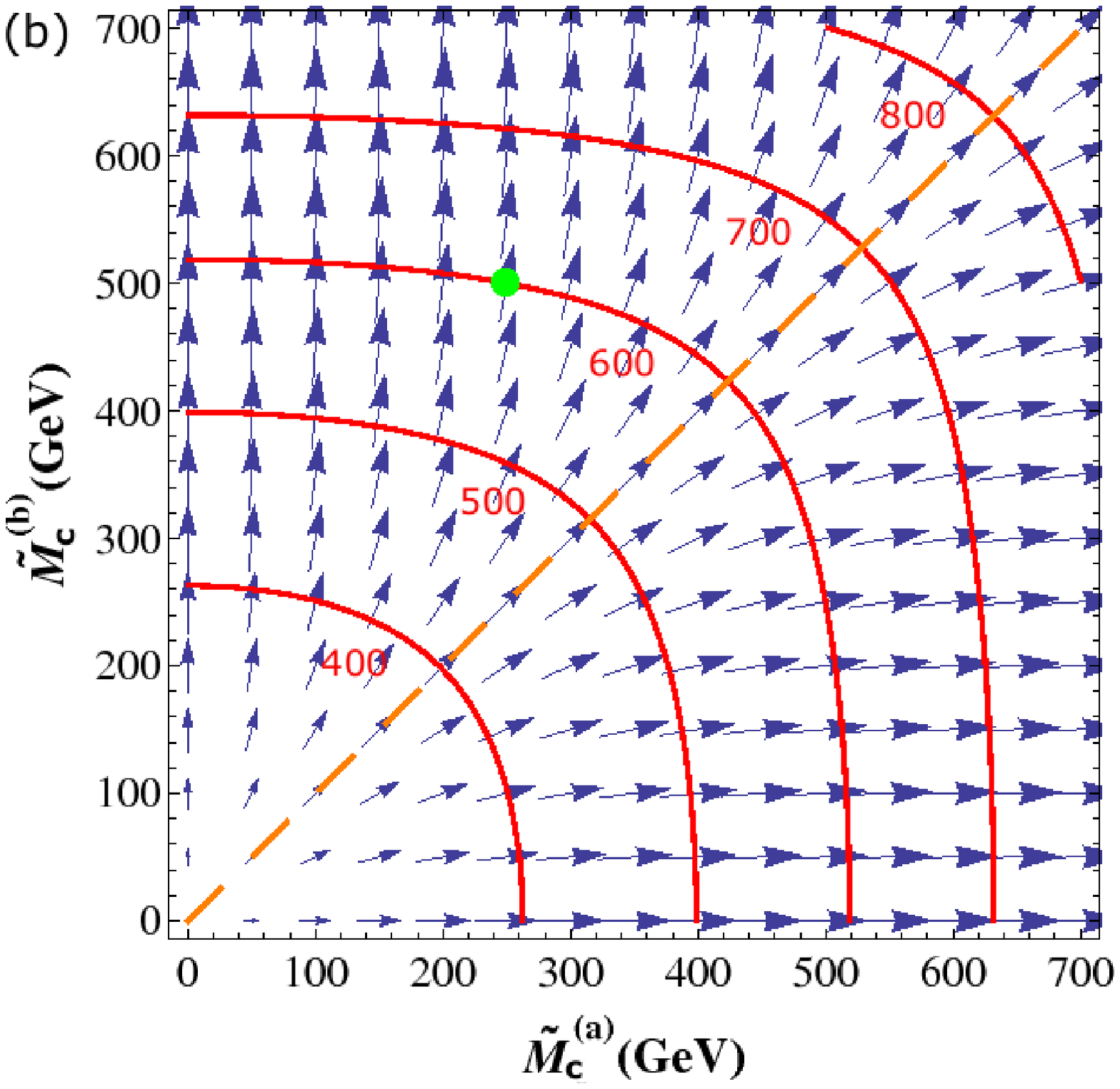, height=6.3cm} 
}
\caption{\it $M_{T2(max)}$
as a function of the two test children masses, 
$\tmaz$ and $\tmbz$, for the event topology of Fig.~\ref{fig:dlsp}(a)
with no upstream momentum ($P_{UTM}=0$), and
the asymmetric mass spectrum I from Table~\ref{tab:mass}:
$(\maz,\mbz,M_p)= (250, 500, 600)$ GeV. We show (a) a three
dimensional view and (b) contour plot projection on the 
$(\tilde M_c^{(a)},\tilde M_c^{(b)})$ plane (red contour lines). The green
dot marks the true values of the children masses.
Panel (b) also shows a gradient plot, where
longer (shorter) arrows imply steeper (gentler) slope. 
A kink structure is absent in this case. 
The symmetric endpoint $M_{T2(max)}(\tilde{M}_c)$ 
of eq.~(\ref{eq:mt2maxlsp}) can be obtained by going along the
diagonal orange line $\tmbz = \tmaz$.}
\label{fig:110}
}
%

We are now ready to present our numerical results for the event
topology of Fig.~\ref{fig:dlsp}(a). We first 
take the asymmetric mass spectrum I from Table~\ref{tab:mass}
and consider the case with no upstream momentum, 
when formula (\ref{mt2max110}) applies.
Fig.~\ref{fig:110} shows the corresponding $M_{T2}$ endpoint 
as a function of the two test children masses $\tilde M_c^{(a)}$ 
and $\tilde M_c^{(b)}$.
In panel (a) we present a three dimensional view, while in panel 
(b) we show a contour plot projection on the 
$(\tilde M_c^{(a)},\tilde M_c^{(b)})$ plane (red contour lines).
On either panel, the green dot marks the true values 
of the children masses, $M_c^{(a)}$ and $M_c^{(b)}$.
Panel (b) also shows a gradient plot, where
longer (shorter) arrows imply steeper (gentler) slope. 
The symmetric endpoint $M_{T2(max)}(\tilde{M}_c,P_{UTM}=0)$ 
of eq.~(\ref{eq:mt2maxlsp}) can be obtained by going along the
diagonal orange line $\tmbz = \tmaz$ in Fig.~\ref{fig:110}(b).
We remind the reader that the endpoint $M_{T2(max)}$
plotted in Fig.~\ref{fig:110} should be interpreted as in
eq.~(\ref{maxint}).

Fig.~\ref{fig:110} illustrates the first basic property of the 
asymmetric $M_{T2}$ variable, which was discussed in Sec.~\ref{sec:mt2dlspprop1}.
The $M_{T2}$ endpoint allows us to find one relation between the two children masses 
$\tilde M_c^{(a)}$ and $\tilde M_c^{(b)}$ and the parent mass
$\tilde M_p=M_{T2(max)}$, and in order to do so,
we do {\em not} have to assume equality of the children masses, 
as is always done in the literature. 
The crucial advantage of our approach, in which we allow the
two children masses to be arbitrary, is its generality and 
model-independence. It allows us to extract the basic information
contained in the $M_{T2}$ endpoint, without muddling it up with 
additional theoretical (and unproven) assumptions.

Unfortunately, to go any further 
and determine each individual mass, we must make 
use of the additional properties discussed in 
Secs.~\ref{sec:kinkdlsp} and \ref{sec:PUTMdlsp}. 
In the case of the simplest event topology of Fig.~\ref{fig:dlsp}(a)
considered here, they both require the presence of some 
upstream momentum \cite{Barr:2007hy,Burns:2008va}.
As a proof of concept, we now reconsider the same type of events, 
but with a fixed upstream momentum of $P_{UTM}=1$ TeV.
(The upstream momentum may be due to initial state radiation, 
or decays of heavier particles upstream.)
The corresponding results are shown in Fig.~\ref{fig:110isr}.

%
\FIGURE[ht]{
\centerline{
\epsfig{file=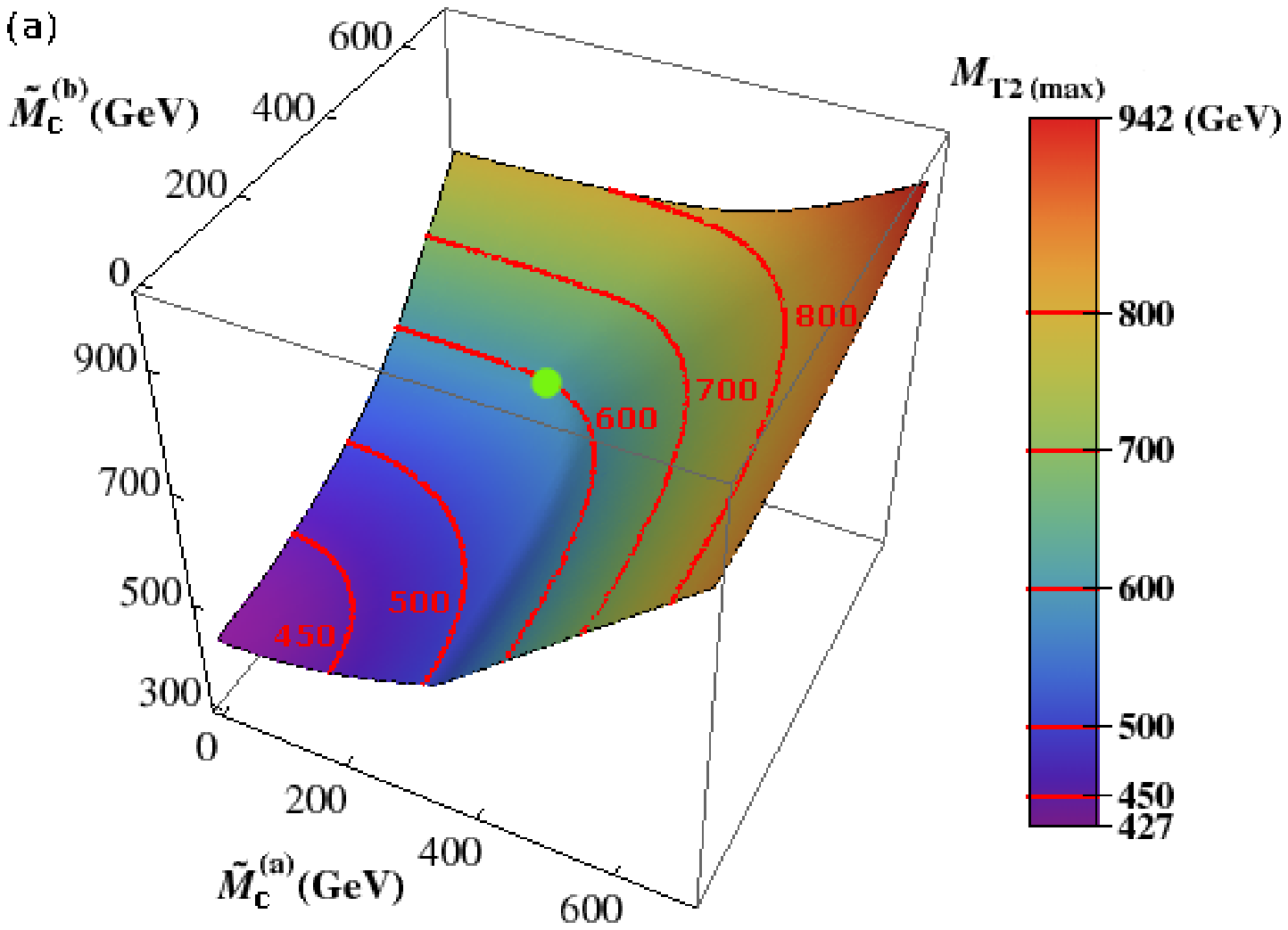,     height=6.3cm}
\epsfig{file=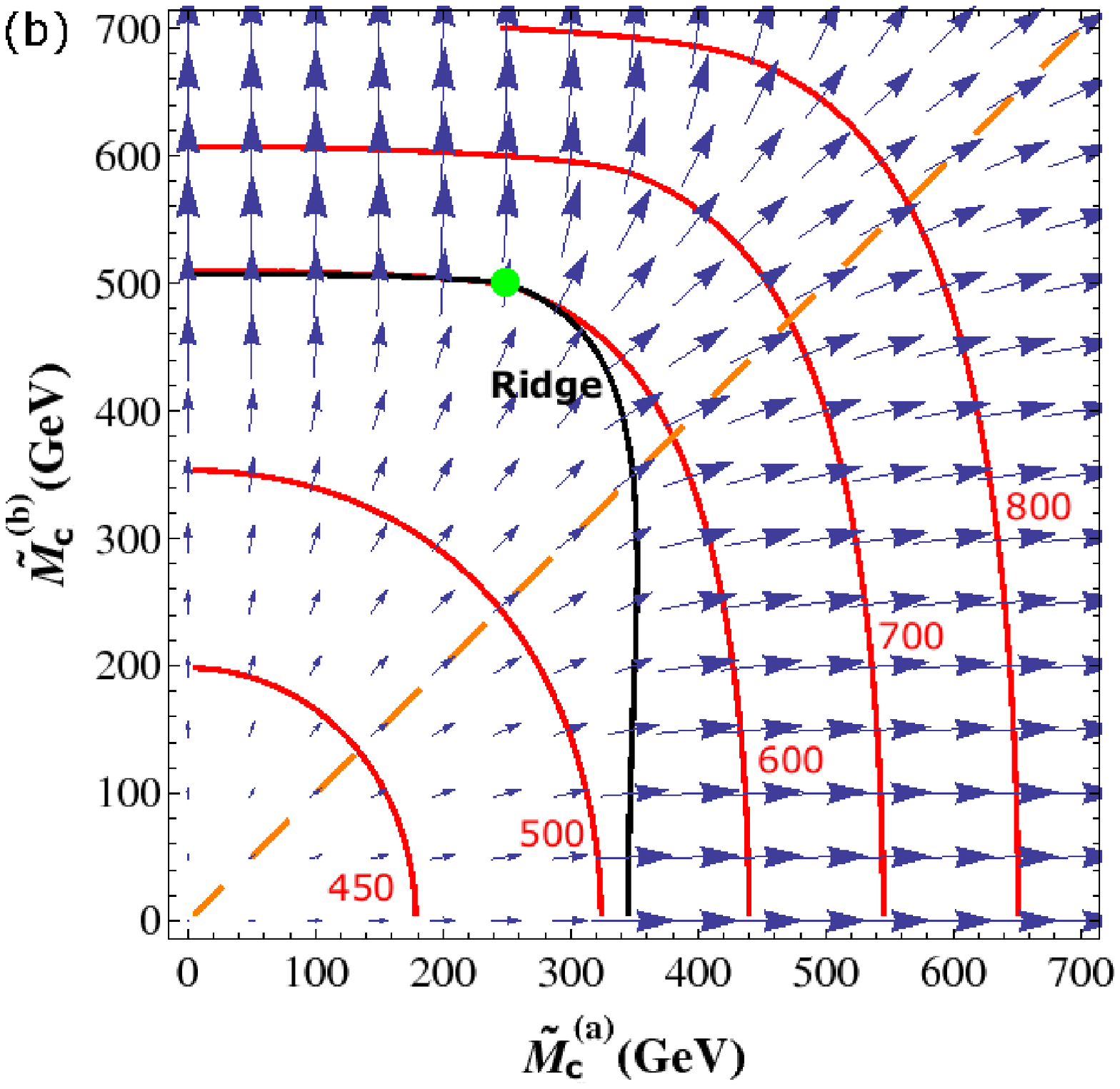,height=6.3cm} }
\caption{\it The same as in Fig.~\ref{fig:110} but with fixed upstream momentum of
$P_{UTM} = 1$ TeV. The ridge structure (shown as the black solid line) 
is revealed by the sudden increase in the slope (gradient) in panel (b).
Notice that the ridge goes through the true values of
the children masses marked by the green dot.}
\label{fig:110isr}
}

Fig.~\ref{fig:110isr} demonstrates the second basic property
of the asymmetric $M_{T2}$ variable discussed in Sec.~\ref{sec:kinkdlsp}.
Unlike the result shown in Fig.~\ref{fig:110}(a), which was perfectly smooth,
this time the $M_{T2(max)}$ function in Fig.~\ref{fig:110isr}(a)
shows a ridge, corresponding to the slope discontinuity
marked with the black solid line in Fig.~\ref{fig:110isr}(b). 
The most important feature of the ridge is 
the fact that it passes through the green dot marking the
true values of the children masses. Notice that applying the
traditional symmetric $M_{T2}$ approach in this case will 
give a completely wrong result. If we were to assume equal 
children masses from the very beginning, we will be constrained
to the diagonal orange line in Fig.~\ref{fig:110isr}(b).
The $M_{T2}$ endpoint will then still exhibit a kink, 
but the kink will be in the wrong location. In the example 
shown in Fig.~\ref{fig:110isr}(b), we will underestimate the parent mass, 
while for the child mass we will find a value which is 
somewhere in between the two true masses $M_c^{(a)}$ and $M_c^{(b)}$.

Using the ridge information,
we now know an additional relation among the children masses,
which allows us to express all three masses in terms of a single
unknown parameter $\theta$, as illustrated in Fig.~\ref{fig:110dlspridge}(a).
\FIGURE[t]{
\centerline{
\epsfig{file=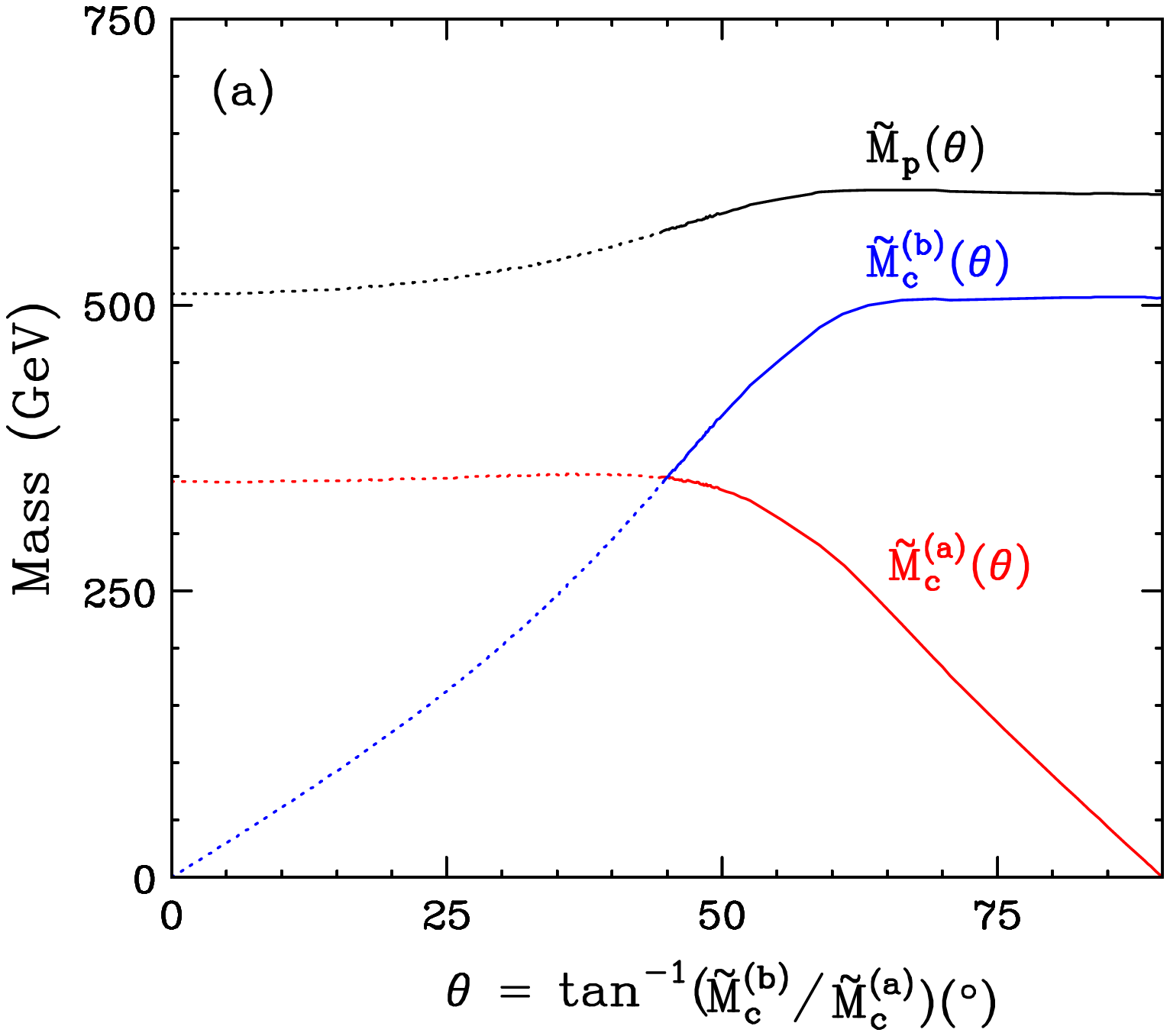,      height=6.2cm}
~~
\epsfig{file=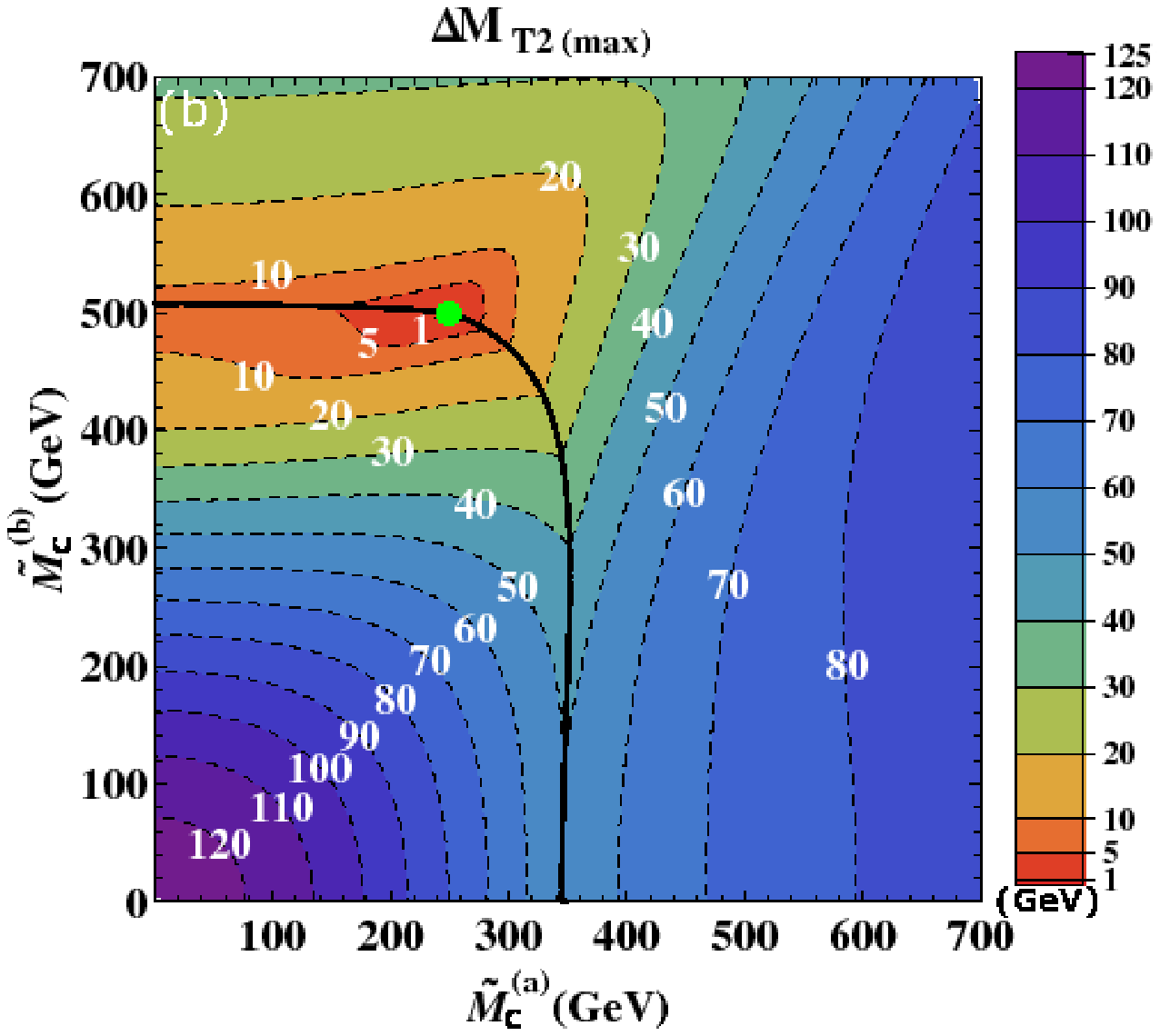,height=6.5cm}}
\caption{\it  (a) Particle masses obtained along the $M_{T2(max)}$
ridge seen in Fig.~\ref{fig:110isr}. The ridge is parametrized by 
the angle $\theta$ defined in (\ref{thetadef}). The two 
children masses $\tilde{M}_c^{(a)}(\theta)$ (in red) and
$\tilde{M}_c^{(b)}(\theta)$ (in blue) as well as the parent mass 
$\tilde M_p$ (in black) are then plotted as a function of $\theta$. 
In our convention (\ref{abconv}) only values of $\theta\ge 45^{\circ}$ 
are physical, and the corresponding masses are shown with solid lines.
Dotted lines show the extrapolation for $\theta<45^{\circ}$.
(b) Contour plot of the quantity 
$\Delta M_{T2(max)}(\tilde M_c^{(a)},\tilde M_c^{(b)},P_{UTM}=1\ {\rm TeV})$ 
defined in eq.~(\ref{deltamt2defdlsp}), in the $(\tilde M_c^{(a)},\tilde M_c^{(b)})$ plane.
This plot is obtained simply by taking the
difference between Fig.~\ref{fig:110isr}(a) and
Fig.~\ref{fig:110}(a).  The solid black curve indicates 
the location of the $M_{T2(max)}$ ridge.  
Only the point corresponding to the true children masses (the green dot) 
satisfies the $P_{UTM}$ invariance condition $\Delta M_{T2(max)}=0$ 
from eq.~(\ref{deltamt2eq0dlsp}).  }
\label{fig:110dlspridge}
}
Let us choose to parametrize the ridge by the polar angle in the
$(\tilde M_c^{(a)},\tilde M_c^{(b)})$ plane:
\begin{equation}
\theta =
\tan^{-1} \left( \frac {\tilde{M}_c^{(b)}} { \tilde{M}_c^{(a)}}\right)\, .
\label{thetadef}
\end{equation}
Using the ridge information from Fig.~\ref{fig:110isr}, we can then find
all three masses as a function of $\theta$. The result is 
shown in Fig.~\ref{fig:110dlspridge}(a). The mass $\tilde M_c^{(a)}$ 
of the lighter child is plotted in red, the mass $\tilde M_c^{(b)}$ 
of the heavier child is plotted in blue, while the parent mass
$\tilde M_p$ is plotted in black.  With our convention 
(\ref{abconv}) for the children masses, only values of 
$\theta\ge 45^{\circ}$ are physical, and the corresponding 
masses are shown with solid lines. The dotted lines in 
Fig.~\ref{fig:110dlspridge}(a) show the extrapolation into 
the unphysical region $\theta<45^{\circ}$.

Fig.~\ref{fig:110dlspridge}(a) has some important and far reaching implications.
For example, one may now start asking the question: 
Are there really any {\em massive} invisible particles in those events,
or is the missing energy simply due to neutrino production \cite{Chang:2009dh}?
The ridge results shown in Fig.~\ref{fig:110dlspridge}(a)
begin to provide the answer to that quite fundamental question.
According to Fig.~\ref{fig:110dlspridge}(a), for {\em any}
value of the (still unknown) parameter $\theta$, 
the two children particles cannot be simultaneously massless.
This means that the missing energy cannot be simply due to neutrinos, 
i.e. there is {\em at least} one new, massive invisible particle 
produced in the missing energy events. At this point, we cannot be 
certain that this is a dark matter particle, but establishing the
production of a WIMP candidate at a collider is by itself 
a tremendously important result. Notice that while we cannot be sure
about the masses of the children, the parent mass $M_p$ is determined 
with a very good precision from Fig.~\ref{fig:110dlspridge}(a): 
the function $\tilde M_p(\theta)$ is almost flat
and rather insensitive to the particular value of $\theta$\footnote{
Interestingly, for the example in Fig.~\ref{fig:110dlspridge}(a),
the maximum value of $\tilde M_p(\theta)$ happens to give the true parent mass $M_p$,
but we have checked that this is a coincidence and does not hold 
in general for other examples which we have studied.}.

Once we have proved that some kind of WIMP production is going on,
the next immediate question is: how many such WIMP particles are present 
in the data -- one or two? Unfortunately, the ridge analysis
of Fig.~\ref{fig:110dlspridge}(a) alone cannot provide the answer 
to this question, since the value of $\theta$ is still undetermined.
If $\theta=90^{\circ}$, one of the missing particles is massless, 
which is consistent with a SM neutrino. Therefore, if $\theta$ were indeed
$90^{\circ}$, the most plausible explanation of this scenario would be that 
only one of the missing particles is a genuine WIMP, while the other is a SM neutrino.
On the other hand, almost any other value of $\theta<90^{\circ}$ would guarantee
that there are {\em two} WIMP candidates in each event.
In that case, the next immediate question is: are they the same or are they different?
Fortunately, our asymmetric approach will allow answering this question 
in a model-independent way. If $\theta$ is determined to be $45^\circ$, 
the two WIMP particles are the same, i.e. we are producing a single 
species of dark matter. On the other hand, if $45^\circ<\theta<90^\circ$, 
then we can be certain that there are not one, but {\em two} different WIMP
particles being produced.

We see that in order to completely understand the physics behind 
the missing energy signal, we must determine the value of $\theta$,
i.e. we must find the exact location of the true children masses along 
the ridge. One of our main results in this paper is that this can be done 
by using the third $M_{T2}$ property  discussed in Sec.~\ref{sec:PUTMdlsp}. 
The idea is illustrated in Fig.~\ref{fig:110dlspridge}(b), 
where we show a contour plot in the $(\tilde M_c^{(a)},\tilde M_c^{(b)})$ plane
of the quantity 
$\Delta M_{T2(max)}(\tilde M_c^{(a)},\tilde M_c^{(b)},P_{UTM})$ 
defined in eq.~(\ref{deltamt2defdlsp}), for a fixed
$P_{UTM}=1$ TeV. 
This plot is obtained simply by taking the
difference between Fig.~\ref{fig:110isr}(a) and
Fig.~\ref{fig:110}(a). (A more practical method for obtaining 
this information was proposed in \cite{Konar:2009wn}.)  
Recall that the function $\Delta M_{T2(max)}$
was introduced in order to quantify the $P_{UTM}$ invariance 
of the $M_{T2}$ endpoint, and it is expected that 
$\Delta M_{T2(max)}$ vanishes at the correct values of the children masses
(see eq.~(\ref{deltamt2eq0dlsp})). This expectation is
confirmed in Fig.~\ref{fig:110dlspridge}(b), where we find the
minimum (zero) of the $\Delta M_{T2(max)}$ function 
exactly at the right spot (marked with the green dot)
along the $M_{T2(max)}$ ridge.  
Thus the $\Delta M_{T2(max)}$ function in  Fig.~\ref{fig:110dlspridge}(b)
completely pins down the spectrum, and in this case would reveal the 
presence of two {\em different} WIMP particles, with unequal masses
$M_c^{(a)}\ne M_c^{(b)}$. Our analysis thus shows that colliders can not 
only produce a WIMP dark matter candidate and measure its mass,
as discussed in the existing literature, but they can do a much more
elaborate dark matter particle spectroscopy, as advertized in the title.
In particular, they can probe the number and type of missing particles,
including particles from subdominant dark matter species, which are 
otherwise unlikely to be discovered experimentally in the usual dark 
matter searches.

\subsection{Symmetric case}
\label{subsec:110ELSP}

While in our approach the two children masses $\tilde M_c^{(a)}$ 
and $\tilde M_c^{(b)}$ are treated as independent inputs, 
this, of course, does not mean that the approach is only valid in 
cases when the children masses are different to begin with. 
The techniques discussed in the previous subsection
remain applicable also in the more conventional case when the
children are identical, i.e.~when colliders produce a single 
dark matter component. In order to illustrate how our method works 
in that case, we shall now work out an example with equal children masses.
We still consider the simplest event topology of Fig.~\ref{fig:dlsp}(a),
but with the symmetric mass spectrum II from Table~\ref{tab:mass}. 
We then repeat the analysis done in 
Figs.~\ref{fig:110},~\ref{fig:110isr}, and ~\ref{fig:110dlspridge} 
and show the corresponding results in Figs.~\ref{fig:110_slsp},
\ref{fig:110_slspisr} and \ref{fig:110elspridge}.

\FIGURE[t]{
\centerline{
\epsfig{file=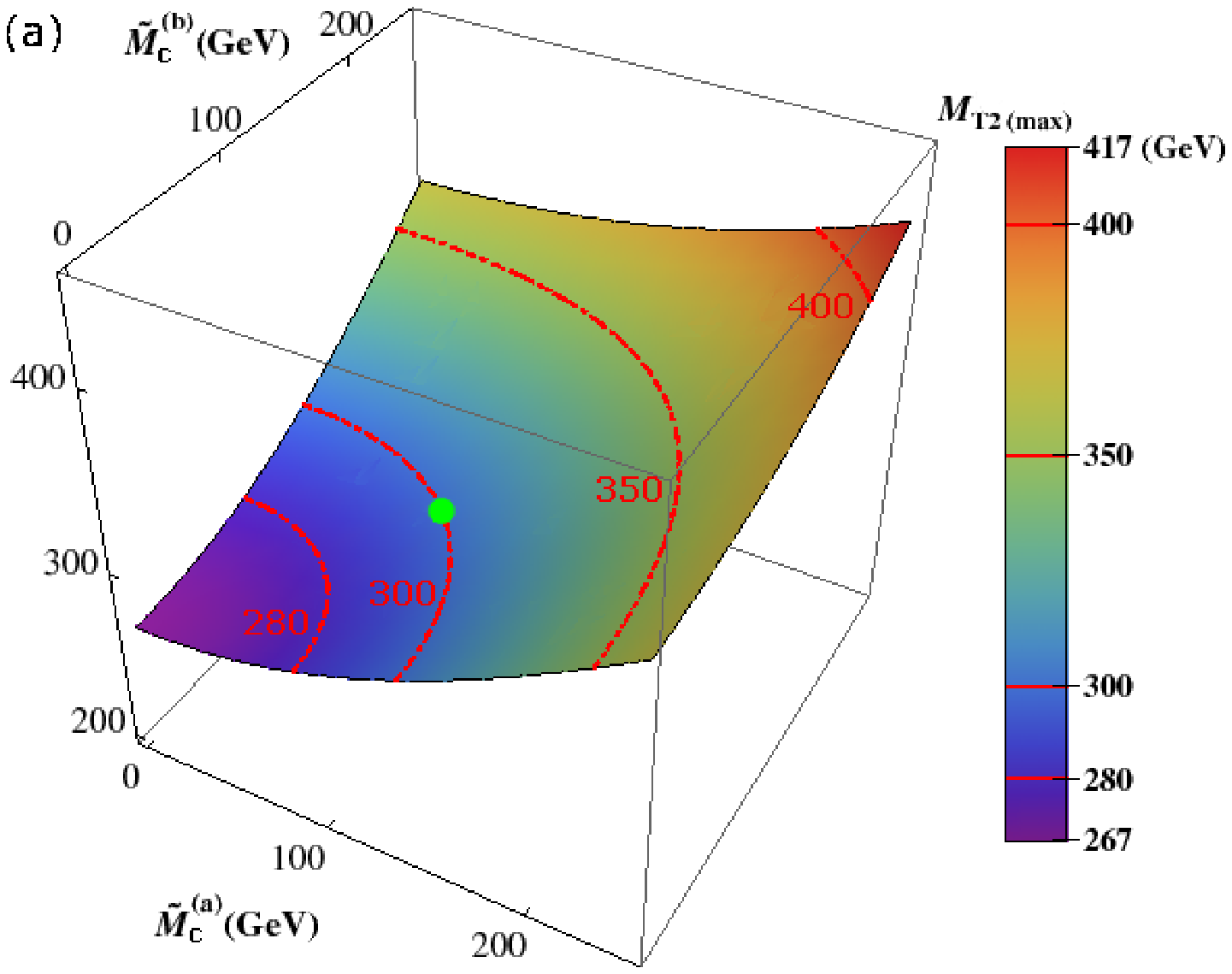,     height=6.3cm}
\epsfig{file=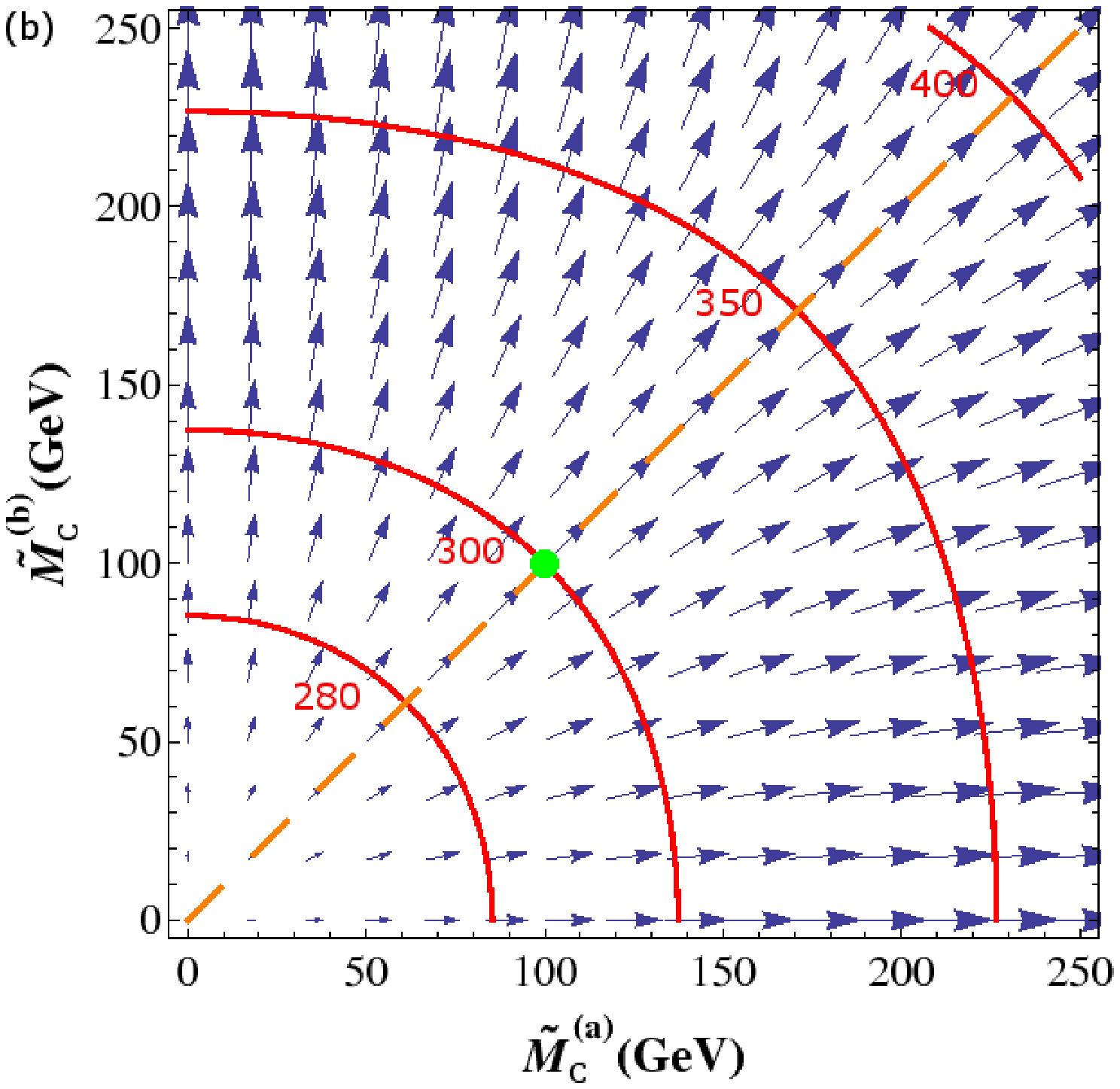,height=6.3cm} }
\caption{\it The same as Fig.~\ref{fig:110}, but for the symmetric
mass spectrum II from Table~\ref{tab:mass}, i.e.
$(\maz, \mbz, M_p)= (100, 100, 300)$ GeV.}
\label{fig:110_slsp}
}

\FIGURE[t]{
\centerline{
\epsfig{file=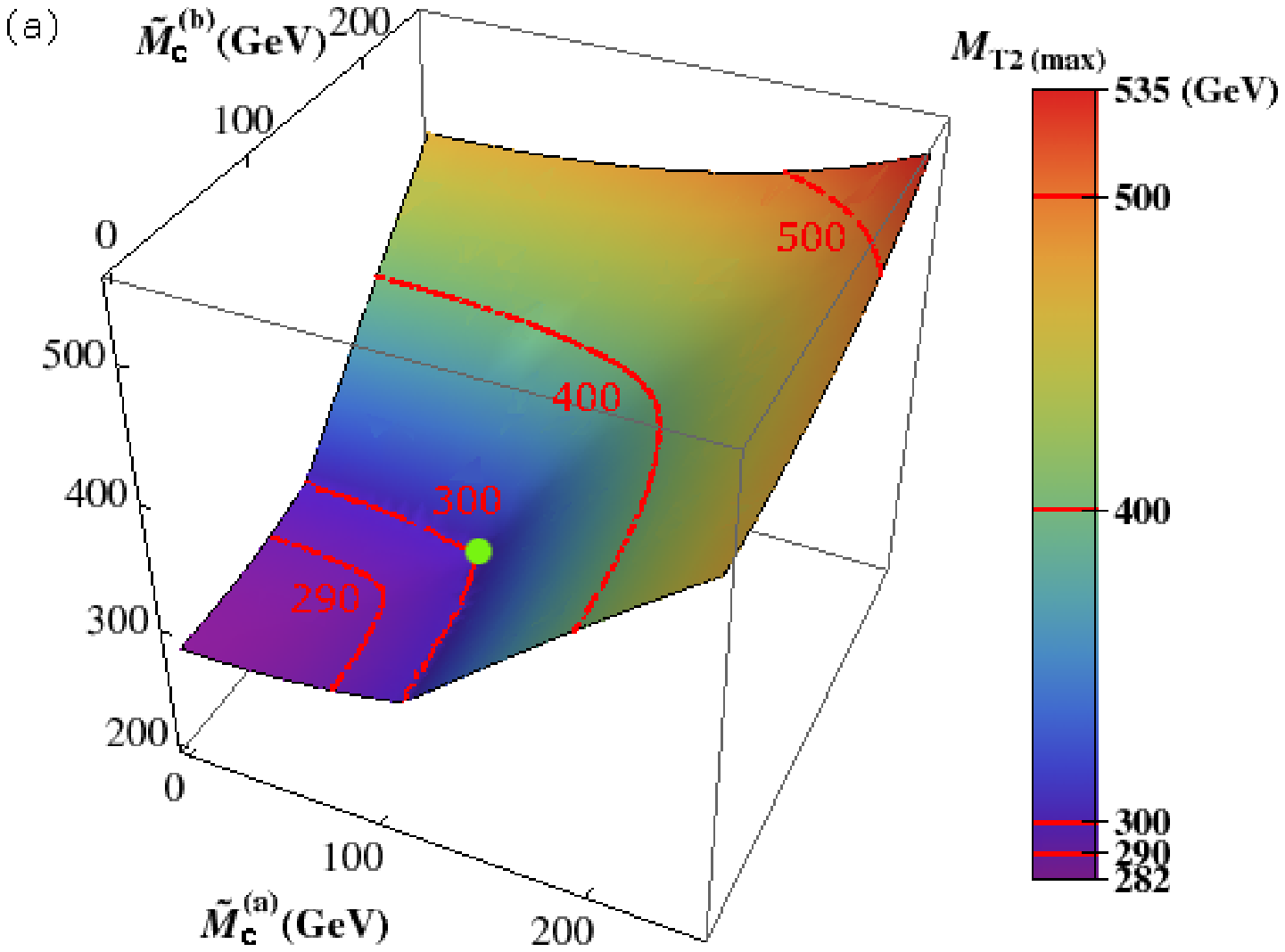,      height=6.3cm}
\epsfig{file=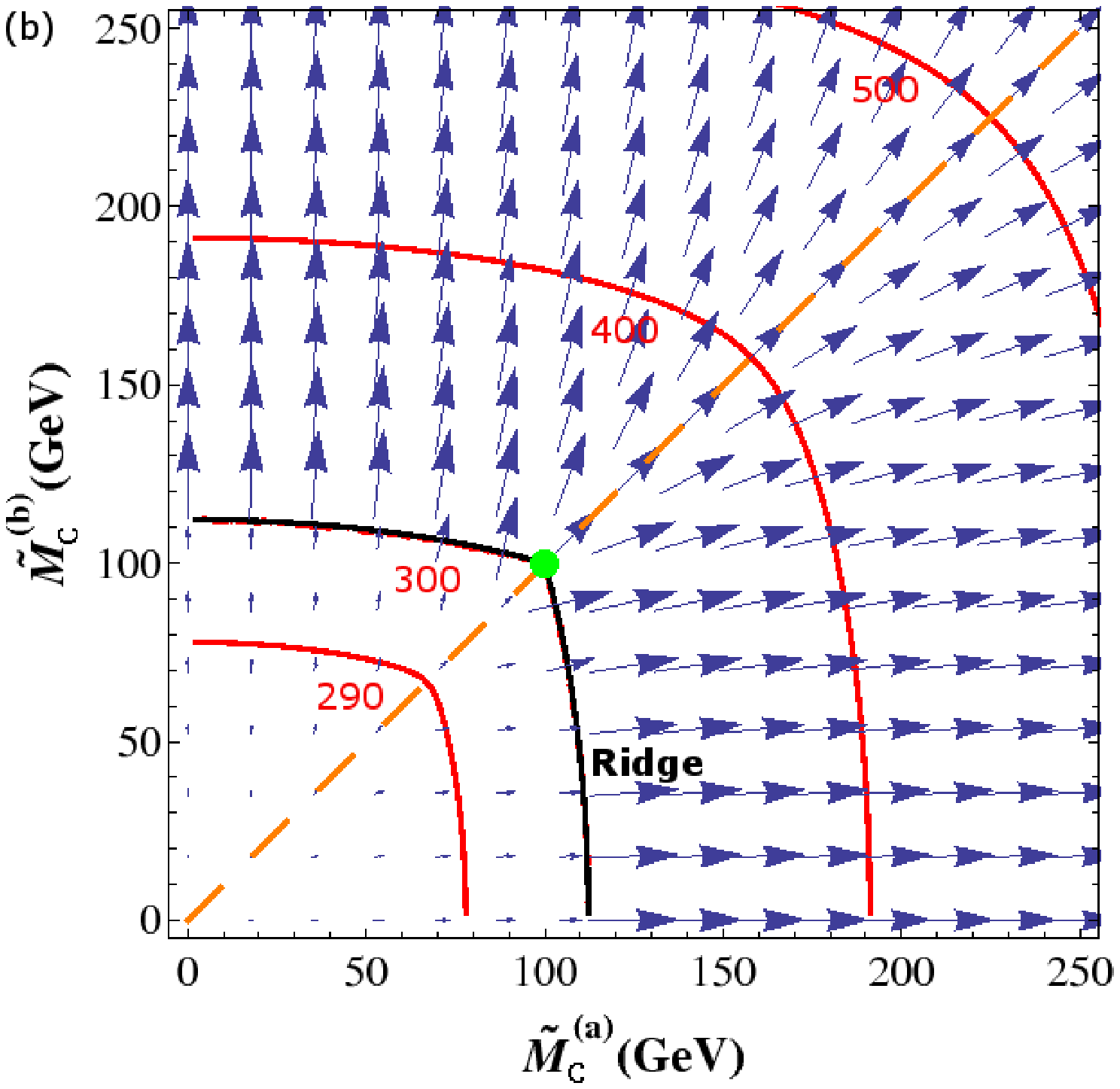,height=6.3cm} }
\caption{\it The same as Fig.~\ref{fig:110isr} but for the symmetric
mass spectrum II from Table~\ref{tab:mass}, i.e.
$(\maz, \mbz, M_p)= (100, 100, 300)$ GeV.}
\label{fig:110_slspisr}
}

The conclusions from this exercise are very similar to what we found 
earlier in Sec.~\ref{subsec:110DLSP} for the asymmetric case.
The $M_{T2}$ endpoint still provides one relation among 
the two children masses $\tilde M_c^{(a)}$ and $\tilde M_c^{(b)}$ 
and the parent mass $\tilde M_p=M_{T2(max)}$. This relation 
is shown in Fig.~\ref{fig:110_slsp} (Fig.~\ref{fig:110_slspisr})
for the case without (with) upstream momentum $P_{UTM}$.
As seen in Fig.~\ref{fig:110_slsp}, in the absence of 
any upstream $P_{UTM}$, the function 
$\tilde M_p (\tilde M_c^{(a)},\tilde M_c^{(b)})$ is smooth
and reveals nothing about the children masses.
However, the presence of upstream momentum significantly 
changes the picture and the function $\tilde M_p (\tilde M_c^{(a)},\tilde M_c^{(b)})$
again develops a ridge, which is clearly visible\footnote{We caution 
the reader that here we are presenting only a proof of concept.
In the actual analysis the ridge may be rather difficult to see, 
for a variety of reasons - detector resolution, finite statistics, 
combinatorial and SM backgrounds, etc. Nevertheless, we expect that
the ridge will be just as easily observable as the traditional kink in the 
symmetric $M_{T2}$ endpoint. If the kink can be seen in the data, 
the ridge can be seen too, and 
there is no reason to make the assumption of equal children masses.
Conversely, if the kink is too difficult to see, 
the ridge will remain hidden as well.} 
in both the three-dimensional view of Fig.~\ref{fig:110_slspisr}(a), 
as well as the gradient plot in Fig.~\ref{fig:110_slspisr}(b). 
The ridge information now further constrains the children masses 
to the black solid line in Fig.~\ref{fig:110_slspisr}(b), leaving
only one unknown degree of freedom. Parametrizing it with the polar angle 
$\theta$ as in (\ref{thetadef}), we obtain the spectrum 
as a function of $\theta$, as shown in Fig.~\ref{fig:110elspridge}(a).
\FIGURE[t]{
\centerline{
\epsfig{file=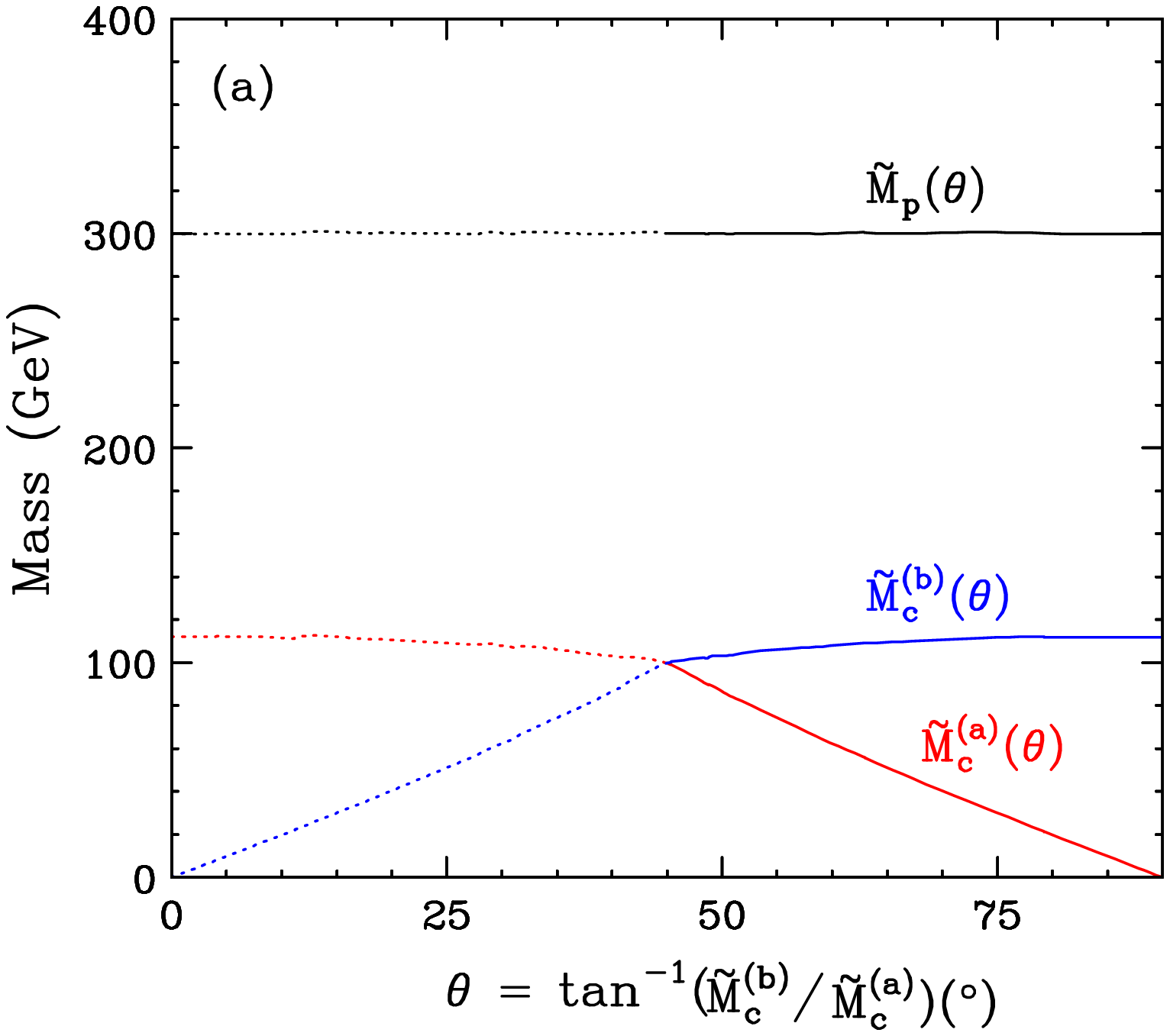,      height=6.2cm}
~~
\epsfig{file=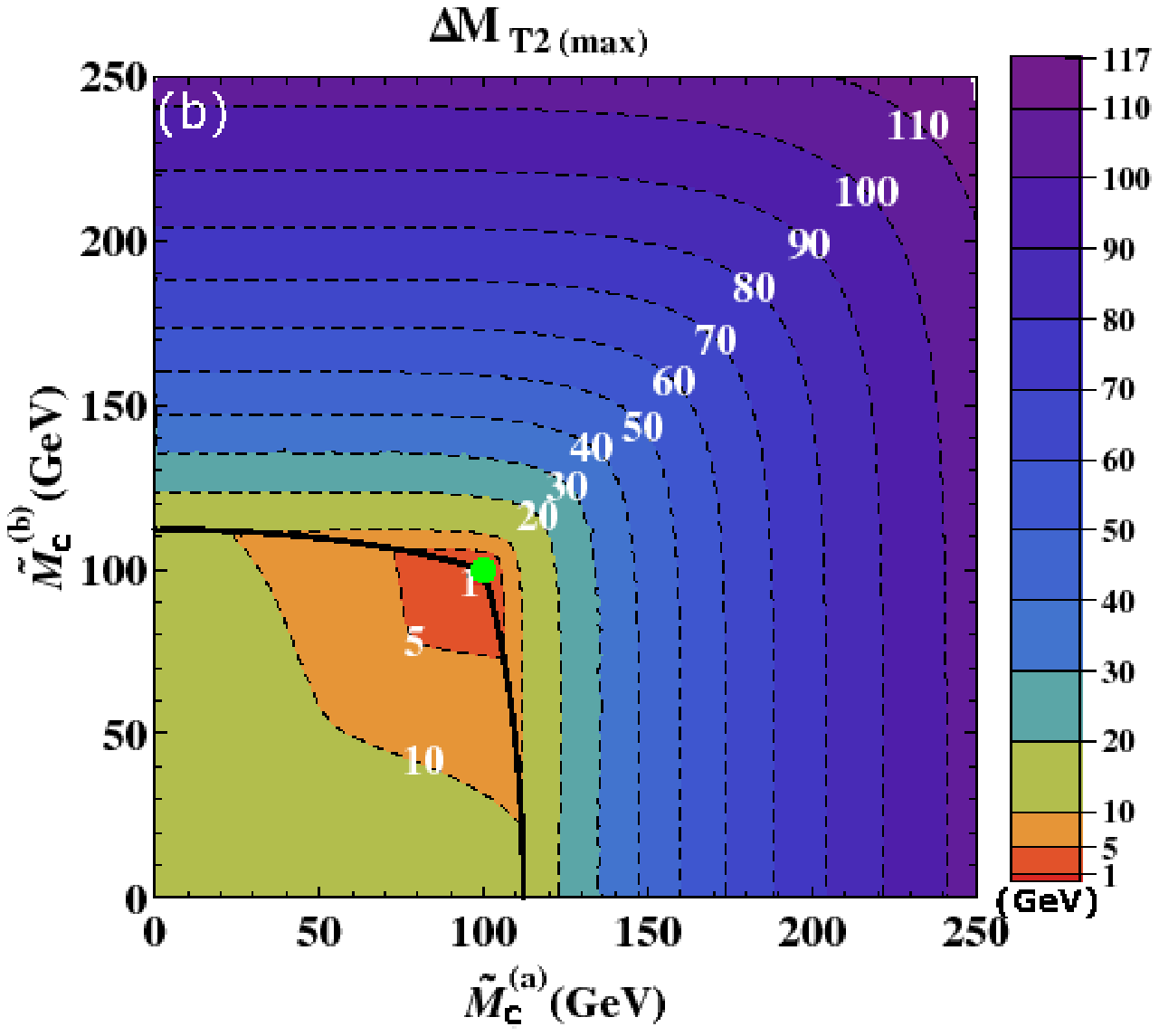,height=6.5cm}}
\caption{\it The same as in Fig.~\ref{fig:110dlspridge} but
for the symmetric
mass spectrum II from Table~\ref{tab:mass}, i.e.
$(\maz, \mbz, M_p)= (100, 100, 300)$ GeV.
Notice that, in contrast to Fig.~\ref{fig:110dlspridge},
the minimum of the $\Delta M_{T2(max)}$ function is 
now obtained at $\tilde M_c^{(a)}=\tilde M_c^{(b)}$, 
implying that the two missing particles are the same.
}
\label{fig:110elspridge}
}
Once again we find the fortuitous result that in spite of the
remaining arbitrariness in the value of $\theta$,
the parent mass $M_p$ is very well determined, since $\tilde M_p(\theta)$
is a very weakly 
varying function of $\theta$. Furthermore, both Fig.~\ref{fig:110_slspisr}(a)
and Fig.~\ref{fig:110_slspisr}(b) exhibit a high degree of symmetry under 
$\tilde M_c^{(a)}\leftrightarrow\tilde M_c^{(b)}$, which
is a good hint that the children are in fact identical.
This suspicion is confirmed in Fig.~\ref{fig:110elspridge}(b),
where we find that the $P_{UTM}$ dependence disappears at
the symmetric point $\tilde M_c^{(a)}=\tilde M_c^{(b)}=100$ GeV,
revealing the true masses of the two children.

In the two examples considered so far in Sections~\ref{subsec:110DLSP} 
and \ref{subsec:110ELSP}, we used a fixed finite value of the upstream 
transverse momentum $P_{UTM}=1$ TeV, which is probably rather extreme ---
in realistic models, one might expect typical values of $P_{UTM}$ on the order of
several hundred GeV. However, things begin to get
much more interesting if one were to consider even {\em larger}
values of $P_{UTM}$. On the one hand, the ridge feature becomes sharper
and easier to observe \cite{Burns:2008va}. More importantly, the 
ridge structure itself is modified, and a second set of ridgelines appears\footnote{A 
keen observer may have already noticed a hint of those in Figs.~\ref{fig:110dlspridge}(b) 
and \ref{fig:110elspridge}(b).} at sufficiently large $P_{UTM}$. 
All ridgelines intersect precisely at the point 
marking the true values of the children masses, thus allowing the 
complete determination of the mass spectrum by the ridge method alone.
This procedure was demonstrated explicitly in Ref.~\cite{Barr:2009jv},
which investigated the extreme case of $P_{UTM}=\infty$ for a study point 
with different parents and identical children. The assumption of
$P_{UTM}=\infty$ justified the use of a ``decoupling argument'', in which the 
two branches $\lambda=a$ and $\lambda=b$ are treated independently, 
allowing the derivation of simple analytical expressions for 
the $M_{T2}$ endpoint \cite{Barr:2009jv}. In Appendix~\ref{app:infpt}
we reproduce the analogous analytical results at $P_{UTM}\to \infty$ 
for the case of interest here (identical parents and different children)
and study in detail the $P_{UTM}$ dependence of the ridgelines.
Unfortunately, we find that the values of $P_{UTM}$ necessary to reveal the
additional ridge structure, are too large to be of any interest experimentally.
On the positive side, the $P_{UTM}$ invariance method discussed in Sec.~\ref{sec:PUTM}
does not require such extremely large values of $P_{UTM}$ and can in principle 
be tested in more realistic experimental conditions.

\subsection{Mixed case}
\label{sec:110:events}

For simplicity, so far in our discussion we have been studying
only one type of missing energy events at a time. In reality, 
the missing energy sample may contain several different 
types of events, and the corresponding $M_{T2}$ measurements 
will first need to be disentangled from each other.

For concreteness, consider the inclusive pair production of 
some parent particle $\chi_p$, which can decay either to a
child particle $\chi_a$ of mass $M_c^{(a)}$, or 
a different child particle $\chi_b$ of mass $M_c^{(b)}$.
Let the corresponding branching fractions be $B_a$ and $B_b$, i.e.
$B_a\equiv B(\chi_p\to\chi_a)$ and $B_b\equiv B(\chi_p\to\chi_b)$.
Furthermore, let $\chi_b$ decay invisibly\footnote{If $\chi_b$
decays visibly, then the respective types of events 
can in principle be sorted by their signature.}
to $\chi_a$. Such a situation can be easily realized in supersymmetry, 
for example, with the parent being a squark, a slepton, or a gluino, 
the heavier child $\chi_b$ being a Wino-like neutralino $\tilde\chi^0_2$ and 
the lighter child $\chi_a$ being a Bino-like neutralino $\tilde\chi^0_1$.
The heavier neutralino has a large invisible decay mode
$\tilde \chi^0_2\to\tilde\chi^0_1\nu\bar{\nu}$, if its mass happens to fall
between the sneutrino mass and the left-handed slepton mass:
$M_{\tilde\nu}<M_{\tilde\chi^0_2}<M_{\tilde\ell_L}$.

Let us start with a certain total number of events $N_{pp}$ in which 
two parent particles $\chi_p$ have been produced. 
Then the missing energy sample will contain 
$N_{bb}=N_{pp} B_b^2$ symmetric events where the two children are $\chi_b$ and $\chi_b$,
$N_{aa}=N_{pp} B_a^2$ symmetric events where the two children are $\chi_a$ and $\chi_a$,
and $N_{ab}=2N_{pp} B_aB_b$ asymmetric events where the two children are $\chi_a$ and $\chi_b$.
How can one analyze such a mixed event sample with a single
$M_{T2}$ variable? 

The black histogram in Fig.~\ref{fig:110ex} shows 
the unit-normalized $M_{T2}$ distribution for the whole 
(mixed) event sample (for convenience, we do not show the zero 
bin \cite{Konar:2009wn}). 
\FIGURE[t]{
\centerline{
\epsfig{file=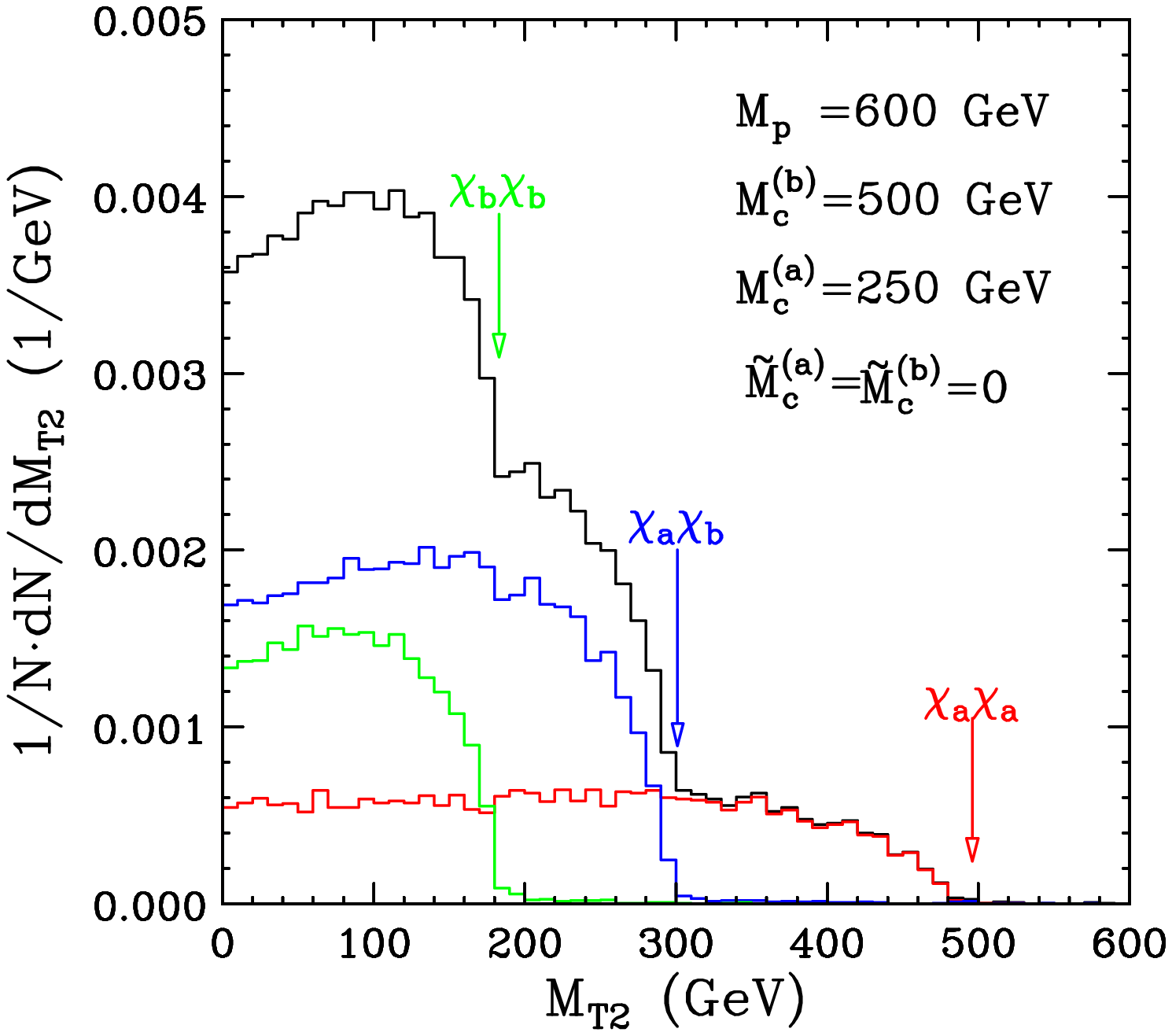,width=9cm}}
\caption{\it Unit-normalized, zero-bin subtracted $M_{T2}$ distribution (black histogram) 
for the full mixed event sample, as well as the individual components
$\chi_a\chi_a$ (red), $\chi_a\chi_b$ (blue) and $\chi_b\chi_b$ (green).
We took zero test masses for the children
$\tilde M^{(a)}_c=\tilde M^{(b)}_c=0$ and
equal branching fraction for the parents $B_a=B_b=50\%$.
The mass spectrum is taken from the asymmetric study point I in 
Table~\ref{tab:mass} with $M^{(a)}_c=250$ GeV, $M^{(b)}_c=500$ GeV and 
$M_p=600$ GeV. The three arrows indicate the expected endpoints for each
individual component in the sample. }
\label{fig:110ex}
}
For this plot, we used the asymmetric mass spectrum I from table~\ref{tab:mass}:
$M^{(a)}_c=250$ GeV, $M^{(b)}_c=500$ GeV and $M_p=600$ GeV,
and chose zero test masses for the children
$\tilde M^{(a)}_c=\tilde M^{(b)}_c=0$.
For definiteness, we fixed equal branching fractions $B_a=B_b=50\%$, so that
the relative normalization of the three individual samples is
$N_{aa}:N_{bb}:N_{ab}=1:1:2$. Fig.~\ref{fig:110ex} shows that
the observable $M_{T2}$ distribution is simply a superposition 
of the $M_{T2}$ distributions of the three individual samples 
$\chi_a\chi_a$, $\chi_a\chi_b$ and $\chi_b\chi_b$, which are shown 
with the red, blue and green histograms, correspondingly.
Each individual sample exhibits its own $M_{T2}$ endpoint,
marked with a vertical arrow, which can also be seen 
in the combined $M_{T2}$ distribution. Using eq.~(\ref{mt2max110}),
the three endpoints are found to be
{\allowdisplaybreaks
\begin{eqnarray}
\chi_a\chi_a &\rightarrow& M_{T2(max)}^{(aa)}(0,0,0)=
M_p\, \left[1 - \left(\frac{M_c^{(a)}}{M_p}\right)^2\right] = 496\ {\rm GeV}, 
\label{mt2aa} \\ [2mm]
\chi_a\chi_b &\rightarrow& M_{T2(max)}^{(ab)}(0,0,0)=
M_p\, \sqrt{\left[1-\left(\frac{M_c^{(a)}}{M_p}\right)^2\right]
            \left[1- \left(\frac{M_c^{(b)}}{M_p}\right)^2\right]}
=301\ {\rm GeV},~~ \label{mt2ab} \\ [2mm]
\chi_b\chi_b &\rightarrow& M_{T2(max)}^{(bb)}(0,0,0)=
M_p\, \left[1 - \left(\frac{M_c^{(b)}}{M_p}\right)^2\right]=183\ {\rm GeV}.
\label{mt2bb} 
\end{eqnarray} 
}

Now suppose that all three endpoints (\ref{mt2aa}-\ref{mt2bb})
are seen in the data. Their interpretation is far from obvious,
and in fact, there will be different competing explanations.
If one insists on the single missing particle hypothesis,
there can be only one type of child particle, and the only way to
get three different endpoints in Fig.~\ref{fig:110ex} 
is to have production of three different pairs of parent 
particles, each of which decays in exactly the same way. 
Since the three parent masses are a priori 
unrelated, one does not expect any particular correlation among the three
observed endpoints (\ref{mt2aa}-\ref{mt2bb}). Now consider an
alternative explanation where we produce a single type of parents,
but have two different children types. This situation also gives rise
to three different event topologies, with three different $M_{T2}$ endpoints,
as we just discussed. However, now there is a predicted relation
among the three $M_{T2}$ endpoints, which follows simply from 
eqs.~(\ref{mt2aa}-\ref{mt2bb}):
\begin{equation}
        M_{T2(max)}^{(ab)}(0,0,0)
= \sqrt{M_{T2(max)}^{(aa)}(0,0,0)\,
        M_{T2(max)}^{(bb)}(0,0,0)}\, .
\label{eqn:mt2maxrel}
\end{equation}
If the parents are the same and the children are different, 
this relation must be satisfied. If the parents are different 
and the children are the same, a priori there is no reason why 
eq.~(\ref{eqn:mt2maxrel}) should hold, and if it does, it must 
be by pure coincidence. The prediction (\ref{eqn:mt2maxrel})
therefore is a direct test of the number of children particles.
Another test can be performed if we could estimate the 
individual event counts $N_{aa}$, $N_{ab}$ and $N_{bb}$,
although this appears rather difficult, due to the unknown 
shape of the $M_{T2}$ distributions in Fig.~\ref{fig:110ex}.
In the asymmetric example discussed here, we have another
prediction, namely
\begin{equation}
N_{ab} = 2 \sqrt{N_{aa} N_{bb}} \, ,
\label{eq:rate}
\end{equation}
which is another test of the different children hypothesis.
Notice that eq.~(\ref{eq:rate}) holds regardless of the
branching fractions $B_a$ and $B_b$, although if one of them dominates, 
the two endpoints which require the other (rare) decay may be
too difficult to observe.

Of course, the ultimate test of the single missing particle
hypothesis is the behavior of the intermediate
$M_{T2}$ endpoint in Fig.~\ref{fig:110ex} corresponding to
the asymmetric events of type $\chi_a\chi_b$.
Applying either one of the two mass determination methods 
discussed earlier in Figs.~\ref{fig:110dlspridge} and \ref{fig:110elspridge},
we should find that $M_{T2(max)}^{(ab)}$
is a result of asymmetric events, indicating the 
simultaneous presence of two {\em different} invisible particles in the data.

\section{A more complex event topology: two SM particles on each side}
\label{sec:220}

In this section, we consider two more examples: 
the off-shell event topology of Fig.~\ref{fig:dlsp}(b) is discussed in
Sec.~\ref{sec:220off}, while the on-shell event topology of 
Fig.~\ref{fig:dlsp}(c) is discussed in Sec.~\ref{sec:220on}.
(For simplicity, we do not consider any $P_{UTM}$ in this section.)
Now there are two visible particles in each leg,
which form a composite visible particle of varying mass 
$m_{(\lambda)}$. In general, by studying the invariant mass 
distribution of $m_{(\lambda)}$, one should be able to observe 
two different invariant mass endpoints, suggesting some type of 
an asymmetric scenario.

\subsection{Off-shell intermediate particle}
\label{sec:220off}

Here we concentrate on the example of Fig.~\ref{fig:dlsp}(b).
Since the intermediate particle is offshell, the maximum 
kinematically allowed value for $m_{(\lambda)}$ is given by
eq.~(\ref{eq:mmaxoff}).

Recall that for the simple topology of Fig.~\ref{fig:dlsp}(a)
discussed in the previous section, the $M_{T2}$ endpoint 
(\ref{mt2max110}) always corresponded to a balanced solution.
More precisely, the $M_{T2}$ variable was maximized for a momentum 
configuration $\vec{p}_T^{~(\lambda)}$ in which $M_{T2}$ was
given by the balanced solution (\ref{mt2b}).
However, in this section we shall find that
for the more complex topologies of Figs.~\ref{fig:dlsp}(b)
and \ref{fig:dlsp}(c), the $M_{T2}$ endpoint may result from one of
four different cases altogether: two different balanced solutions, 
which we shall label as $B$ and $B'$, or the unbalanced solutions
$Ua$ and $Ub$ discussed in Sec.~\ref{sec:mt2dlspcomp}. 
Depending on the type of solution giving the endpoint $M_{T2(max)}$,
the $(\tilde M_c^{(a)},\tilde M_c^{(b)})$ parameter plane divides
into the three regions\footnote{The fourth case of the $B'$ balanced 
solution happens to coincide with the two unbalanced solutions along the  
boundary between $Ua$ and $Ub$.}
shown in Fig.~\ref{fig:offshelldiag}.
\FIGURE[t]{
\centerline{
\hspace{0.5cm}
\epsfig{file=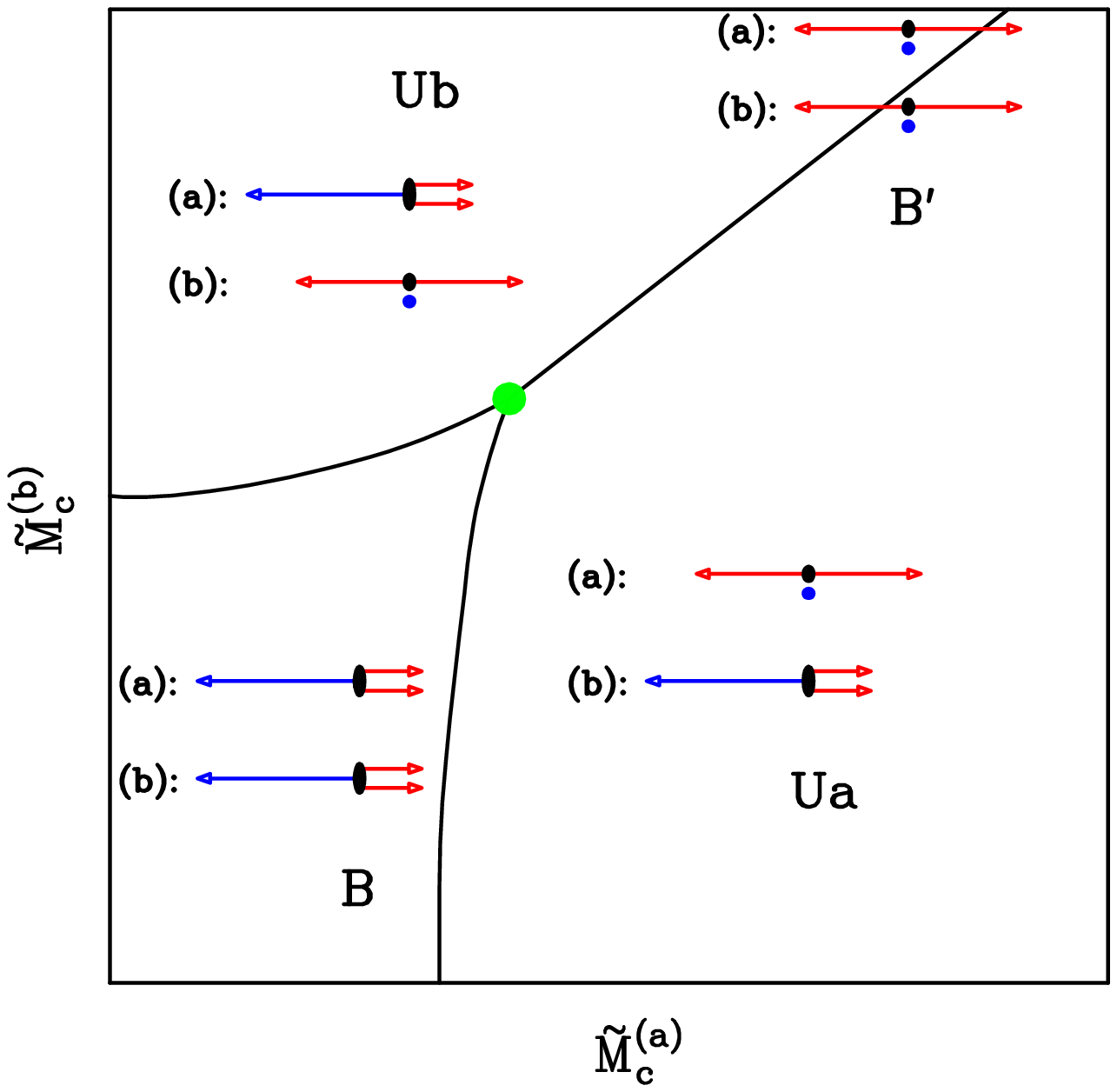, width=7.5cm}  }
\caption{\it  The four regions in the $(\tilde M_c^{(a)},\tilde M_c^{(b)})$ 
parameter plane leading to the four different types of solutions for the $M_{T2}$ 
endpoint, for the off-shell event topology of Fig.~\ref{fig:dlsp}(b). 
The green dot marks the true location of the two children masses.
Within each region, we indicate the relevant momentum configuration 
for the visible particles (red arrows) and the children 
particles (blue arrows) in each leg ($a$ or $b$). 
The momenta are quoted in the ``back-to-back
boosted" (BB) frame \cite{Cho:2007dh}, in which the two parents are at rest. 
A blue dot implies that the corresponding daughter is at rest and 
therefore the two visible particles are emitted back-to-back. 
The two balanced solutions are denoted as $B$ and $B'$, 
while the two unbalanced solutions are $Ua$ and $Ub$. The black
solid lines represent phase changes between different solution types
and delineate the expected locations of the ridges in the $M_{T2(max)}$
function shown in Fig.~\ref{fig:off3d}. }
\label{fig:offshelldiag}
}
The green dot in Fig.~\ref{fig:offshelldiag} denotes the
true children masses in this parameter space. 
Within each region, we show the relevant momentum configuration 
for the visible particles (red arrows) and the children 
particles (blue arrows) in each leg ($a$ or $b$). 
The momenta are quoted in the ``back-to-back
boosted" (BB) frame \cite{Cho:2007dh}, in which the two parents are at rest. 
The length of an arrow is indicative of the magnitude of the momentum. 
A blue dot implies that the corresponding daughter is at rest and 
therefore the two visible particles are emitted back-to-back. 
The two balanced solutions are denoted as $B$ and $B'$, 
while the two unbalanced solutions are $Ua$ and $Ub$. The black
solid lines represent phase changes between different solution types
and delineate the expected locations of the ridges in the $M_{T2(max)}$
function shown in Fig.~\ref{fig:off3d} below.
Perhaps the most striking feature of Fig.~\ref{fig:offshelldiag}
is that the three (in fact, all four) regions come together 
precisely at the green dot marking 
the true values of the two children masses. The boundaries of the 
regions shown in Fig.~\ref{fig:offshelldiag} will manifest themselves
as the locations of the ridges (i.e. gradient discontinuities) in 
the $M_{T2(max)}$ function. Therefore, we expect that by studying the 
ridge structure and finding its ``triple'' point, one 
will be able to completely determine the mass spectrum.

We shall now give analytical formulas for the $M_{T2}$ endpoint
in each of the four regions of Fig.~\ref{fig:offshelldiag}.
We begin with the two balanced solutions $B$ and $B'$, 
for which the event-by-event balanced solution for $M_{T2}$ 
is given by eq.~(\ref{mt2b}).
In the parameter space region of Fig.~\ref{fig:offshelldiag} which is 
adjacent to the origin, we find the balanced configuration $B$,
in which all visible particles have the same direction in the BB frame.
As a result, we have
\begin{equation}
m_{(a)} = m_{(b)} = 0 \, \, \label{220offb1}
\end{equation}
and
\begin{equation}
A_T = \frac{(M_p^2 - \mazsq)(M_p^2 - \mbzsq)}{2 M_p^2} \, .
\label{220offb2}
\end{equation}
Substituting eqs.~(\ref{220offb1}) and (\ref{220offb2}) in the balanced
$M_{T2}$ solution (\ref{mt2b}), where we should take the plus sign, we obtain
\begin{equation}
\left[M^B_{T2(max)} (\tmaz,\tmbz)\right]^2 
= 2\bar\mu^2_{ppc} +\tilde{M}_+^2 
+ \sqrt{ 4\,\bar\mu^2_{ppc}(\bar\mu_{ppc}^2+\tilde M_+^2) + \tilde M_-^4} \, ,
\label{mt2b220}
\end{equation}
which we recognize as the balanced solution (\ref{mt2max110})
found for the decay topology of Fig.~\ref{fig:dlsp}(a).

Moving away from the origin in Fig.~\ref{fig:offshelldiag}, 
we find a second balanced solution $B'$ along the boundary of the 
unbalanced regions $Ua$ and $Ub$. In this case the visible particles are 
back-to-back, and their invariant mass is maximized:
\begin{equation}
m_{(\lambda)} = M_p - M_c^{(\lambda)} \, , 
\label{220offbp1}
\end{equation}
and correspondingly
\begin{equation}
A_T = \left(M_p - M_c^{(a)}\right)
      \left(M_p - M_c^{(b)}\right)\, .
\label{220offbp2}
\end{equation}
Substituting eqs.~(\ref{220offbp1}) and (\ref{220offbp2}) in the balanced
$M_{T2}$ solution (\ref{mt2b}), we obtain the $B'$-type $M_{T2}$ endpoint as
\begin{equation}
\left[M^{B'}_{T2(max)} (\tmaz,\tmbz)\right]^2 
= \left(M_p - M_c^{(a)}\right)\left(M_p - M_c^{(b)}\right)
+ \tilde M_+^2
+ \frac{2M_p-M_c^{(a)}-M_c^{(b)}}{M_c^{(b)}-M_c^{(a)}}\, \tilde M_-^2.
\label{mt2bp220}
\end{equation}

The corresponding formulas for the unbalanced cases $Ua$ and $Ub$
are obtained by taking the maximum value for the invariant 
mass of the visible particles in the corresponding decay chain:
\begin{eqnarray}
m_{(a)} &=& m_{(a)}^{max}= M_p - \maz ~~~\text{for ~region ~(Ua)} \, ,\\
m_{(b)} &=& m_{(b)}^{max}= M_p - \mbz ~~~\text{for ~region ~(Ub)} \, .
\end{eqnarray}
The corresponding formula for $M_{T2(max)}$ is then given by
\begin{eqnarray}
M_{T2(max)}^{Ua}(\tmaz)&=& M_p - \maz  + \tmaz  \, , 
\label{220u1}\\
M_{T2(max)}^{Ub}(\tmbz)&=& M_p - \mbz  + \tmbz  \, .
\label{220u2}
\end{eqnarray}

One can now use the analytical results (\ref{mt2b220}),
(\ref{mt2bp220}), (\ref{220u1}) and (\ref{220u2})
to understand the ridge structure shown in Fig.~\ref{fig:offshelldiag}.
For example, the boundary between the $B$ and $Ua$ regions is 
parametrically given by the condition 
\begin{equation}
M_{T2(max)}^{B}(\tmaz,\tmbz) = M_{T2(max)}^{Ua}(\tmaz)\, ,
\label{boundaryBUa}
\end{equation}
while the boundary between the $B$ and $Ub$ regions is 
parametrically given by 
\begin{equation}
M_{T2(max)}^{B}(\tmaz,\tmbz) = M_{T2(max)}^{Ub}(\tmbz)\, .
\label{boundaryBUb}
\end{equation}
On the other hand, the boundary 
\begin{equation}
M_{T2(max)}^{Ua}(\tmaz) = M_{T2(max)}^{Ub}(\tmbz)
\label{boundaryUaUb}
\end{equation}
between the two unbalanced regions $Ua$ and $Ub$ is 
quite interesting. The parametric equation 
(\ref{boundaryUaUb}) is nothing but a straight line
in the $(\tmaz,\tmbz)$ plane:
\begin{equation}
\tmbz = M_c^{(b)}-M_c^{(a)}+ \tmaz\, ,
\end{equation}
as seen in Fig.~\ref{fig:offshelldiag}.

It is now easy to understand the triple point structure in
Fig.~\ref{fig:offshelldiag}. The triple point
is obtained by the merging of all three boundaries
(\ref{boundaryBUa}), (\ref{boundaryBUb}) and
(\ref{boundaryUaUb}), i.e. when
\begin{equation}
  M_{T2(max)}^{B }(\tmaz,\tmbz) 
= M_{T2(max)}^{B'}(\tmaz,\tmbz)
= M_{T2(max)}^{Ua}(\tmaz)
= M_{T2(max)}^{Ub}(\tmbz) 
\, .
\end{equation}
It is easy to check that $\tilde M_c^{(a)}=M_c^{(a)}$ and
$\tilde M_c^{(b)}=M_c^{(b)}$ identically satisfy
these equations, thereby proving that 
the triple intersection of the boundaries 
seen in Fig.~\ref{fig:offshelldiag} indeed takes place 
at the true values of the children masses.

These results are confirmed in our numerical simulations.
\FIGURE[ht]{
\centerline{
\epsfig{file=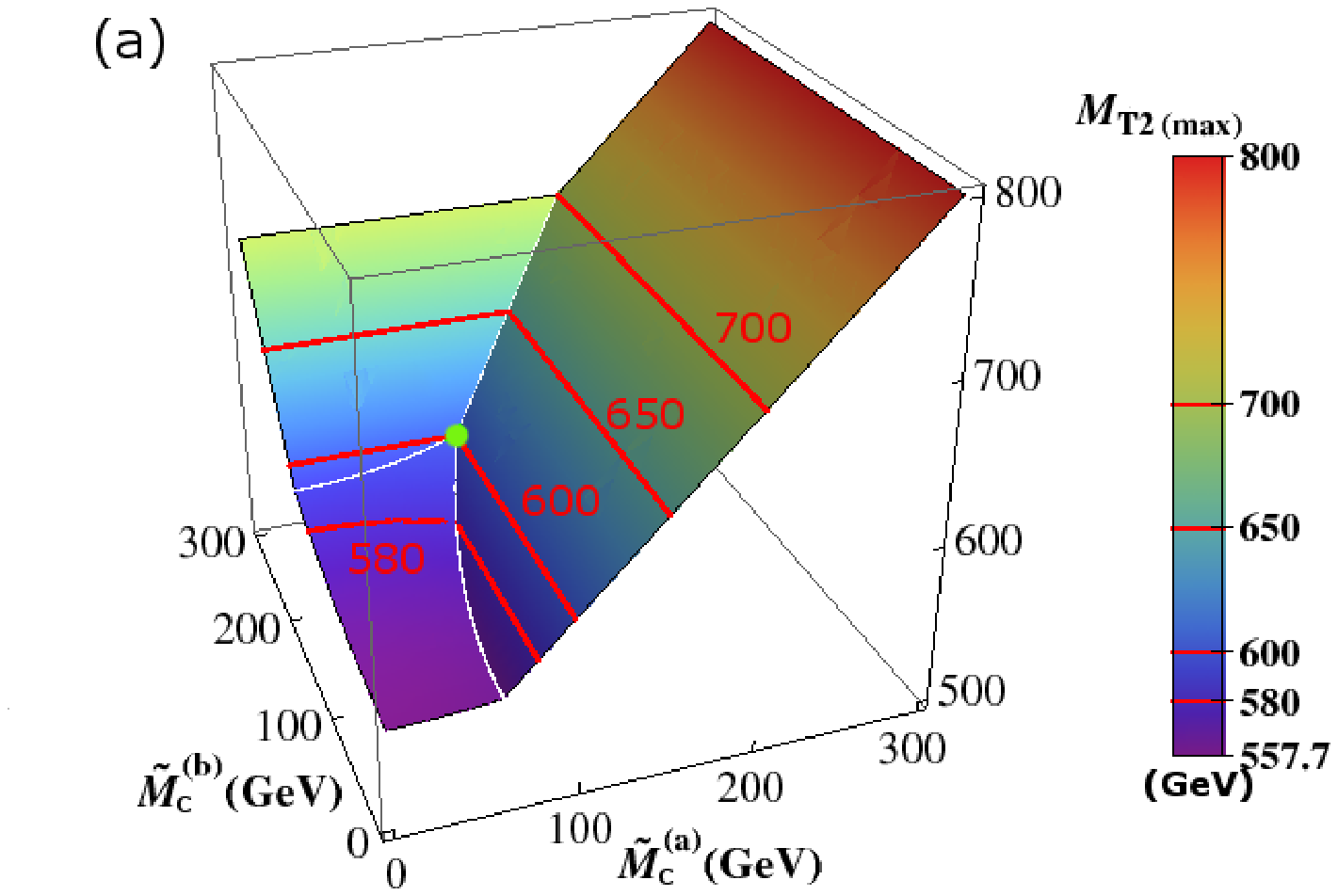,      height=5.7cm}
\epsfig{file=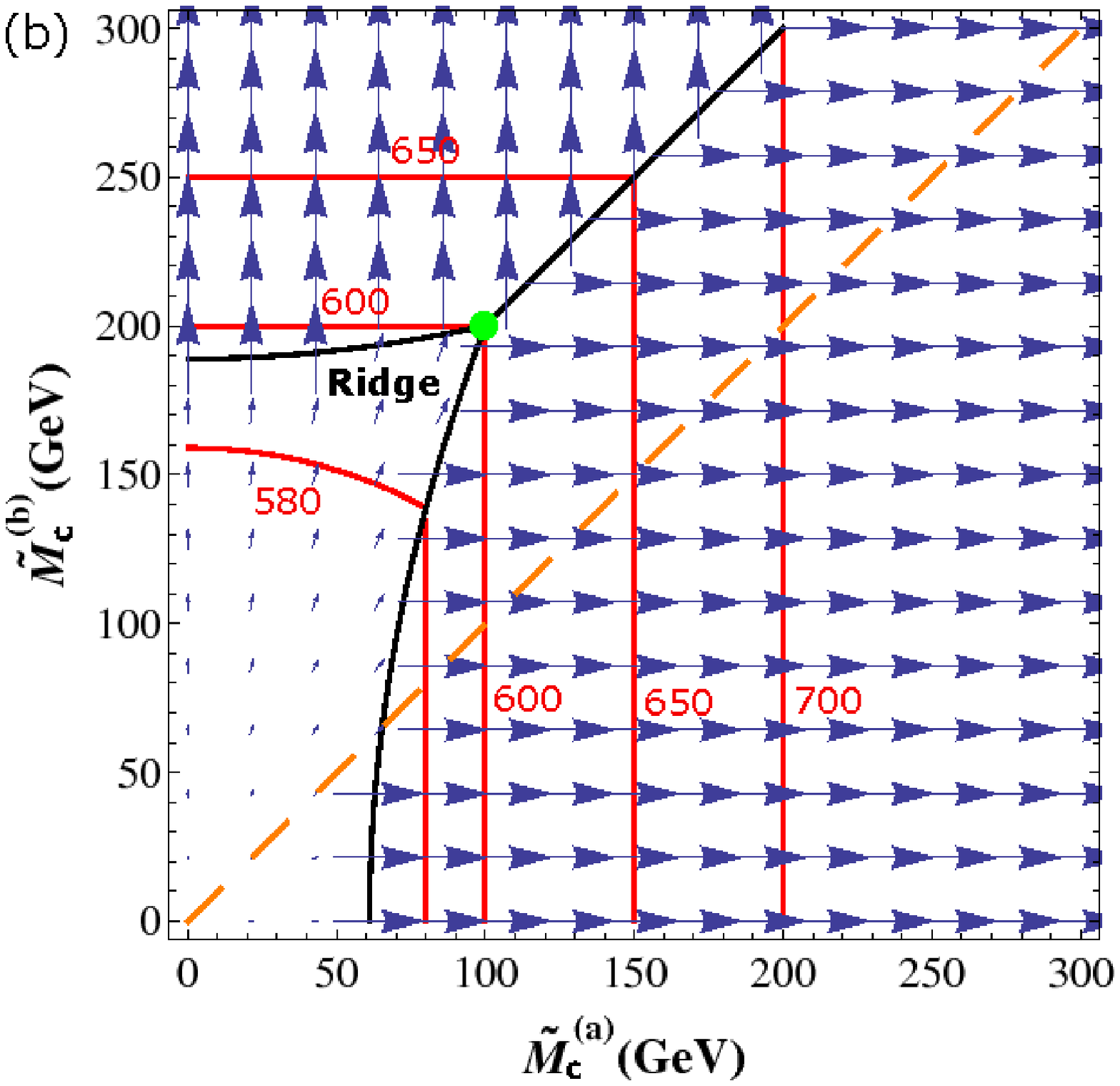, height=5.7cm} }
\caption{\it The same as Fig.~\ref{fig:110}, but for 
the off-shell event topology of Fig.~\ref{fig:dlsp}(b). 
We use the mass spectrum from the example in Fig.~\ref{fig:offshellpairing}:
$M_c^{(a)}=100$ GeV, $M_c^{(b)}=200$ GeV and $M_p=600$ GeV
and for simplicity consider only events with $P_{UTM}=0$.}
\label{fig:off3d}
}
In Fig.~\ref{fig:off3d} we present (a) a three dimensional view 
and (b) a gradient plot of the ridge structure found in 
events with the off-shell topology of Fig.~\ref{fig:dlsp}(b).
The mass spectrum for this study point was fixed as in
Fig.~\ref{fig:offshellpairing}, namely
$M_c^{(a)}=100$ GeV, $M_c^{(b)}=200$ GeV and $M_p=600$ GeV.
Since the ridge structure for this topology 
does not require the presence of upstream momentum,
for simplicity we consider only events with $P_{UTM}=0$.
The ridge pattern is clearly evident in Fig.~\ref{fig:off3d}(a),
which shows a three-dimensional view of the $M_{T2}$ endpoint
function $M_{T2(max)}(\tilde M_c^{(a)},\tilde M_c^{(b)})$.
It is even more apparent in Fig.~\ref{fig:off3d}(b),
where one can see a sharp gradient change along the ridge lines:
in regions $Ua$ and $Ub$, the corresponding gradient vectors 
point in trivial directions (either horizontally or vertically), 
in accord with eqs.~(\ref{220u1})-(\ref{220u2}). On the other hand,
the gradient in region $B$ is very small, and 
the $M_{T2}$ endpoint function is rather flat.
The green dot marks the location of the true children masses 
($M_c^{(a)}=100$ GeV, $M_c^{(b)}=200$ GeV) and is indeed the
intersection point of the three ridgelines.
As expected, the corresponding $M_{T2(max)}$ at that point is
the true parent particle mass $M_p=600$ GeV.

At this point, it is interesting to ask the question, what 
would be the outcome of this exercise if one were to
make the usual assumption of identical children, and apply 
the traditional symmetric $M_{T2}$ to this situation.
The answer can be deduced from Fig.~\ref{fig:off3d}(b),
where the diagonal orange dotdashed line corresponds to the
usual assumption of $\tilde M_c^{(a)}=\tilde M_c^{(b)}$.
In that case, one still finds a kink, but at the wrong location:
in Fig.~\ref{fig:off3d}(b) the intersection of the diagonal orange line 
and the solid black ridgeline occurs at $\tilde M_c^{(a)}=\tilde M_c^{(b)}=65.3$ GeV 
and the corresponding parent mass is $\tilde M_p=565.3$ GeV.
Therefore, the traditional kink method can easily lead to
a wrong mass measurement. Then the only way to know that
there was something wrong with the measurement would be to 
study the effect of the upstream momentum and see that the 
observed kink is not invariant under $P_{UTM}$.

We should note that, depending on the actual mass spectrum, 
the two-dimensional ridge pattern seen in Figs.~\ref{fig:offshelldiag}
and \ref{fig:off3d}(b) may look very differently. For example,
the balanced region $B$ may or may not include the origin. 
One can show that if 
\begin{equation}
M_p < 
\frac{\mb}{4 \ma}\left( \mb + \sqrt{8 \masq + \mbsq}\right)\, ,
\end{equation} 
the boundary between $B$ and $Ua$ does not cross the $\tilde M_c^{(a)}$
axis. In this case the diagonal line in Fig.~\ref{fig:off3d}(b)
does not cross any ridgelines and the traditional $M_{T2}$ 
approach will not produce any kink structure, in contradiction 
with one's expectations. This exercise teaches us that the failsafe 
approach to measuring the masses in missing energy events is to
apply from the very beginning the asymmetric $M_{T2}$ concept 
advertized in this paper.

\subsection{On-shell intermediate particle}
\label{sec:220on}

Our final example is the on-shell event topology 
illustrated in Fig.~\ref{fig:dlsp}(c).
Now there is an additional parameter which enters
the game --- the mass $M_i^{(\lambda)}$ of the
intermediate particle in the $\lambda$-th decay chain.
As a result, the allowed range of invariant masses for the 
visible particle pair on each side is limited from above by 
eq.~(\ref{eq:mmaxon}).

In this case we find that the $M_{T2}$ endpoint exhibits a similar 
phase structure as the one shown in Fig.~\ref{fig:offshelldiag}.
\FIGURE[t]{
\centerline{
\hspace{0.5cm}
\epsfig{file=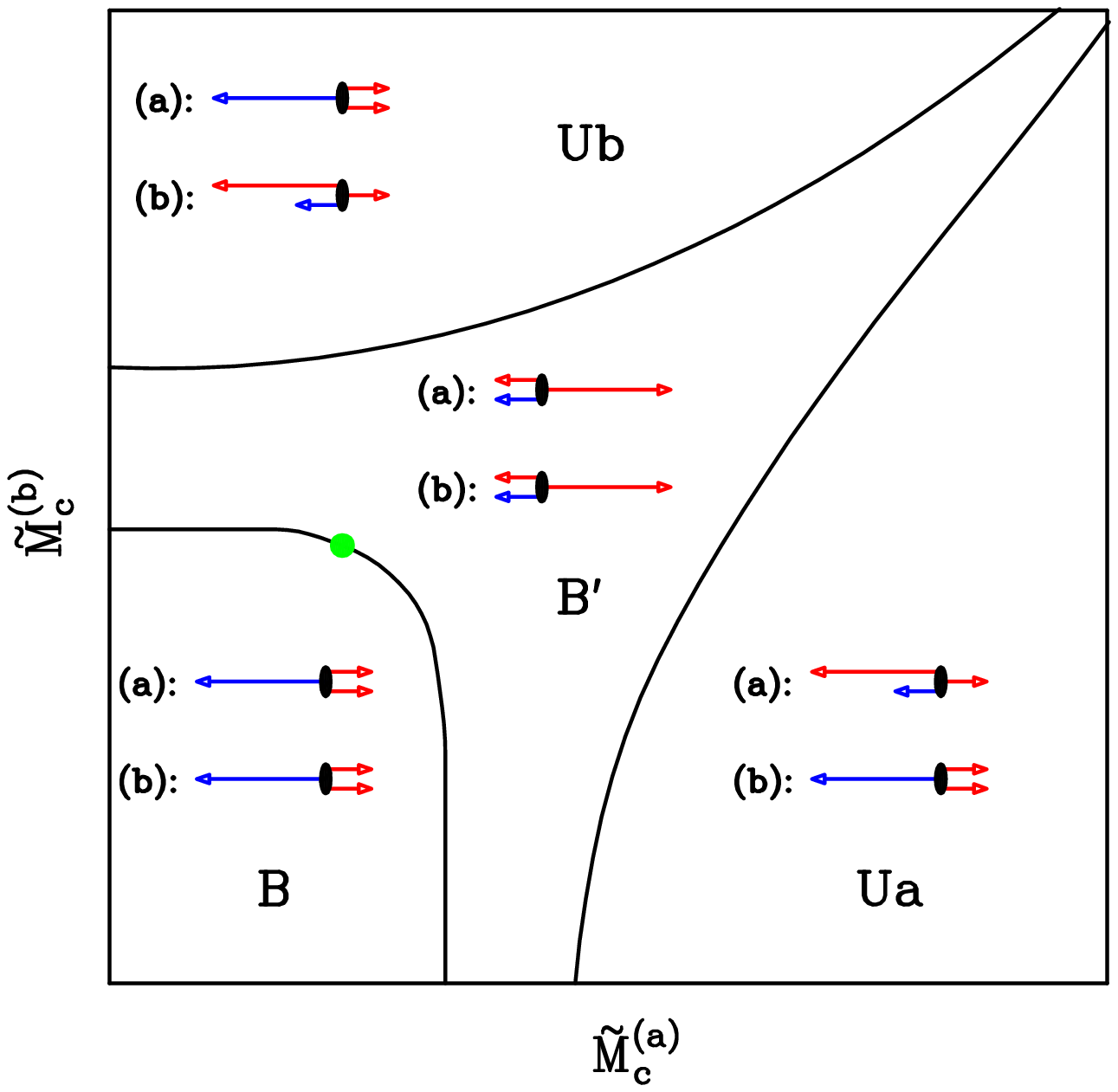, width=7.5cm}  }
\caption{\it 
The same as Fig.~\ref{fig:offshelldiag} but for the onshell
scenario illustrated in Fig.~\ref{fig:dlsp}(c). }
\label{fig:onshelldiag}
}
One particular pattern is illustrated in Fig.~\ref{fig:onshelldiag}, which 
exhibits the same four regions 
$B$, $B'$, $Ua$ and $Ub$ seen in Fig.~\ref{fig:offshelldiag}.
The difference now is that region $B'$ is considerably expanded,
and as a result, region $B$ does not have a common border 
with regions $Ua$ and $Ub$ any more.
The triple point of Fig.~\ref{fig:offshelldiag}
has now disappeared and the correct values
of the children masses now lie somewhere on the border between
regions $B$ and $B'$, but their exact location along this ridgeline 
is at this point unknown. 

Just like we did for the off-shell case in Sec.~\ref{sec:220off},
we shall now present analytical formulas for the $M_{T2}$ endpoint
in each region of Fig.~\ref{fig:onshelldiag}. In the balanced 
region $B$, we find the same results (\ref{220offb1}-\ref{mt2b220})
as in the off-shell case considered in the previous Section \ref{sec:220off}.
The other balanced region $B'$ is characterized by
\begin{equation}
m_{(\lambda)} = m_{(\lambda)}^{max}\, ,
\label{mvis220on}
\end{equation}
where $m_{(\lambda)}^{max}$ is given by eq.~(\ref{eq:mmaxon}), and
\begin{eqnarray}
A_T
&=&
\frac{M_p^2}{4} \, 
\left[\, 2-\left(\frac{M_i^{(a)}}{M_p}\right)^2 -\left(\frac{M_c^{(a)}}{M_i^{(a)}}\right)^2  \right]
\left[\, 2-\left(\frac{M_i^{(b)}}{M_p}\right)^2 -\left(\frac{M_c^{(b)}}{M_i^{(b)}}\right)^2  \right] \nonumber \\ [2mm]
&+&
\frac{M_p^2}{4} \, 
\left| 
\left[\left(\frac{M_c^{(a)}}{M_i^{(a)}}\right)^2-\left(\frac{M_i^{(a)}}{M_p}\right)^2\right]  
\left[\left(\frac{M_c^{(b)}}{M_i^{(b)}}\right)^2-\left(\frac{M_i^{(b)}}{M_p}\right)^2\right]  
\right|
\, .
\label{mt2220on}
\end{eqnarray}
The formula for the endpoint $M_{T2(max)}^{B'}$ in region $B'$ 
is then simply obtained by substituting (\ref{mvis220on})
and (\ref{mt2220on}) into the balanced solution (\ref{mt2b}).

Finally, the $M_{T2}$ endpoint in the unbalanced
regions $Ua$ and $Ub$ is given by
\begin{eqnarray}
M_{T2(max)}^{Ua}(\tmaz)&=& m_{(a)}^{max}  + \tmaz  \, , 
\label{220onua}\\
M_{T2(max)}^{Ub}(\tmbz)&=& m_{(b)}^{max}  + \tmbz  \, ,
\label{220onub}
\end{eqnarray}
where $m_{(a)}^{max}$ and $m_{(b)}^{max}$ 
are given by eq.~(\ref{eq:mmaxon}).

\FIGURE[t]{
\centerline{\hspace*{-1cm}
\epsfig{file=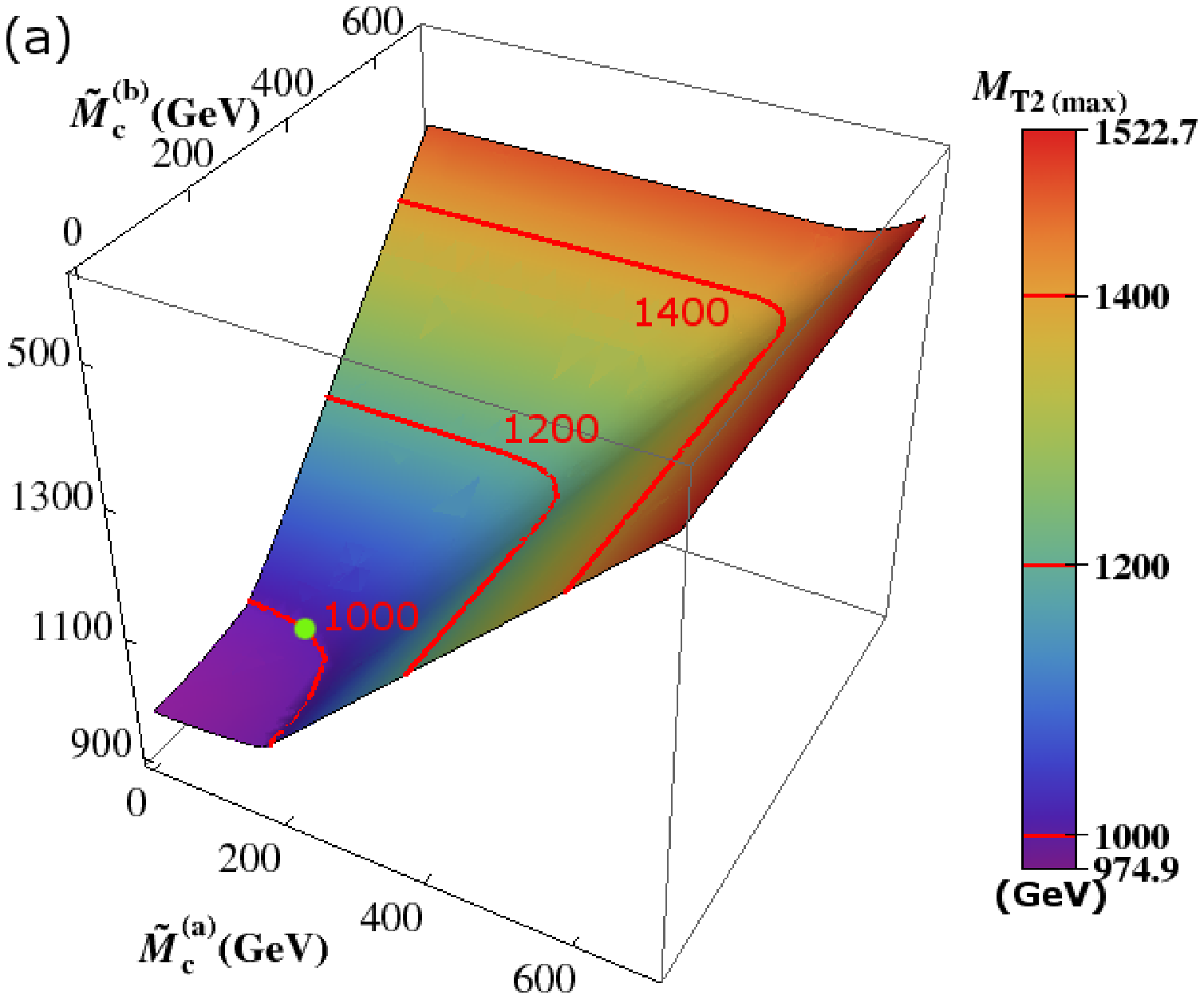,     height=6.3cm}
\epsfig{file=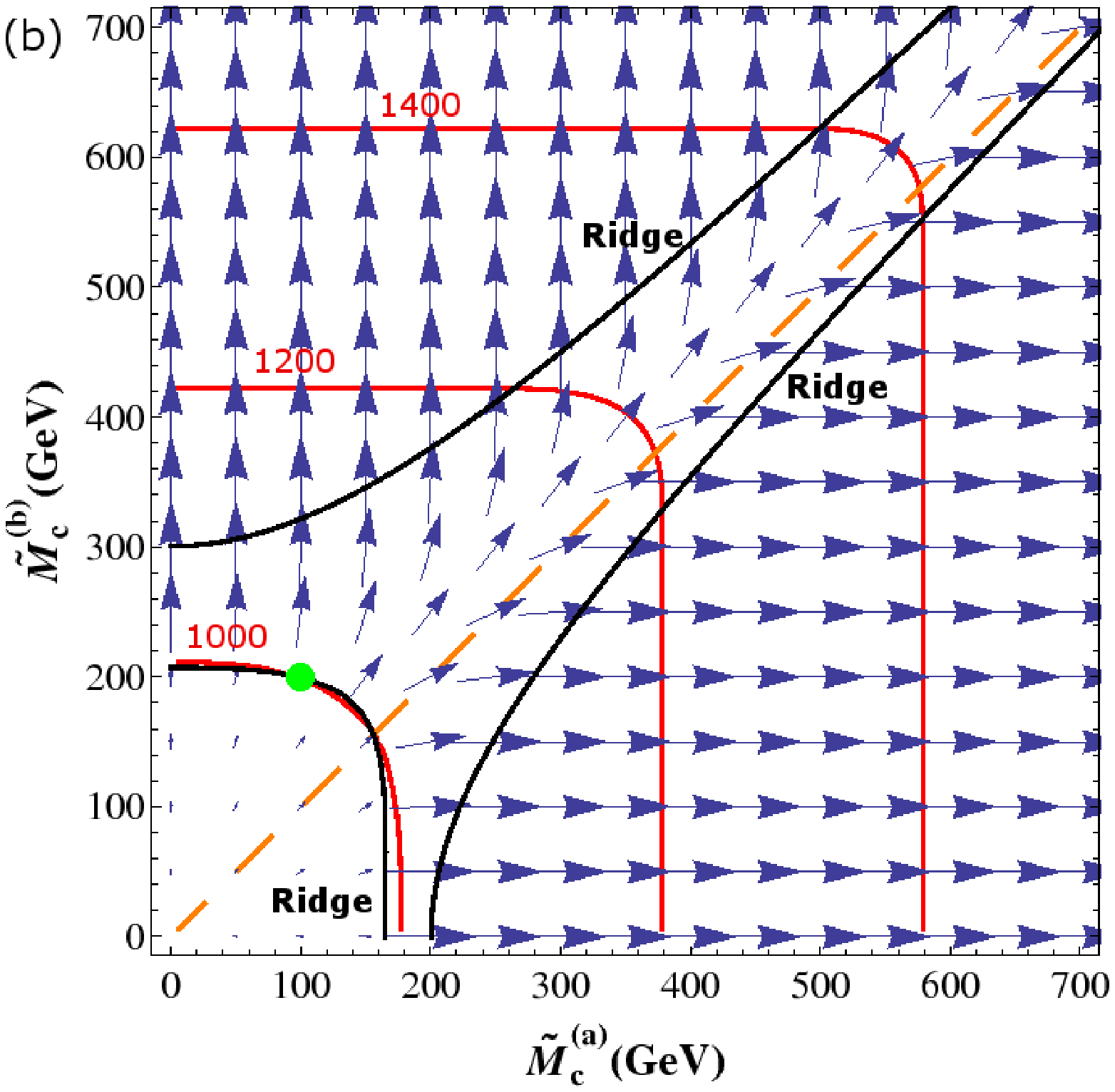,height=6.3cm} 
}
\caption{ The same as in Fig.~\ref{fig:off3d} but for the onshell scenario
of Fig.~\ref{fig:dlsp}(c), with a mass spectrum 
$\maz=100$ GeV, $\mbz=200$ GeV,
$M_i^{(a)}=M_i^{(b)}=550$ GeV and $M_p=1$ TeV.
}
\label{fig:onshell}
}

In Fig.~\ref{fig:onshell} we present our numerical results in this
on-shell scenario. The mass spectrum is fixed as:
$\maz=100$ GeV, $\mbz=200$ GeV,
$M_i^{(a)}=M_i^{(b)}=550$ GeV and $M_p=1$ TeV, and we still do not include 
the effects of any upstream momentum.
Fig.~\ref{fig:onshell}(a) shows the three-dimensional view of the 
$M_{T2}$ endpoint function $M_{T2(max)}(\tilde M_c^{(a)},\tilde M_c^{(b)})$,
which exhibits three different sets of ridges, which 
are more easily seen in the gradient plot of Fig.~\ref{fig:onshell}(b).
As usual, the green dot marks the true children masses.
Fig.~\ref{fig:onshell}(b) shows that the ridgeline 
separating the two balanced regions $B$ and $B'$ 
does go through the green dot and thus reveals 
a relationship between the two children masses,
leaving the ridgeline parameter $\theta$ as 
the only remaining unknown degree of freedom.
However, unlike the off-shell case of Sec.~\ref{sec:220off},
now there is no special point on this ridgeline, 
and we cannot completely pin down the masses by the ridge 
method. Thus, in order to determine all masses 
in the problem, one must use an additional piece of information, 
for example the visible invariant mass endpoint 
(\ref{eq:mmaxon}) or the $P_{UTM}$ invariance method
suggested in Sec.~\ref{sec:PUTMdlsp}.

\section{Summary and conclusions}
\label{sec:conclusions}

Cosmological observations hint towards the existence of one 
or more hypothetical dark matter particles. The start
of the Large Hadron Collider may offer an unique opportunity to
produce and study dark matter in a high-energy experimental 
laboratory. Unfortunately, the dark matter signatures 
at colliders always involve missing transverse energy.
Such events will be quite challenging to fully 
reconstruct and/or interpret. All previous studies have 
made (either explicitly or implicitly) the assumption that
each event has two {\em identical} missing particles.
Our main point in this paper is that this assumption is unnecessary,
and by suitable modifications of the existing analysis techniques
one can in principle test both the number and the type of 
missing particles in the data. Our proposal here was to
modify the Cambridge $M_{T2}$ variable \cite{Lester:1999tx}
by treating each children mass as an independent input parameter.
In this approach, one obtains the $M_{T2}$ endpoint 
$M_{T2(max)}$ as a function of the two children masses
$\tilde M_c^{(a)}$ and $\tilde M_c^{(b)}$,
and proceeds to study its properties. 
The two most important features of the thus obtained function
$M_{T2(max)}(\tilde M_c^{(a)},\tilde M_c^{(b)})$, 
identified in this paper, were the following:
\begin{itemize}
\item The function $M_{T2(max)}(\tilde M_c^{(a)},\tilde M_c^{(b)})$
exhibits a ridge structure (i.e.~a gradient discontinuity),
as illustrated with specific examples in Figs.~\ref{fig:110isr},
\ref{fig:110_slspisr}, \ref{fig:off3d} and \ref{fig:onshell}. 
The point corresponding to the correct children masses 
always lies on a ridgeline, thus the ridgelines 
provide a model-independent constraint among the children masses,
just like the $M_{T2}$ endpoint provides a model-independent 
constraint on the masses of the child(ren) and the parent.
\item In general, the $M_{T2}$ endpoint function
also depends on the value of the upstream transverse momentum in the event:
$M_{T2(max)}(\tilde M_c^{(a)},\tilde M_c^{(b)},P_{UTM})$.
However, the $P_{UTM}$ dependence disappears completely 
for precisely the right values of the children masses, 
as seen in the examples of
Figs.~\ref{fig:110dlspridge}(b) and \ref{fig:110elspridge}(b).
This provides a second, quite general and model-independent,
method for measuring the {\em individual} particle masses in such 
missing energy events.
\end{itemize}

Before we conclude, we shall discuss a few other possible
applications of the asymmetric $M_{T2}$ idea, 
besides the examples already considered in the paper.
\begin{enumerate}
\item {\em Invisible decays of the next-to-lightest particle.} 
Most new physics models introduce some new massive and neutral
particle which plays the role of a dark matter candidate.
Often the very same models also contain other, heavier particles,
which for collider purposes behave just like a dark matter candidate:
they decay invisibly and result in missing energy in the detector.
For example, in supersymmetry one may find an invisibly decaying
sneutrino $\tilde{\nu}_\ell\to \nu_\ell \tilde\chi_1^0$, in UED one finds
an invisibly decaying KK neutrino $\nu_1 \to \nu \gamma_1$, etc.
These scenarios can easily generate an asymmetric event topology.
For example, consider the strong production of a squark 
($\tilde{q}$) pair, as illustrated in Fig.~\ref{fig:dlspex}(a). 
One of the squarks subsequently decays to 
the second lightest neutralino $\tilde\chi^0_2$, which in turn
decays to the lightest neutralino $\tilde\chi^0_1$ by emitting two 
SM fermions $\tilde{\chi}_2^0 \to \ell^+ \ell^- \tilde{\chi}_1^0$ 
(or $\tilde{\chi}_2^0 \to j j \tilde{\chi}_1^0$). 
The other squark decays to a chargino $\tilde\chi^\pm_1$,
which then decays to a sneutrino as $\tilde{\chi}_1^\pm \to \ell^\pm
\tilde{\nu}_\ell$. Since $\tilde{\nu}_\ell$ can only decay invisibly,
we obtain the asymmetric event topology outlined with the blue box
in Fig.~\ref{fig:dlspex}(a).
The two squarks are the parents, the lightest neutralino $\tilde\chi^0_1$
is the first child, and the sneutrino $\tilde{\nu}_\ell$ is the second child.
%
\FIGURE[ht]{
{
\unitlength=1.5 pt
\SetScale{1.4}
\SetWidth{1.0}      
\normalsize    
{} \qquad\allowbreak
\hspace*{-4.5cm}
\begin{picture}(700,110)(35,-10)
%
\SetColor{Gray}
\Line(130,65)(170,65)
\Line(130,35)(170,35)
\Line(150,65)(170,95)
\Line(150,35)(170, 5)
\Line(200,65)(220,95)
\Line(230,65)(250,95)
\Line(260,65)(280,95)
\Line    (200,35)(220, 5)
\Line    (230,35)(250, 5)
\DashLine(260,35)(280, 5){1}
\SetColor{Red}
\SetWidth{1.2}      
\Line    (170,65)(260,65)
\DashLine(260,65)(290,65){1}
\Line    (170,35)(230,35)
\DashLine(230,35)(290,35){1}
\Text(133,67)[c]{\Black{$p(\bar{p})$}}
\Text(133,39)[c]{\Black{$p(\bar{p})$}}
\Text(155,89)[r]{\Black{ISR}}
\Text(155, 4)[r]{\Black{ISR}}
\Text(183,67)[c]{\Red{$\tilde{q}$}}
\Text(205,67)[c]{\Red{$\tilde{\chi}_2^0$}}
\Text(233,67)[c]{\Red{$\tilde{\ell}^\pm$}}
\Text(260,67)[c]{\Red{$\tilde{\chi}_1^0$}}
\Text(183,39)[c]{\Red{$\tilde{q}$}}  
\Text(205,39)[c]{\Red{$\tilde{\chi}_1^\pm$}}
\Text(233,39)[c]{\Red{$\tilde{\nu}_\ell$}}
\Text(260,39)[c]{\Red{$\tilde{\chi}_1^0$}}
\Text(200,89)[c]{\Black{$q$}}
\Text(228,89)[c]{\Black{$\ell^\mp$}}
\Text(257,89)[c]{\Black{$\ell^\pm$}}
\Text(200, 4)[c]{\Black{$q$}}
\Text(228, 4)[c]{\Black{$\ell^\pm$}}
\Text(257, 4)[c]{\Black{$\tilde{\nu}_\ell$}}
\COval(170,50)(30,10)(0){Blue}{Green}
\SetColor{Blue}
\Line(187,100)(187,  0)
\Line(187,  0)(255,  0)
\Line(255,  0)(255, 50)
\Line(255, 50)(285, 50)
\Line(285, 50)(285,100)
\Line(285,100)(187,100)
\SetColor{Gray}
\Line(310,65)(350,65)
\Line(310,35)(350,35)
\Line(330,65)(350,95)
\Line(330,35)(350, 5)
\Line(380,65)(400,95)
\Line(410,65)(430,95)
\Line    (380,35)(400, 5)
\DashLine(410,35)(430, 5){1}
\SetColor{Red}
\SetWidth{1.2}      
\Line    (350,65)(410,65)
\DashLine(410,65)(440,65){1}
\Line    (350,35)(380,35)
\DashLine(380,35)(440,35){1}
\Text(302,67)[c]{\Black{$p(\bar{p})$}}
\Text(302,39)[c]{\Black{$p(\bar{p})$}}
\Text(324,89)[r]{\Black{ISR}}
\Text(324, 4)[r]{\Black{ISR}}
\Text(352,66)[c]{\Red{$t$}}
\Text(373,66)[c]{\Red{$W^+$}}
\Text(400,66)[c]{\Red{$\nu_\ell$}}
\Text(352,38)[c]{\Red{$\bar{t}$}}
\Text(372,38)[c]{\Red{$W^-$}}
\Text(368,89)[c]{\Black{$b$}}
\Text(397,89)[c]{\Black{$\ell^+$}}
\Text(367, 5)[c]{\Black{$\bar{b}$}}
\COval(350,50)(30,10)(0){Blue}{Green}
\SetColor{Blue}
\Line(367,100)(367,  0)
\Line(367,  0)(405,  0)
\Line(405,  0)(405, 50)
\Line(405, 50)(435, 50)
\Line(435, 50)(435,100)
\Line(435,100)(367,100)
\Text(200, -15)[c]{\bf \Black{(a)}}
\Text(360, -15)[c]{\bf \Black{(b)}}
\end{picture}
}
\caption{\it Event topology for the two examples discussed in 
Section~\ref{sec:conclusions}.
The black solid lines represent SM particles which are visible in
the detector while red solid lines represent particles at 
intermediate sages. The missing particles are denoted by dotted
lines. (a) Squark pair production with decay chains terminating
in two different invisible particles ($\tilde{\chi}_1^0$ and
$\tilde{\nu}_\ell$, correspondingly). In this case $\tilde{\nu}_\ell$ 
decays invisibly.
(b) The subsystem $M_{T2}$ variable applied to $t\bar{t}$ events. 
The $W$-boson in the lower leg is treated as a child particle
and can decay either hadronically or leptonically.}
\label{fig:dlspex}
}
\item {\em Applying $M_{T2}$ to an asymmetric subsystem.}
One can also apply the $M_{T2}$ idea even to events in which there is only one 
(or even no) missing particles to begin with. Such an example is shown 
in Fig.~\ref{fig:dlspex}(b), where we consider $t\bar{t}$
production in the dilepton or semi-leptonic channel. 
In the first leg we can take $b\ell$ as our visible system 
and the neutrino $\nu_\ell$ as the invisible particle, 
while in the other leg we can treat the $b$-jet
as the visible system and the $W$-boson as the child particle.
In this case, there still should be a ridge structure revealing 
the true $t$, $W$ and $\nu$ masses.
\item {\em Multi-component dark matter}.
Of course, the model may contain two (or more) different 
genuine dark matter particles \cite{Boehm:2003ha,Ma:2006uv,Hur:2007ur,Cao:2007fy,Lee:2008pc,%
Feng:2008ya,SungCheon:2008ts,Fairbairn:2008fb,Zurek:2008qg}, 
whose production in various combinations will
inevitably lead at times to asymmetric event topologies.
\end{enumerate}

In conclusion, our work shows that the $M_{T2}$ concept can be easily
generalized to decay chains terminating in two different daughter particles.
Nevertheless, the methods discussed in this paper allow 
to extract all masses involved in the decays, at least as a matter of principle.
We believe that such methods will prove extremely useful, 
if a missing energy signal of new physics is seen at the Tevatron or the LHC.

\bigskip

\acknowledgments
We are grateful to A.~J.~Barr, B.~Gripaios, C.~G.~Lester, and L.~Pape
for their insightful and stimulating comments.
All authors would like to thank the Fermilab Theoretical Physics Department for
warm hospitality and support at various stages during the completion of this work. 
This work is supported in part by a US Department of Energy grant
DE-FG02-97ER41029. KK is supported in part by the DOE under contract DE-AC02-76SF00515.

\appendix
\section{Appendix: \ The asymmetric $M_{T2}$ in the limit of infinite $P_{UTM}$}
\label{app:infpt}
\allowdisplaybreaks
\renewcommand{\theequation}{A.\arabic{equation}}
\setcounter{equation}{0}

In this appendix we revisit our previous two examples from Sections
\ref{subsec:110DLSP} and \ref{subsec:110ELSP}, this time considering
the infinitely large $P_{UTM}$ limit \cite{Barr:2009jv}. 
While this situation is impossible to achieve in a real experiment,
its advantage is that it can be treated by analytical means. 
In the $P_{UTM}\to\infty$ limit, the ``decoupling argument'' of 
Ref.~\cite{Barr:2009jv} holds, and one finds the following analytical 
expression for the $M_{T2}$ endpoint as a function of the two test
children masses $\tilde{M}_c^{(a)}$ and $\tilde{M}_c^{(b)}$:
\begin{equation}
M_{T2(max)}(\tilde{M}_c^{(a)},\tilde{M}_c^{(b)},\infty)=
\left\{
\begin{array}{cl}
\sqrt{M_p^2-(M_c^{(a)})^2+(\tilde{M}_c^{(a)})^2}, & ~
	\text{if }  (\tilde{M}_c^{(a)}, \tilde{M}_c^{(b)}) \in {\cal R}_1,\\[4mm]
\sqrt{M_p^2-(M_c^{(b)})^2+(\tilde{M}_c^{(b)})^2}, & ~
	\text{if }  (\tilde{M}_c^{(a)}, \tilde{M}_c^{(b)}) \in {\cal R}_2,\\[4mm]
\frac{\tilde{M}_c^{(b)}}{M_c^{(b)}}\, M_p,          & ~
	\text{if }  (\tilde{M}_c^{(a)}, \tilde{M}_c^{(b)}) \in {\cal R}_3,\\[4mm]
\frac{\tilde{M}_c^{(a)}}{M_c^{(a)}}\, M_p,          & ~ 
	\text{if }  (\tilde{M}_c^{(a)}, \tilde{M}_c^{(b)}) \in {\cal R}_4,
\end{array} \right.
\label{mt2maxPTinf}
\end{equation}
where the four defining regions ${\cal R}_i$, $(i=1,\ldots,4)$ 
are shown in Fig.~\ref{fig:kink_regions} and are defined as follows:
%
\FIGURE[ht]{
\centerline{
\epsfig{file=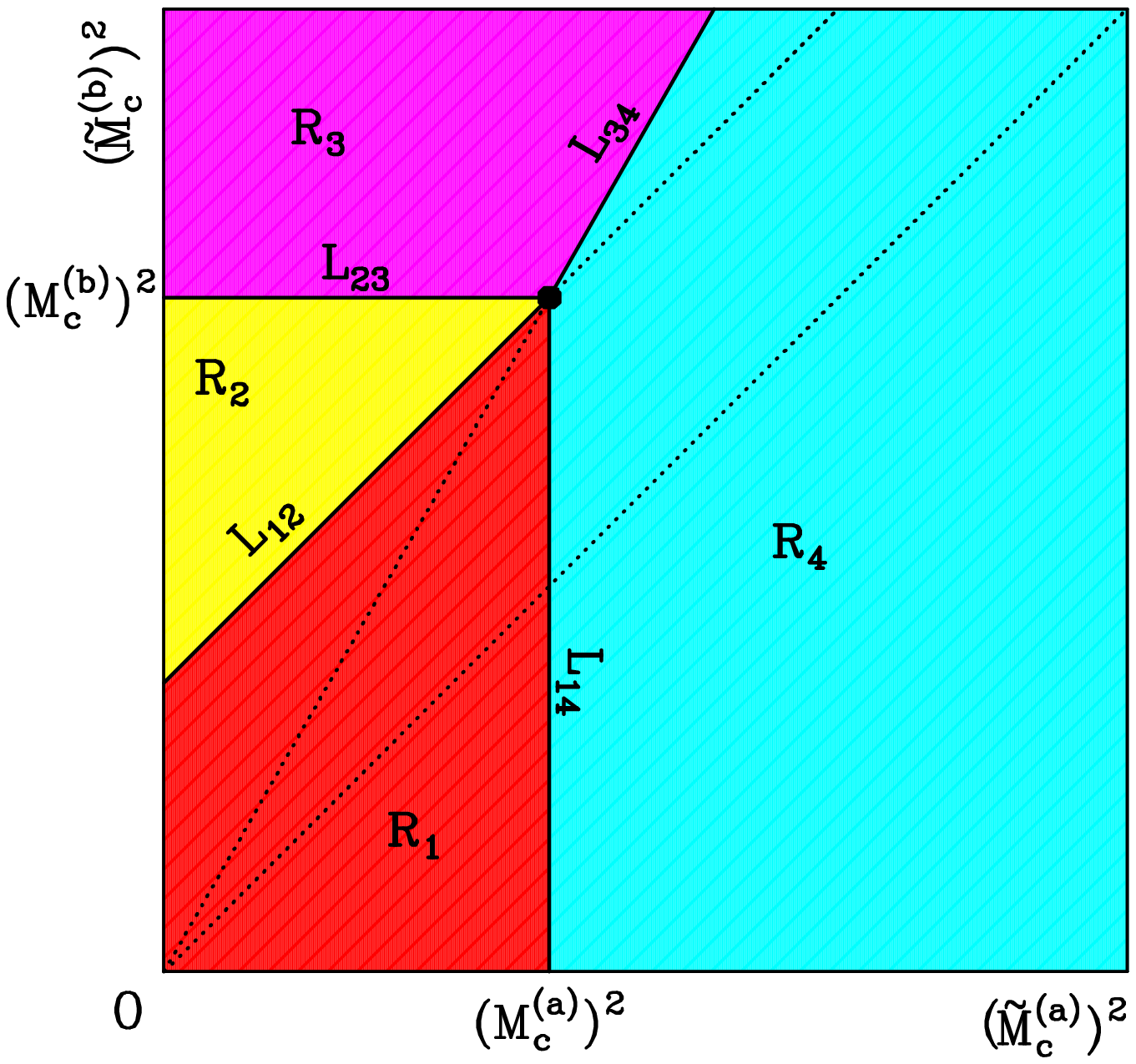,       height=7.0cm} }
\caption{\it The parameter plane of test children masses squared, 
divided into the four different regions ${\cal R}_i$ 
used to define the $M_{T2}$ endpoint function (\ref{mt2maxPTinf}).
Their common boundaries $L_{ij}$ are 
parametrically defined in eqs.~(\ref{L12}-\ref{L14}). 
The black dot corresponds to the true values of the children masses.
}
\label{fig:kink_regions}
}
\begin{eqnarray}
{\cal R}_1&:& 
\tilde{M}_c^{(b)}< \sqrt{(M_c^{(b)})^2-(M_c^{(a)})^2+(\tilde{M}_c^{(a)})^2}
\ \wedge \  
\tilde{M}_c^{(a)}<M_c^{(a)}\, , \\ [2mm]
{\cal R}_2&:& 
\sqrt{(M_c^{(b)})^2-(M_c^{(a)})^2+(\tilde{M}_c^{(a)})^2} <\tilde{M}_c^{(b)}<M_c^{(b)}\, ,
\\ [2mm]
{\cal R}_3&:& 
M_c^{(b)}<\tilde{M}_c^{(b)}
\ \wedge \  
\tilde{M}_c^{(a)}<\left(\frac{M_c^{(a)}}{M_c^{(b)}}\right)\tilde{M}_c^{(b)}\, ,
\\ [2mm]
{\cal R}_4&:& 
M_c^{(a)}<\tilde{M}_c^{(a)}
\ \wedge \  
\tilde{M}_c^{(b)}<\left(\frac{M_c^{(b)}}{M_c^{(a)}}\right)\tilde{M}_c^{(a)}\, .
\end{eqnarray}
Since the functional expression for $M_{T2(max)}$ within each region ${\cal R}_i$ 
is different, there is in general a gradient discontinuity when crossing from 
one region into the next. Therefore, the ridges on the $M_{T2(max)}$ hypersurface
will appear along the common boundaries of the four regions ${\cal R}_i$.
Let us denote by $L_{ij}$ the boundary between regions ${\cal R}_i$ and ${\cal R}_j$. 
As indicated in Fig.~\ref{fig:kink_regions}, each $L_{ij}$ is a 
straight line in the parameter space of the children test masses {\em squared}
and is given by
\begin{eqnarray}
L_{12}&:&
(\tilde{M}_c^{(b)})^2 = (M_c^{(b)})^2-(M_c^{(a)})^2+(\tilde{M}_c^{(a)})^2\, , \tilde{M}_c^{(a)}\le M_c^{(a)} \, ;
\label{L12} \\ [2mm]
L_{23}&:&
\tilde{M}_c^{(b)}=M_c^{(b)}\, , \tilde{M}_c^{(a)}\le M_c^{(a)}\, ;
\label{L23} \\ [2mm]
L_{34}&:&
\tilde{M}_c^{(b)} = 
\frac{ M_c^{(b)}}{ M_c^{(a)}}\,  \tilde M_c^{(a)}\, , \tilde{M}_c^{(a)} \ge M_c^{(a)}\, ;
\label{L34} \\ [2mm]
L_{14}&:&
\tilde{M}_c^{(a)}=M_c^{(a)}\, , \tilde{M}_c^{(b)}\le M_c^{(b)}\, .
\label{L14} 
\end{eqnarray}
As seen in Fig.~\ref{fig:kink_regions}, all four lines $L_{ij}$ meet 
at the true children mass point $\tilde{M}_c^{(a)}= M_c^{(a)}$,
$\tilde{M}_c^{(b)}= M_c^{(b)}$, where in turn the $M_{T2}$ endpoint 
$M_{T2(max)}$
gives the true parent mass $M_p$, in accordance with eq.~(\ref{Mptruedlsp}).

With those preliminaries, we are now in a position to revisit our two examples 
from Sections \ref{subsec:110DLSP} and \ref{subsec:110ELSP}.
Figs.~\ref{fig:110isrinf} and \ref{fig:110_slspisrinf} are the
corresponding analogues of Figs.~\ref{fig:110isr} and \ref{fig:110_slspisr}
in the case of infinite $P_{UTM}$. Comparing with our earlier results, we notice 
both quantitative and qualitative changes in the ridge structure.
First, the smooth ridge in Fig.~\ref{fig:110isr}(b) 
(Fig.~\ref{fig:110_slspisr}(b)) has now been 
deformed into two straight line segments, 
one horizontal ($L_{23}$) and the other vertical ($L_{14}$),
which meet at an angle of $90^\circ$ precisely at the true values of the children masses. 
More importantly, Figs.~\ref{fig:110isrinf} and \ref{fig:110_slspisrinf}
now exhibit another pair of ridges $L_{12}$ and $L_{34}$ (plotted in red in
Figs.~\ref{fig:110isrinf}(b) and \ref{fig:110_slspisrinf}(b)), 
which were absent from the earlier figures in Section~\ref{sec:110}.
The system of four ridges seen in Figs.~\ref{fig:110isrinf}(a) and 
\ref{fig:110_slspisrinf}(a) is very similar to the crease structure 
observed in Ref.~\cite{Barr:2009jv}. We thus confirm the result of 
Ref.~\cite{Barr:2009jv} that in the infinite $P_{UTM}$ limit
there exist four different ridges, whose common intersection point 
reveals the true masses of the parent and children particles.

At this point it is instructive to contrast the two sets of ridgelines:
$L_{23}$ and $L_{14}$ (shown in Figs.~\ref{fig:110isrinf}(b) 
and \ref{fig:110_slspisrinf}(b) in black) versus
$L_{12}$ and $L_{34}$ (shown in Figs.~\ref{fig:110isrinf}(b) 
and \ref{fig:110_slspisrinf}(b) in red).
The boundaries $L_{23}$ and $L_{14}$
separate the union of regions ${\cal R}_1$ and ${\cal R}_2$
from the union of regions ${\cal R}_3$ and ${\cal R}_4$.
Along those boundaries, we observe a transition in the configuration of 
visible momenta which yields the maximum possible value of $M_{T2}$. 
More precisely, in regions ${\cal R}_1$ and ${\cal R}_2$ we find 
that the visible momenta $\vec{p}_T^{~(\lambda)}$ for $M_{T2(max)}$
are parallel to the direction of the upstream momentum
$\vec{P}_{UTM}$, while in regions ${\cal R}_3$ and ${\cal R}_4$ we find that 
$\vec{p}_T^{~(\lambda)}$ are anti-parallel to $\vec{P}_{UTM}$.
This fact remains true even at finite values of $P_{UTM}$, which
is why the ridgelines $L_{23}$ and $L_{14}$ could also be seen in 
the earlier plots from Sec.~\ref{sec:110} at finite $P_{UTM}=1$ TeV.

%
\FIGURE[ht]{
\centerline{
\epsfig{file=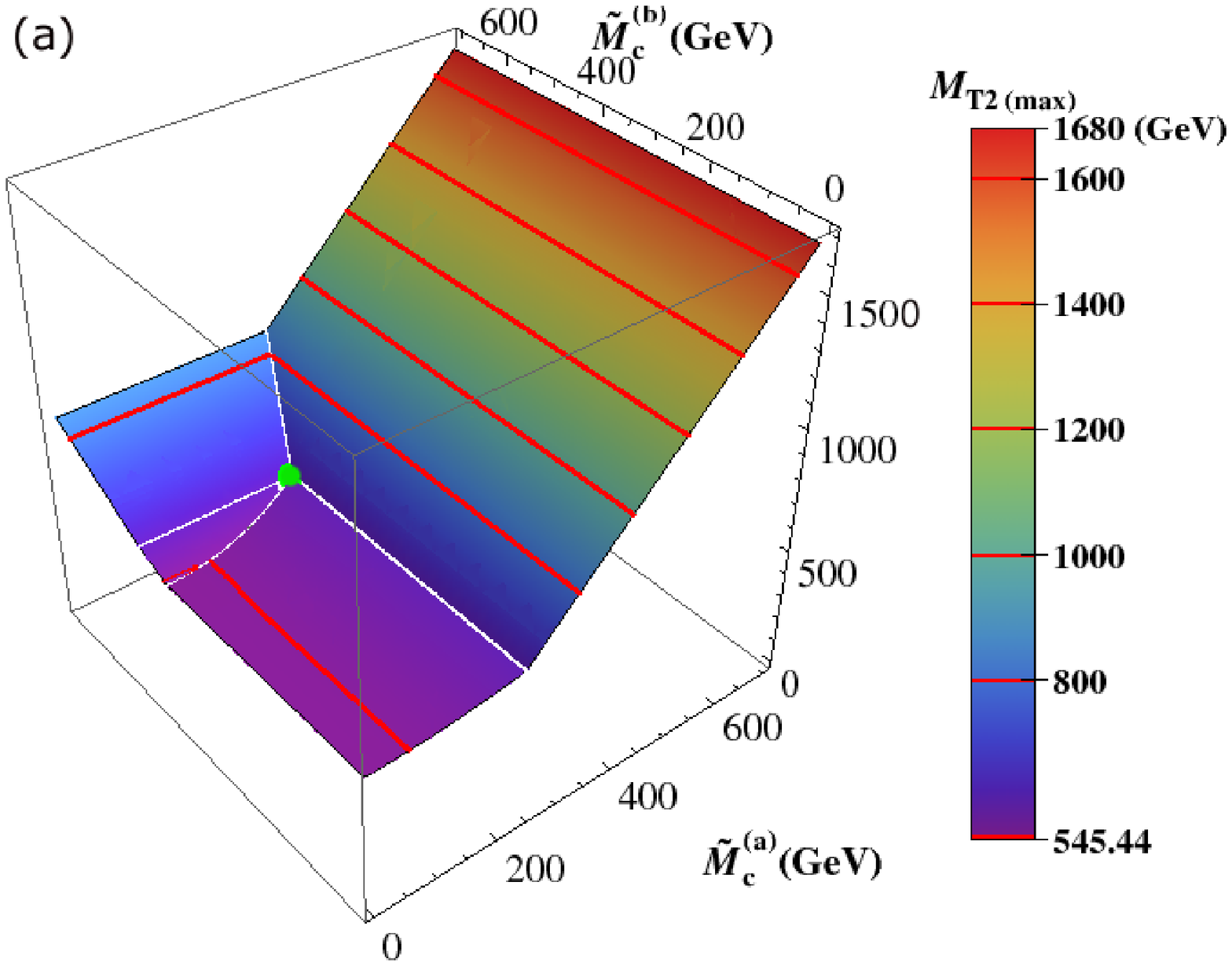,       height=6.3cm}~
\epsfig{file=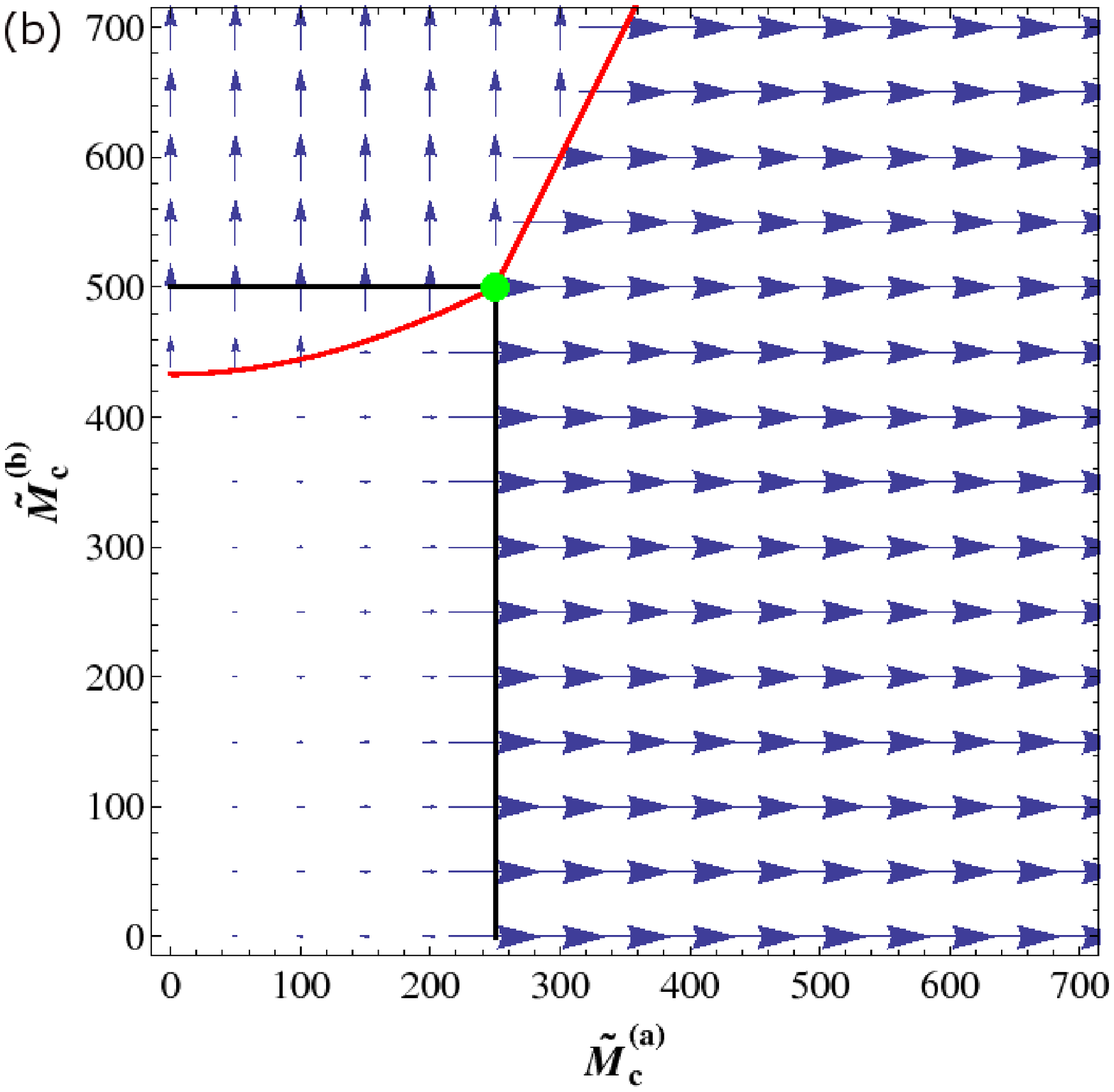,height=6.3cm} }
\caption{\it The same as Fig.~\ref{fig:110isr} but for
$P_{UTM} \to \infty$. 
}
\label{fig:110isrinf}
}

%
\FIGURE[ht]{
\centerline{
\epsfig{file=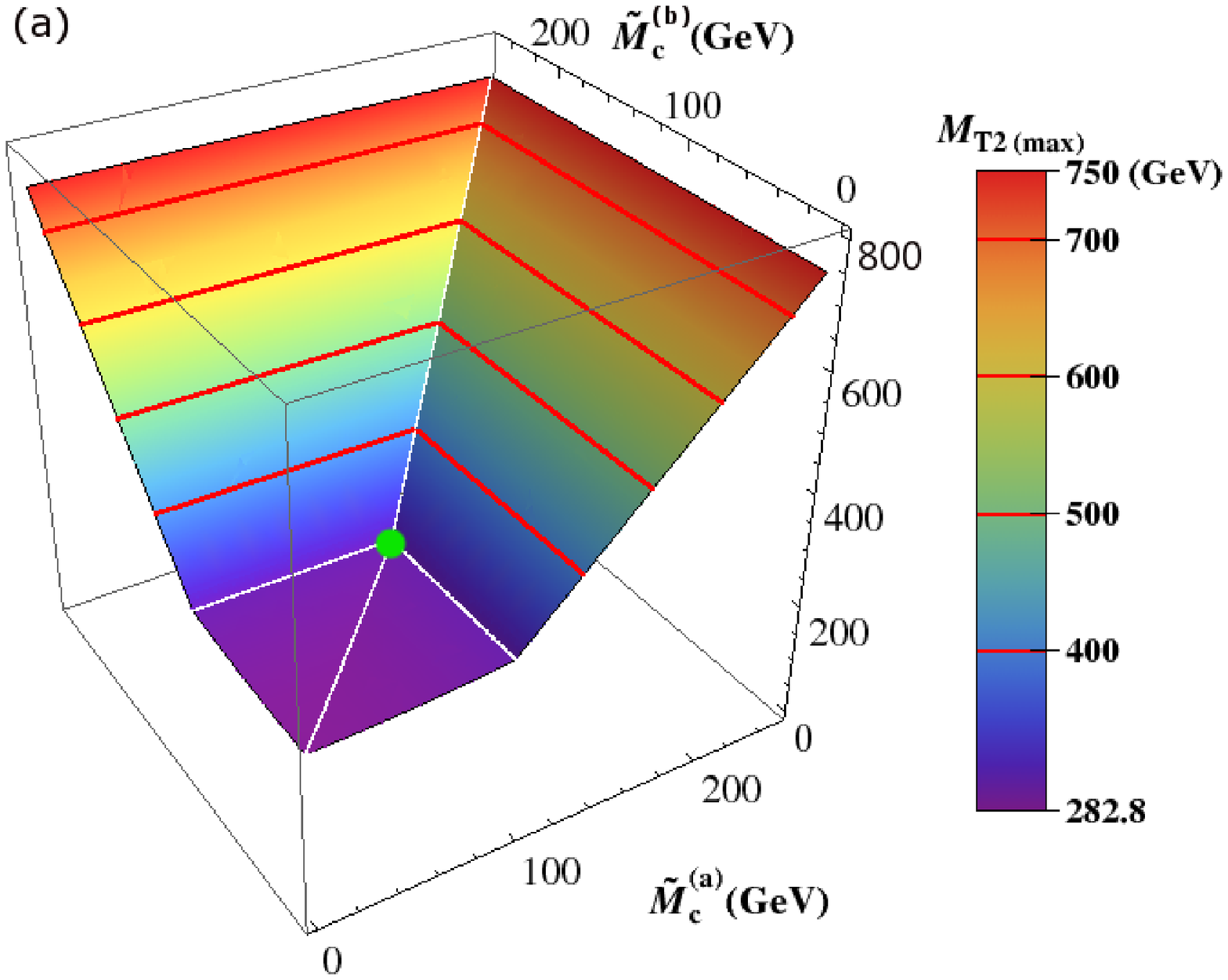,       height=6.3cm}~
\epsfig{file=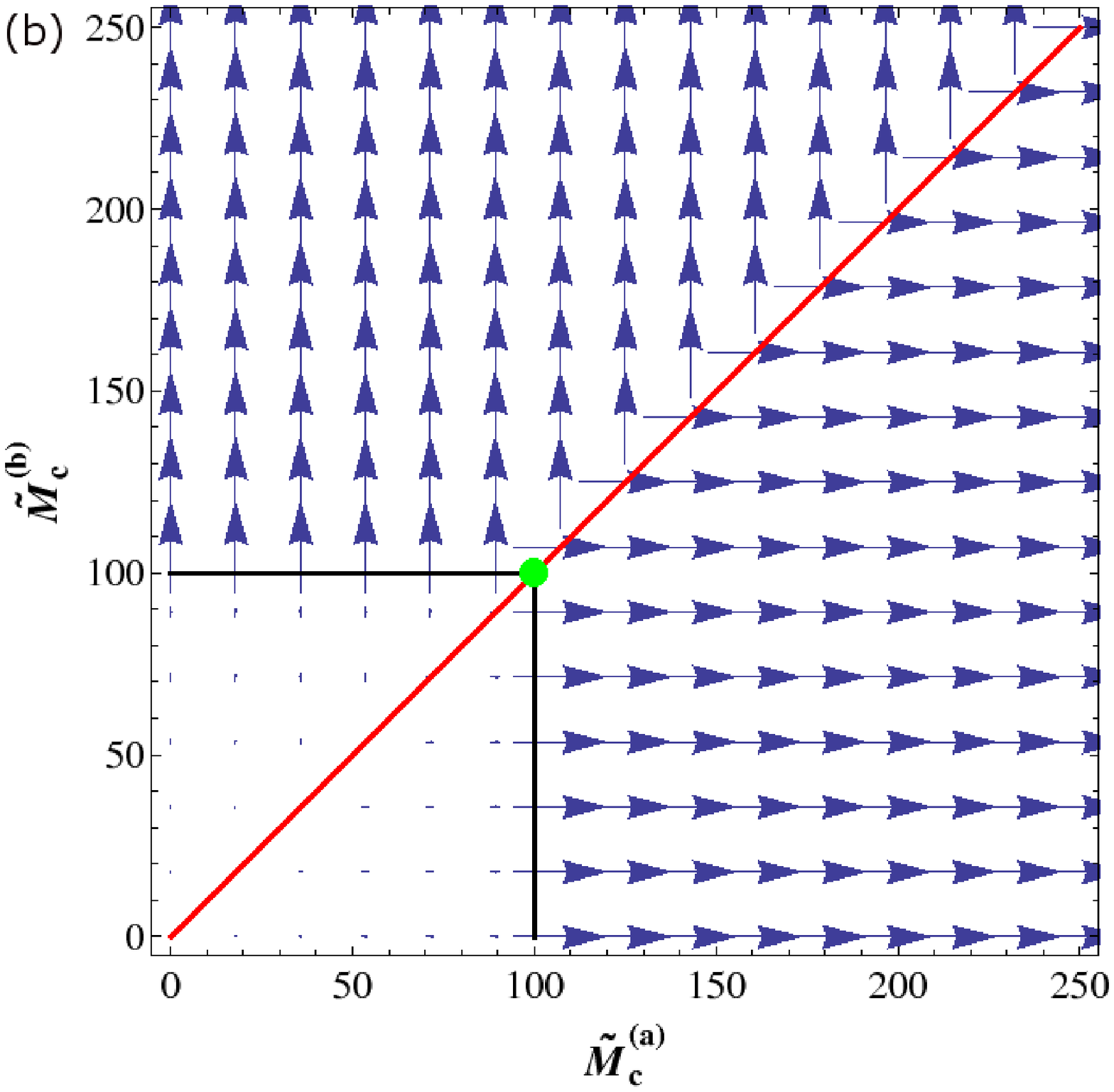,height=6.3cm} }
\caption{\it The same as Fig.~\ref{fig:110_slspisr} but for
$P_{UTM} \to \infty$. 
}
\label{fig:110_slspisrinf}
}

On the other hand, the ridgelines $L_{12}$ and $L_{34}$
shown in red in Figs.~\ref{fig:110isrinf}(b) and \ref{fig:110_slspisrinf}(b)
are due to the ``decoupling argument'' \cite{Barr:2009jv}, 
which is strictly valid only in the infinite $P_{UTM}$ limit. 
This is why these ridges become apparent only at very 
large values of $P_{UTM}$, and are gradually smeared out 
at smaller $P_{UTM}$. 

%
\FIGURE[ht]{
\centerline{
\epsfig{file=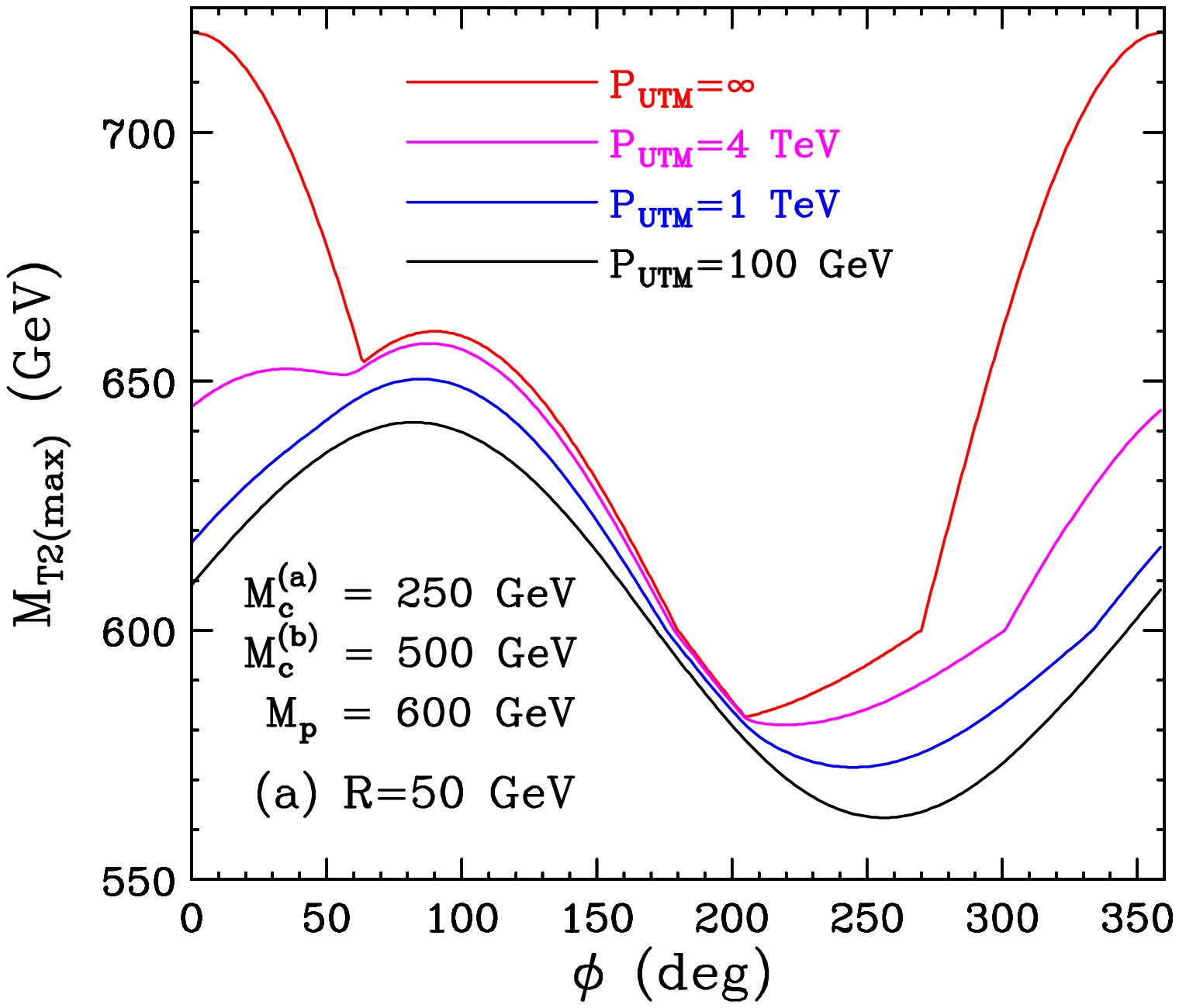,height=6.3cm}~
\epsfig{file=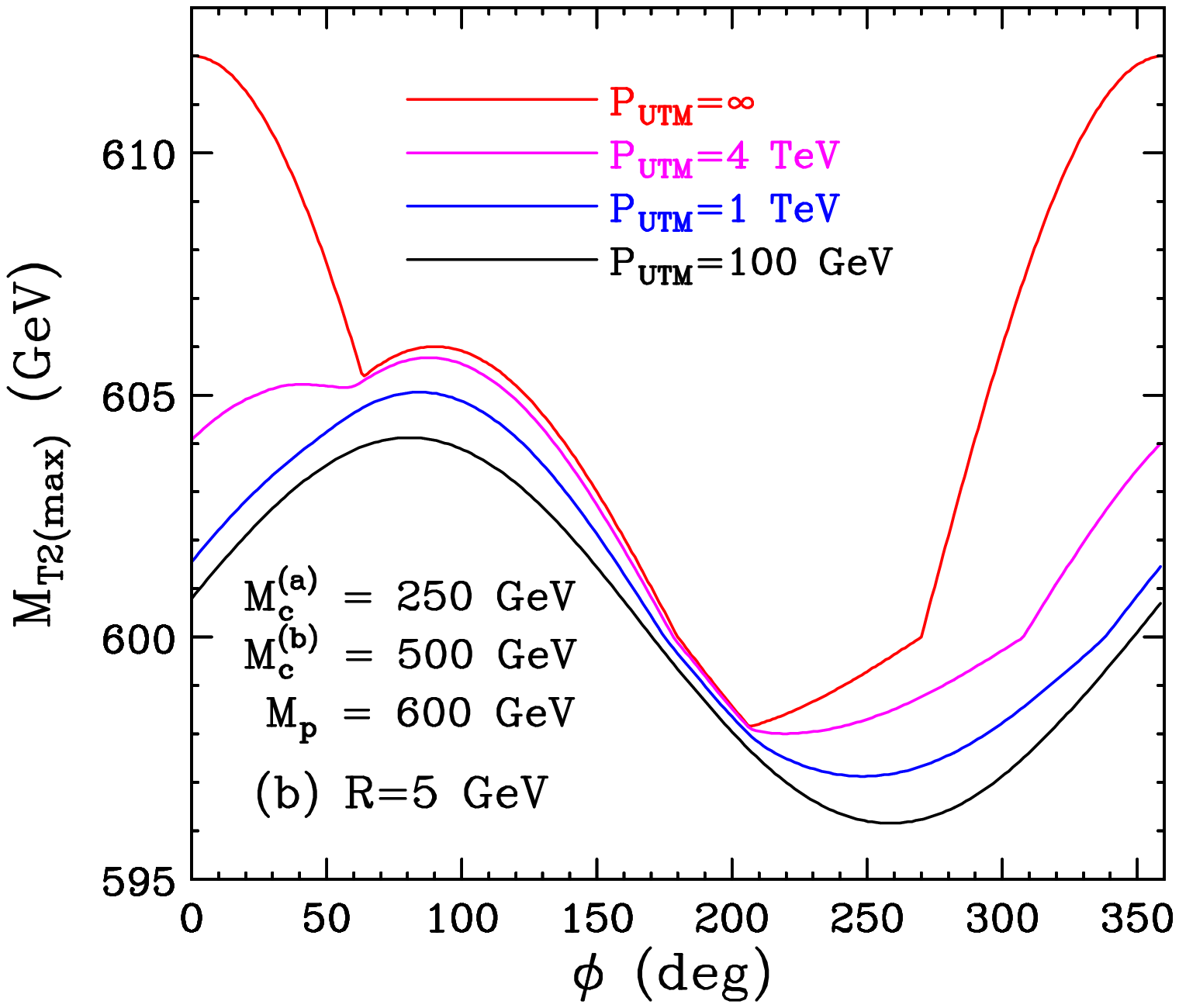,height=6.3cm} }
\caption{\it A study of the sharpness of the $M_{T2}$ 
ridge for the example considered in Sec.~\ref{subsec:110DLSP}.
The event topology is that of Fig.~\ref{fig:dlsp}(a)
and the mass spectrum is $M_c^{(a)}=250$ GeV,
$M_c^{(b)}=500$ GeV and $M_p=600$ GeV.
We plot the asymmetric $M_{T2}$ endpoint 
$M_{T2(max)}(\tilde M_c^{(a)}(\phi),\tilde M_c^{(b)}(\phi),P_{UTM})$,
as a function of the angular variable $\phi$ 
parameterizing the circle of radius $R$ 
defined in eqs.~(\ref{cosphi},\ref{sinphi}).
The radius $R$ of the circle is taken to be 
$R=50$ GeV in panel (a) and $R=5$ GeV in panel (b).
We present results for four different choices of the
upstream momentum $P_{UTM}$ as labelled in the plot.
}
\label{fig:kinksdlsp}
}
%
\FIGURE[ht]{
\centerline{
\epsfig{file=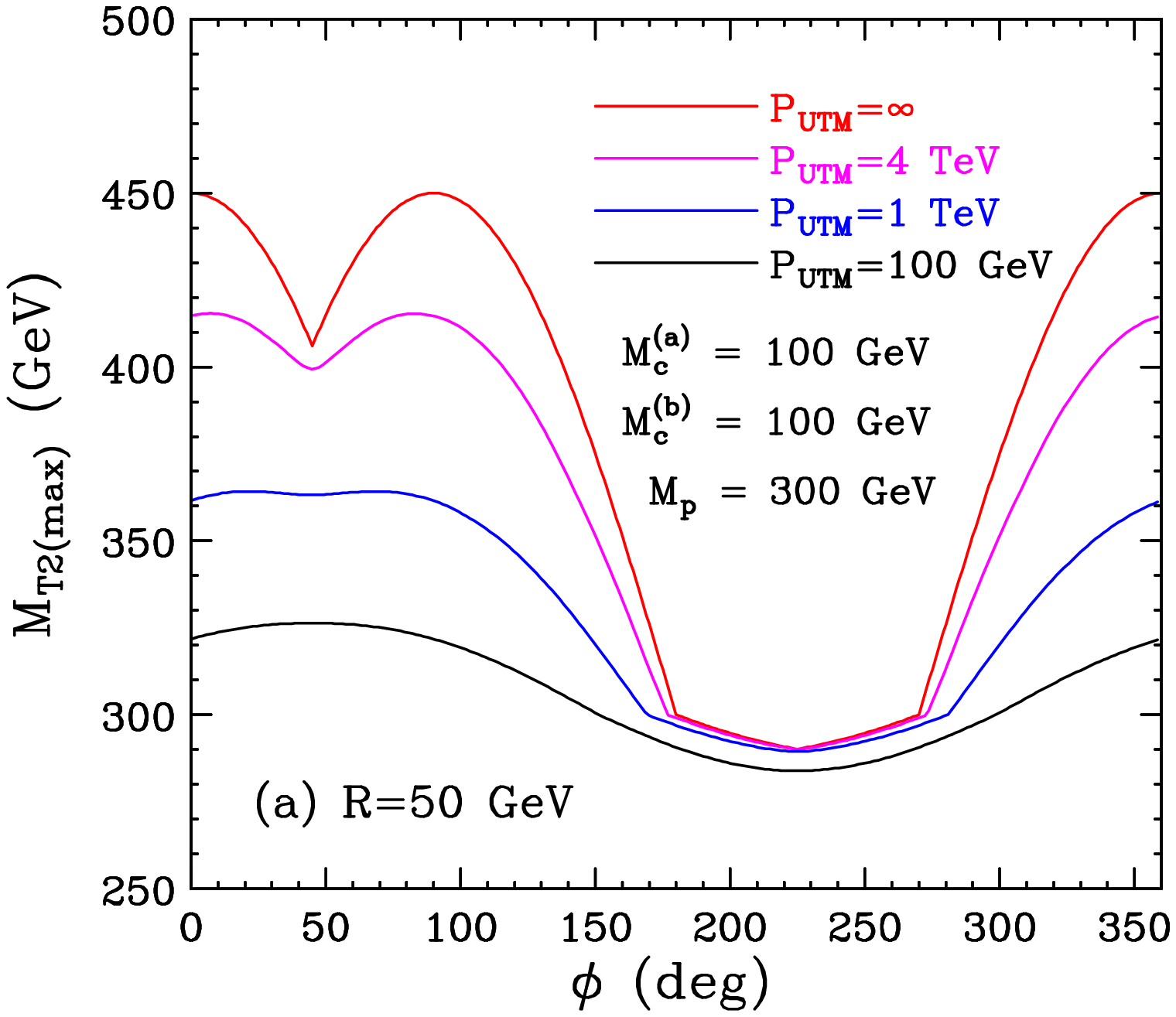,height=6.3cm}~
\epsfig{file=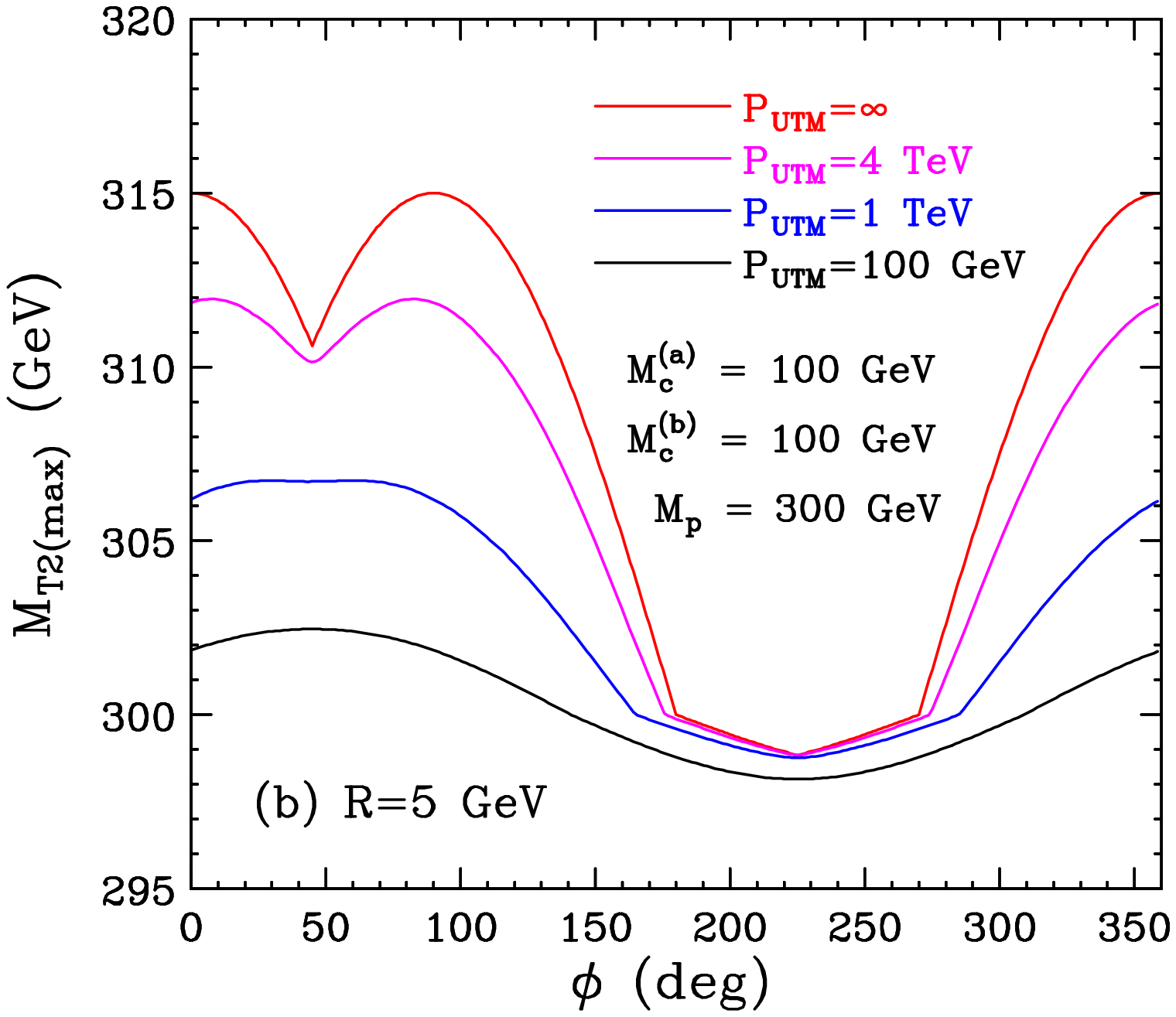,height=6.3cm} }
\caption{\it The same as Fig.~\ref{fig:kinksdlsp},
but for the example considered in Sec.~\ref{subsec:110ELSP},
where the input mass spectrum is fixed as
$M_c^{(a)}=100$ GeV,
$M_c^{(b)}=100$ GeV and $M_p=300$ GeV.
}
\label{fig:kinkselsp}
}

The evolution of the ridge structure as a function of $P_{UTM}$ is
shown in Figs.~\ref{fig:kinksdlsp} and \ref{fig:kinkselsp}.
In order to compare the sharpness of the four ridges,
we choose to vary the test children masses $\tilde M_c^{(a)}$
and $\tilde M_c^{(b)}$ along a circle centered on their true values
and with a fixed radius $R$. Such a circle is guaranteed to cross 
all four ridges, and can be parameterized in terms of an angular 
coordinate $\phi$ as follows
\begin{eqnarray}
\tilde M_c^{(a)}(\phi)&=&M_c^{(a)}+R\cos\phi, \label{cosphi} \\
\tilde M_c^{(b)}(\phi)&=&M_c^{(b)}+R\sin\phi. \label{sinphi}
\end{eqnarray}
Then in Fig.~\ref{fig:kinksdlsp} (Fig.~\ref{fig:kinkselsp})
we plot the asymmetric $M_{T2}$ endpoint 
$M_{T2(max)}(\tilde M_c^{(a)}(\phi),\tilde M_c^{(b)}(\phi),P_{UTM})$,
as a function of the angular variable $\phi$, for the 
case of mass spectrum I studied in Section~\ref{subsec:110DLSP}
(mass spectrum II studied in Section~\ref{subsec:110ELSP}).
The radius $R$ is taken to be 
$R=50$ GeV in panels (a) and $R=5$ GeV in panels (b).
We present results for four different choices of the
upstream momentum: $P_{UTM}=100$ GeV (black lines),
$P_{UTM}=1$ TeV (blue lines), $P_{UTM}=4$ TeV (magenta lines), 
and $P_{UTM}=\infty$ (red lines). Notice that the red lines 
at $P_{UTM}=\infty$ in Figs.~\ref{fig:kinksdlsp} and \ref{fig:kinkselsp}
are directly correlated to the three-dimensional plots of 
Figs.~\ref{fig:110isrinf} and \ref{fig:110_slspisrinf}, while
the blue lines at $P_{UTM}=1$ TeV in Figs.~\ref{fig:kinksdlsp} 
and \ref{fig:kinkselsp} are directly correlated to the 
three-dimensional plots of Figs.~\ref{fig:110isr} and \ref{fig:110_slspisr}.

Each one of the previously discussed ridges manifests
itself as a kink in Figs.~\ref{fig:kinksdlsp} and \ref{fig:kinkselsp}. 
Indeed, the red lines for $P_{UTM}=\infty$ reveal four clear kinks, which 
(from left to right) correspond to the ridgelines 
$L_{34}$, $L_{23}$, $L_{12}$, and $L_{14}$.
Using eqs.~(\ref{L12}-\ref{L14}), it is easy to find the expected
location of each kink in the $P_{UTM}\to \infty$ limit:
$\phi=\{63.4^\circ, 180^\circ, 204.9^\circ, 270^\circ\}$ for Fig.~\ref{fig:kinksdlsp}(a),
$\phi=\{63.4^\circ, 180^\circ, 206.4^\circ, 270^\circ\}$ for Fig.~\ref{fig:kinksdlsp}(b),
and $\phi=\{45^\circ, 180^\circ, 225^\circ, 270^\circ\}$ for Figs.~\ref{fig:kinkselsp}(a)
and \ref{fig:kinkselsp}(b).
However, as the upstream momentum is lowered to more realistic values, 
the kinks gradually wash out, albeit to a different degree.
As anticipated from our earlier results, the smearing effect is quite 
severe for $L_{34}$ and $L_{12}$, and by the time we reach 
$P_{UTM}=1$ TeV, those two kinks have completely disappeared.
On the other hand, $L_{23}$ and $L_{14}$ are affected to a lesser 
degree by the smearing effect and are still visible at $P_{UTM}=1$ TeV,
but by $P_{UTM}=100$ GeV they are essentially gone as well.
Notice that the variation in $P_{UTM}$ affects not only the sharpness
of the kinks, but also their location. This was to be expected, 
since we already saw that the shape of the ridge is different 
at $P_{UTM}=1$ TeV and $P_{UTM}=\infty$: compare the black ridge lines
in Figs.~\ref{fig:110isr}(b) and \ref{fig:110_slspisr}(b) 
to those in Figs.~\ref{fig:110isrinf}(b) and \ref{fig:110_slspisrinf}(b).
Finally, as a curious fact we notice that the 
results shown in panels (a) and panels (b) 
of Figs.~\ref{fig:kinksdlsp} and \ref{fig:kinkselsp} are approximately 
related by a simple scaling with a constant factor.

\listoftables           
\listoffigures          



\begin{thebibliography}{999}

\bibitem{Jungman:1995df}
  G.~Jungman, M.~Kamionkowski and K.~Griest,
  ``Supersymmetric dark matter,''
  Phys.\ Rept.\  {\bf 267}, 195 (1996)
  [arXiv:hep-ph/9506380].

\bibitem{Servant:2002aq}
  G.~Servant and T.~M.~P.~Tait,
  ``Is the lightest Kaluza-Klein particle a viable dark matter candidate?,''
  Nucl.\ Phys.\  B {\bf 650}, 391 (2003)
  [arXiv:hep-ph/0206071].

\bibitem{Burnell:2005hm}
  F.~Burnell and G.~D.~Kribs,
  ``The abundance of Kaluza-Klein dark matter with coannihilation,''
  Phys.\ Rev.\  D {\bf 73}, 015001 (2006)
  [arXiv:hep-ph/0509118].

\bibitem{Kong:2005hn}
  K.~Kong and K.~T.~Matchev,
  ``Precise calculation of the relic density of Kaluza-Klein dark matter in
  universal extra dimensions,''
  JHEP {\bf 0601}, 038 (2006)
  [arXiv:hep-ph/0509119].

\bibitem{Cheng:2003ju}
  H.~C.~Cheng and I.~Low,
  ``TeV symmetry and the little hierarchy problem,''
  JHEP {\bf 0309}, 051 (2003)
  [arXiv:hep-ph/0308199].

\bibitem{Birkedal:2006fz}
  A.~Birkedal, A.~Noble, M.~Perelstein and A.~Spray,
  ``Little Higgs dark matter,''
  Phys.\ Rev.\  D {\bf 74}, 035002 (2006)
  [arXiv:hep-ph/0603077].

\bibitem{Birkedal:2004xn}
  A.~Birkedal, K.~Matchev and M.~Perelstein,
  ``Dark matter at colliders: A model-independent approach,''
  Phys.\ Rev.\  D {\bf 70}, 077701 (2004)
  [arXiv:hep-ph/0403004].

\bibitem{Feng:2005gj}
  J.~L.~Feng, S.~Su and F.~Takayama,
  ``Lower limit on dark matter production at the Large Hadron Collider,''
  Phys.\ Rev.\ Lett.\  {\bf 96}, 151802 (2006)
  [arXiv:hep-ph/0503117].

\bibitem{Baltz:2006fm}
  E.~A.~Baltz, M.~Battaglia, M.~E.~Peskin and T.~Wizansky,
  ``Determination of dark matter properties at high-energy colliders,''
  Phys.\ Rev.\  D {\bf 74}, 103521 (2006)
  [arXiv:hep-ph/0602187].

\bibitem{Chung:2007cn}
  D.~J.~H.~Chung, L.~L.~Everett, K.~Kong and K.~T.~Matchev,
  ``Connecting LHC, ILC, and Quintessence,''
  JHEP {\bf 0710}, 016 (2007)
  [arXiv:0706.2375 [hep-ph]].

\bibitem{Berger:2007yu}
  C.~F.~Berger, J.~S.~Gainer, J.~L.~Hewett, T.~G.~Rizzo and B.~Lillie,
  ``The LHC Inverse Problem, Supersymmetry, and the ILC,''
  Phys.\ Lett.\  B {\bf 677}, 48 (2009)
  [arXiv:0711.1374 [hep-ph]].

\bibitem{Baer:2008uu}
  H.~Baer and X.~Tata,
  ``Dark matter and the LHC,''
  arXiv:0805.1905 [hep-ph].

\bibitem{Arrenberg:2008wy}
  S.~Arrenberg, L.~Baudis, K.~Kong, K.~T.~Matchev and J.~Yoo,
  ``Kaluza-Klein Dark Matter: Direct Detection vis-a-vis LHC,''
  Phys.\ Rev.\  D {\bf 78}, 056002 (2008)
  [arXiv:0805.4210 [hep-ph]].

\bibitem{Berger:2008cq}
  C.~F.~Berger, J.~S.~Gainer, J.~L.~Hewett and T.~G.~Rizzo,
  ``Supersymmetry Without Prejudice,''
  JHEP {\bf 0902}, 023 (2009)
  [arXiv:0812.0980 [hep-ph]].

\bibitem{Baer:2009bu}
  H.~Baer, E.~K.~Park and X.~Tata,
  ``Collider, direct and indirect detection of supersymmetric dark matter,''
  New J.\ Phys.\  {\bf 11}, 105024 (2009)
  [arXiv:0903.0555 [hep-ph]].

\bibitem{Roszkowski:2009ye}
  L.~Roszkowski, R.~Ruiz de Austri and R.~Trotta,
  ``Efficient reconstruction of CMSSM parameters from LHC data - A case
  study,''
  arXiv:0907.0594 [hep-ph].


\bibitem{Birkedal:2005jq}
  A.~Birkedal {\it et al.},
  ``Testing cosmology at the ILC,''
{\it In the Proceedings of 2005 International Linear Collider Workshop (LCWS 2005), Stanford, California, 18-22 Mar 2005, pp 0708}
  [arXiv:hep-ph/0507214].

\bibitem{Belanger:2008yc}
  G.~Belanger, O.~Kittel, S.~Kraml, H.~U.~Martyn and A.~Pukhov,
  ``Neutralino relic density from ILC measurements in the CPV MSSM,''
  Phys.\ Rev.\  D {\bf 78}, 015011 (2008)
  [arXiv:0803.2584 [hep-ph]].

\bibitem{Bernal:2008zk}
  N.~Bernal, A.~Goudelis, Y.~Mambrini and C.~Munoz,
  ``Determining the WIMP mass using the complementarity between direct and
  indirect searches and the ILC,''
  JCAP {\bf 0901}, 046 (2009)
  [arXiv:0804.1976 [hep-ph]].

\bibitem{Konar:2009ae}
  P.~Konar, K.~Kong, K.~T.~Matchev and M.~Perelstein,
  ``Shedding Light on the Dark Sector with Direct WIMP Production,''
  New J.\ Phys.\  {\bf 11}, 105004 (2009)
  [arXiv:0902.2000 [hep-ph]].


\bibitem{Chang:2009dh}
  S.~Chang and A.~de Gouvea,
  ``Neutrino Alternatives For Missing Energy Events At Colliders,''
  Phys.\ Rev.\  D {\bf 80}, 015008 (2009)
  [arXiv:0901.4796 [hep-ph]].


\bibitem{Barr:2004ze}
  A.~J.~Barr,
  ``Using lepton charge asymmetry to investigate the spin of supersymmetric particles at the LHC,''
  Phys.\ Lett.\  B {\bf 596}, 205 (2004)
  [arXiv:hep-ph/0405052].

\bibitem{Smillie:2005ar}
  J.~M.~Smillie and B.~R.~Webber,
  ``Distinguishing spins in supersymmetric and universal extra dimension
  models at the Large Hadron Collider,''
  JHEP {\bf 0510}, 069 (2005)
  [arXiv:hep-ph/0507170].

\bibitem{Athanasiou:2006ef}
  C.~Athanasiou, C.~G.~Lester, J.~M.~Smillie and B.~R.~Webber,
  ``Distinguishing spins in decay chains at the Large Hadron Collider,''
  JHEP {\bf 0608}, 055 (2006)
  [arXiv:hep-ph/0605286].

\bibitem{Athanasiou:2006hv}
  C.~Athanasiou, C.~G.~Lester, J.~M.~Smillie and B.~R.~Webber,
  ``Addendum to 'Distinguishing spins in decay chains at the Large Hadron Collider',''
  arXiv:hep-ph/0606212.

\bibitem{Goto:2004cpa}
  T.~Goto, K.~Kawagoe and M.~M.~Nojiri,
  ``Study of the slepton non-universality at the CERN Large Hadron  Collider,''
  Phys.\ Rev.\  D {\bf 70}, 075016 (2004)
  [Erratum-ibid.\  D {\bf 71}, 059902 (2005)]
  [arXiv:hep-ph/0406317].

\bibitem{Battaglia:2005zf}
  M.~Battaglia, A.~Datta, A.~De Roeck, K.~Kong and K.~T.~Matchev,
  ``Contrasting supersymmetry and universal extra dimensions at the CLIC
  multi-TeV e+ e- collider,''
  JHEP {\bf 0507}, 033 (2005)
  [arXiv:hep-ph/0502041].

\bibitem{Battaglia:2005ma}
  M.~Battaglia, A.~K.~Datta, A.~De Roeck, K.~Kong and K.~T.~Matchev,
  ``Contrasting supersymmetry and universal extra dimensions at colliders,''
{\it In the Proceedings of 2005 International Linear Collider Workshop (LCWS 2005), Stanford, California, 18-22 Mar 2005, pp 0302}
  [arXiv:hep-ph/0507284].

\bibitem{Datta:2005zs}
  A.~Datta, K.~Kong and K.~T.~Matchev,
  ``Discrimination of supersymmetry and universal extra dimensions at  hadron
  colliders,''
  Phys.\ Rev.\  D {\bf 72}, 096006 (2005)
  [Erratum-ibid.\  D {\bf 72}, 119901 (2005)]
  [arXiv:hep-ph/0509246].

\bibitem{Datta:2005vx}
  A.~Datta, G.~L.~Kane and M.~Toharia,
  ``Is it SUSY?,''
  arXiv:hep-ph/0510204.

\bibitem{Barr:2005dz}
  A.~J.~Barr,
  ``Measuring slepton spin at the LHC,''
  JHEP {\bf 0602}, 042 (2006)
  [arXiv:hep-ph/0511115].

\bibitem{Meade:2006dw}
  P.~Meade and M.~Reece,
  ``Top partners at the LHC: Spin and mass measurement,''
  Phys.\ Rev.\  D {\bf 74}, 015010 (2006)
  [arXiv:hep-ph/0601124].

\bibitem{Alves:2006df}
  A.~Alves, O.~Eboli and T.~Plehn,
  ``It's a gluino,''
  Phys.\ Rev.\  D {\bf 74}, 095010 (2006)
  [arXiv:hep-ph/0605067].

\bibitem{Wang:2006hk}
  L.~T.~Wang and I.~Yavin,
  ``Spin Measurements in Cascade Decays at the LHC,''
  JHEP {\bf 0704}, 032 (2007)
  [arXiv:hep-ph/0605296].

\bibitem{SA:2006jm}
  S.~Abdullin {\it et al.}  [TeV4LHC Working Group],
  ``Tevatron-for-LHC report: Preparations for discoveries,''
  arXiv:hep-ph/0608322.

\bibitem{Smillie:2006cd}
  J.~M.~Smillie,
  ``Spin Correlations in Decay Chains Involving W Bosons,''
  Eur.\ Phys.\ J.\  C {\bf 51}, 933 (2007)
  [arXiv:hep-ph/0609296].

\bibitem{Kong:2006pi}
  K.~Kong and K.~T.~Matchev,
  ``Phenomenology of universal extra dimensions,''
  AIP Conf.\ Proc.\  {\bf 903}, 451 (2007)
  [arXiv:hep-ph/0610057].

\bibitem{Kilic:2007zk}
  C.~Kilic, L.~T.~Wang and I.~Yavin,
  ``On the Existence of Angular Correlations in Decays with Heavy Matter Partners,''
  JHEP {\bf 0705}, 052 (2007)
  [arXiv:hep-ph/0703085].

\bibitem{Alves:2007xt}
  A.~Alves and O.~Eboli,
  ``Unravelling the sbottom spin at the CERN LHC,''
  Phys.\ Rev.\  D {\bf 75}, 115013 (2007)
  [arXiv:0704.0254 [hep-ph]].

\bibitem{Csaki:2007xm}
  C.~Csaki, J.~Heinonen and M.~Perelstein,
  ``Testing Gluino Spin with Three-Body Decays,''
  JHEP {\bf 0710}, 107 (2007)
  [arXiv:0707.0014 [hep-ph]].

\bibitem{Datta:2007xy}
  A.~Datta, P.~Dey, S.~K.~Gupta, B.~Mukhopadhyaya and A.~Nyffeler,
  ``Distinguishing the Littlest Higgs model with T-parity from supersymmetry at
  the LHC using trileptons,''
  Phys.\ Lett.\  B {\bf 659}, 308 (2008)
  [arXiv:0708.1912 [hep-ph]].

\bibitem{Buckley:2007th}
  M.~R.~Buckley, H.~Murayama, W.~Klemm and V.~Rentala,
  ``Discriminating spin through quantum interference,''
  Phys.\ Rev.\  D {\bf 78}, 014028 (2008)
  [arXiv:0711.0364 [hep-ph]].

\bibitem{Buckley:2008pp}
  M.~R.~Buckley, B.~Heinemann, W.~Klemm and H.~Murayama,
  ``Quantum Interference Effects Among Helicities at LEP-II and Tevatron,''
  Phys.\ Rev.\  D {\bf 77}, 113017 (2008)
  [arXiv:0804.0476 [hep-ph]].

\bibitem{Kane:2008kw}
  G.~L.~Kane, A.~A.~Petrov, J.~Shao and L.~T.~Wang,
  ``Initial determination of the spins of the gluino and squarks at LHC,''
  arXiv:0805.1397 [hep-ph].

\bibitem{Burns:2008cp}
  M.~Burns, K.~Kong, K.~T.~Matchev and M.~Park,
  ``A General Method for Model-Independent Measurements of Particle Spins,
  Couplings and Mixing Angles in Cascade Decays with Missing Energy at Hadron
  Colliders,''
  JHEP {\bf 0810}, 081 (2008), 
  arXiv:0808.2472 [hep-ph].

\bibitem{Gedalia:2009ym}
  O.~Gedalia, S.~J.~Lee and G.~Perez,
  ``Spin Determination via Third Generation Cascade Decays,''
  Phys.\ Rev.\  D {\bf 80}, 035012 (2009)
  [arXiv:0901.4438 [hep-ph]].

\bibitem{Boudjema:2009fz}
  F.~Boudjema and R.~K.~Singh,
  ``A model independent spin analysis of fundamental particles using azimuthal
  asymmetries,''
  JHEP {\bf 0907}, 028 (2009)
  [arXiv:0903.4705 [hep-ph]].


\bibitem{Ehrenfeld:2009rt}
  W.~Ehrenfeld, A.~Freitas, A.~Landwehr and D.~Wyler,
  ``Distinguishing spins in decay chains with photons at the Large Hadron
  Collider,''
  JHEP {\bf 0907}, 056 (2009)
  [arXiv:0904.1293 [hep-ph]].


\bibitem{Hinchliffe:1996iu}
  I.~Hinchliffe, F.~E.~Paige, M.~D.~Shapiro, J.~Soderqvist and W.~Yao,
  ``Precision SUSY measurements at LHC,''
  Phys.\ Rev.\  D {\bf 55}, 5520 (1997)
  [arXiv:hep-ph/9610544].

\bibitem{Lester:1999tx}
  C.~G.~Lester and D.~J.~Summers,
   ``Measuring masses of semi-invisibly decaying particles pair produced at
  hadron colliders,''
  Phys.\ Lett.\  B {\bf 463}, 99 (1999)
  [arXiv:hep-ph/9906349].

\bibitem{Bachacou:1999zb}
  H.~Bachacou, I.~Hinchliffe and F.~E.~Paige,
  ``Measurements of masses in SUGRA models at LHC,''
  Phys.\ Rev.\  D {\bf 62}, 015009 (2000)
  [arXiv:hep-ph/9907518].

\bibitem{Hinchliffe:1999zc}
  I.~Hinchliffe and F.~E.~Paige,
  ``Measurements in SUGRA models with large tan(beta) at LHC,''
  Phys.\ Rev.\  D {\bf 61}, 095011 (2000)
  [arXiv:hep-ph/9907519].

\bibitem{Allanach:2000kt}
  B.~C.~Allanach, C.~G.~Lester, M.~A.~Parker and B.~R.~Webber,
   ``Measuring sparticle masses in non-universal string inspired models at  the
  LHC,''
  JHEP {\bf 0009}, 004 (2000)
  [arXiv:hep-ph/0007009].

\bibitem{Barr:2003rg}
  A.~Barr, C.~Lester and P.~Stephens,
  ``m(T2): The truth behind the glamour,''
  J.\ Phys.\ G {\bf 29}, 2343 (2003)
  [arXiv:hep-ph/0304226].

\bibitem{Nojiri:2003tu}
  M.~M.~Nojiri, G.~Polesello and D.~R.~Tovey,
   ``Proposal for a new reconstruction technique for SUSY processes at the
  LHC,''
  arXiv:hep-ph/0312317.

\bibitem{Kawagoe:2004rz}
  K.~Kawagoe, M.~M.~Nojiri and G.~Polesello,
  ``A new SUSY mass reconstruction method at the CERN LHC,''
  Phys.\ Rev.\  D {\bf 71}, 035008 (2005)
  [arXiv:hep-ph/0410160].

\bibitem{Gjelsten:2004ki}
  B.~K.~Gjelsten, D.~J.~Miller and P.~Osland,
  ``Measurement of SUSY masses via cascade decays for SPS 1a,''
  JHEP {\bf 0412}, 003 (2004)
  [arXiv:hep-ph/0410303].

\bibitem{Gjelsten:2005aw}
  B.~K.~Gjelsten, D.~J.~Miller and P.~Osland,
  ``Measurement of the gluino mass via cascade decays for SPS 1a,''
  JHEP {\bf 0506}, 015 (2005)
  [arXiv:hep-ph/0501033].

\bibitem{Birkedal:2005cm}
  A.~Birkedal, R.~C.~Group and K.~Matchev,
  ``Slepton mass measurements at the LHC,''
{\it In the Proceedings of 2005 International Linear Collider Workshop (LCWS 2005), Stanford, California, 18-22 Mar 2005, pp 0210}
  [arXiv:hep-ph/0507002].

\bibitem{Miller:2005zp}
  D.~J.~Miller, P.~Osland and A.~R.~Raklev,
  ``Invariant mass distributions in cascade decays,''
  JHEP {\bf 0603}, 034 (2006)
  [arXiv:hep-ph/0510356].

\bibitem{Gjelsten:2006tg}
  B.~K.~Gjelsten, D.~J.~Miller, P.~Osland and A.~R.~Raklev,
  ``Mass determination in cascade decays using shape formulas,''
  AIP Conf.\ Proc.\  {\bf 903}, 257 (2007)
  [arXiv:hep-ph/0611259].

\bibitem{Matsumoto:2006ws}
  S.~Matsumoto, M.~M.~Nojiri and D.~Nomura,
  ``Hunting for the top partner in the littlest Higgs model with T-parity at
  the LHC,''
  Phys.\ Rev.\  D {\bf 75}, 055006 (2007)
  [arXiv:hep-ph/0612249].

\bibitem{Cheng:2007xv}
  H.~C.~Cheng, J.~F.~Gunion, Z.~Han, G.~Marandella and B.~McElrath,
  ``Mass Determination in SUSY-like Events with Missing Energy,''
  JHEP {\bf 0712}, 076 (2007)
  [arXiv:0707.0030 [hep-ph]].

\bibitem{Lester:2007fq}
  C.~Lester and A.~Barr,
  ``MTGEN : Mass scale measurements in pair-production at colliders,''
  JHEP {\bf 0712}, 102 (2007)
  [arXiv:0708.1028 [hep-ph]].

\bibitem{Cho:2007qv}
  W.~S.~Cho, K.~Choi, Y.~G.~Kim and C.~B.~Park,
  ``Gluino Stransverse Mass,''
  Phys.\ Rev.\ Lett.\  {\bf 100}, 171801 (2008)
  [arXiv:0709.0288 [hep-ph]].

\bibitem{Gripaios:2007is}
  B.~Gripaios,
  ``Transverse Observables and Mass Determination at Hadron Colliders,''
  JHEP {\bf 0802}, 053 (2008)
  [arXiv:0709.2740 [hep-ph]].

\bibitem{Barr:2007hy}
  A.~J.~Barr, B.~Gripaios and C.~G.~Lester,
   ``Weighing Wimps with Kinks at Colliders: Invisible Particle Mass
  Measurements from Endpoints,''
  JHEP {\bf 0802}, 014 (2008)
  [arXiv:0711.4008 [hep-ph]].

\bibitem{Cho:2007dh}
  W.~S.~Cho, K.~Choi, Y.~G.~Kim and C.~B.~Park,
   ``Measuring superparticle masses at hadron collider using the transverse mass
  kink,''
  JHEP {\bf 0802}, 035 (2008)
  [arXiv:0711.4526 [hep-ph]].

\bibitem{Ross:2007rm}
  G.~G.~Ross and M.~Serna,
  ``Mass Determination of New States at Hadron Colliders,''
  Phys.\ Lett.\  B {\bf 665}, 212 (2008)
  [arXiv:0712.0943 [hep-ph]].

\bibitem{Nojiri:2007pq}
  M.~M.~Nojiri, G.~Polesello and D.~R.~Tovey,
   ``A hybrid method for determining SUSY particle masses at the LHC with fully
  identified cascade decays,''
  JHEP {\bf 0805}, 014 (2008)
  [arXiv:0712.2718 [hep-ph]].

\bibitem{Huang:2008ae}
  P.~Huang, N.~Kersting and H.~H.~Yang,
  ``Hidden Thresholds: A Technique for Reconstructing New Physics Masses at
  Hadron Colliders,''
  arXiv:0802.0022 [hep-ph].

\bibitem{Nojiri:2008hy}
  M.~M.~Nojiri, Y.~Shimizu, S.~Okada and K.~Kawagoe,
  ``Inclusive transverse mass analysis for squark and gluino mass
  determination,''
  JHEP {\bf 0806}, 035 (2008)
  [arXiv:0802.2412 [hep-ph]].

\bibitem{Tovey:2008ui}
  D.~R.~Tovey,
  ``On measuring the masses of pair-produced semi-invisibly decaying particles
  at hadron colliders,''
  JHEP {\bf 0804}, 034 (2008)
  [arXiv:0802.2879 [hep-ph]].


\bibitem{Nojiri:2008ir}
  M.~M.~Nojiri and M.~Takeuchi,
  ``Study of the top reconstruction in top-partner events at the LHC,''
  JHEP {\bf 0810}, 025 (2008)
  [arXiv:0802.4142 [hep-ph]].

\bibitem{Cheng:2008mg}
  H.~C.~Cheng, D.~Engelhardt, J.~F.~Gunion, Z.~Han and B.~McElrath,
  ``Accurate Mass Determinations in Decay Chains with Missing Energy,''
  Phys.\ Rev.\ Lett.\  {\bf 100}, 252001 (2008)
  [arXiv:0802.4290 [hep-ph]].

\bibitem{Cho:2008cu}
  W.~S.~Cho, K.~Choi, Y.~G.~Kim and C.~B.~Park,
  ``Measuring the top quark mass with $m_{T2}$ at the LHC,''
  Phys.\ Rev.\  D {\bf 78}, 034019 (2008)
  [arXiv:0804.2185 [hep-ph]].

\bibitem{Serna:2008zk}
  M.~Serna,
  ``A short comparison between $m_{T2}$ and $m_{CT}$,''
  JHEP {\bf 0806}, 004 (2008)
  [arXiv:0804.3344 [hep-ph]].

\bibitem{Bisset:2008hm}
  M.~Bisset, R.~Lu and N.~Kersting,
  ``Improving SUSY Spectrum Determinations at the LHC with Wedgebox and Hidden
  Threshold Techniques,''
  arXiv:0806.2492 [hep-ph].

\bibitem{Barr:2008ba}
  A.~J.~Barr, G.~G.~Ross and M.~Serna,
  ``The Precision Determination of Invisible-Particle Masses at the LHC,''
  Phys.\ Rev.\  D {\bf 78}, 056006 (2008)
  [arXiv:0806.3224 [hep-ph]].


\bibitem{Kersting:2008qn}
  N.~Kersting,
  ``On Measuring Split-SUSY Gaugino Masses at the LHC,''
  Eur.\ Phys.\ J.\  C {\bf 63}, 23 (2009)
  [arXiv:0806.4238 [hep-ph]].

\bibitem{Nojiri:2008vq}
  M.~M.~Nojiri, K.~Sakurai, Y.~Shimizu and M.~Takeuchi,
  ``Handling jets + missing $E_T$ channel using inclusive mT2,''
  JHEP {\bf 0810}, 100 (2008)
  [arXiv:0808.1094 [hep-ph]].

\bibitem{Cho:2008tj}
  W.~S.~Cho, K.~Choi, Y.~G.~Kim and C.~B.~Park,
  ``$M_{T2}$-assisted on-shell reconstruction of missing momenta and its
  application to spin measurement at the LHC,''
  Phys.\ Rev.\  D {\bf 79}, 031701 (2009)
  [arXiv:0810.4853 [hep-ph]].


\bibitem{Barr:2008hv}
  A.~J.~Barr, A.~Pinder and M.~Serna,
  ``Precision Determination of Invisible-Particle Masses at the CERN LHC: II,''
  Phys.\ Rev.\  D {\bf 79}, 074005 (2009)
  [arXiv:0811.2138 [hep-ph]].

\bibitem{Cheng:2008hk}
  H.~C.~Cheng and Z.~Han,
  ``Minimal Kinematic Constraints and MT2,''
  JHEP {\bf 0812}, 063 (2008)
  [arXiv:0810.5178 [hep-ph]].


\bibitem{Burns:2008va}
  M.~Burns, K.~Kong, K.~T.~Matchev and M.~Park,
  ``Using Subsystem MT2 for Complete Mass Determinations in Decay Chains with
  Missing Energy at Hadron Colliders,''
  JHEP {\bf 0903}, 143 (2009)
  [arXiv:0810.5576 [hep-ph]].

\bibitem{Burns:2009zi}
  M.~Burns, K.~T.~Matchev and M.~Park,
  ``Using kinematic boundary lines for particle mass measurements and
  disambiguation in SUSY-like events with missing energy,''
  JHEP {\bf 0905}, 094 (2009)
  [arXiv:0903.4371 [hep-ph]].

\bibitem{Konar:2008ei}
  P.~Konar, K.~Kong and K.~T.~Matchev,
  ``$\sqrt{\hat{s}}_{min}$ : A Global inclusive variable for determining the
  mass scale of new physics in events with missing energy at hadron
  colliders,''
  JHEP {\bf 0903}, 085 (2009)
  [arXiv:0812.1042 [hep-ph]].

\bibitem{Cheng:2009fw}
  H.~C.~Cheng, J.~F.~Gunion, Z.~Han and B.~McElrath,
  ``Accurate Mass Determinations in Decay Chains with Missing Energy: II,''
  Phys.\ Rev.\  D {\bf 80}, 035020 (2009)
  [arXiv:0905.1344 [hep-ph]].


\bibitem{Matchev:2009iw}
  K.~T.~Matchev, F.~Moortgat, L.~Pape and M.~Park,
  ``Precise reconstruction of sparticle masses without ambiguities,''
 JHEP {\bf 0908}, 104 (2009)
  [arXiv:0906.2417 [hep-ph]].

\bibitem{Han:2009ss}
  T.~Han, I.~W.~Kim and J.~Song,
  ``Kinematic Cusps: Determining the Missing Particle Mass at the LHC,''
  arXiv:0906.5009 [hep-ph].

\bibitem{Barr:2009wu}
  A.~J.~Barr and C.~Gwenlan,
  ``The race for supersymmetry: using mT2 for discovery,''
  Phys.\ Rev.\  D {\bf 80}, 074007 (2009)
  [arXiv:0907.2713 [hep-ph]].

\bibitem{Webber:2009vm}
  B.~Webber,
  ``Mass determination in sequential particle decay chains,''
  JHEP {\bf 0909}, 124 (2009)
  [arXiv:0907.5307 [hep-ph]].

\bibitem{Kim:2009nq}
  S.~G.~Kim, N.~Maekawa, K.~I.~Nagao, M.~M.~Nojiri and K.~Sakurai,
  ``LHC signature of supersymmetric models with non-universal sfermion
  masses,''
  JHEP {\bf 0910}, 005 (2009)
  [arXiv:0907.4234 [hep-ph]].

\bibitem{Kang:2009sk}
  Z.~Kang, N.~Kersting, S.~Kraml, A.~R.~Raklev and M.~J.~White,
  ``Neutralino Reconstruction at the LHC from Decay-frame Kinematics,''
  arXiv:0908.1550 [hep-ph].

\bibitem{MPtalk}
M.~Park, Fermilab theory seminar,
http://theory.fnal.gov/seminars/slides/2009/MPark.pdf.

\bibitem{Barr:2009jv}
  A.~J.~Barr, B.~Gripaios and C.~G.~Lester,
  ``Transverse masses and kinematic constraints: from the boundary to the
  crease,''
  JHEP {\bf 0911}, 096 (2009)
  [arXiv:0908.3779 [hep-ph]].

\bibitem{Matchev:2009fh}
  K.~T.~Matchev, F.~Moortgat, L.~Pape and M.~Park,
  ``Precision sparticle spectroscopy in the inclusive same-sign dilepton
  channel at LHC,''
  arXiv:0909.4300 [hep-ph].

\bibitem{Polesello:2009rn}
  G.~Polesello and D.~R.~Tovey,
  ``Supersymmetric particle mass measurement with the boost-corrected
  contransverse mass,''
  arXiv:0910.0174 [hep-ph].

\bibitem{Matchev:2009ad}
  K.~T.~Matchev and M.~Park,
  ``A general method for determining the masses of semi-invisibly decaying
  particles at hadron colliders,''
  arXiv:0910.1584 [hep-ph].

\bibitem{Konar:2009wn}
  P.~Konar, K.~Kong, K.~T.~Matchev and M.~Park,
  ``Superpartner mass measurements with 1D decomposed MT2,''
  arXiv:0910.3679 [hep-ph].

\bibitem{Autermann:2009js}
  C.~Autermann, B.~Mura, C.~Sander, H.~Schettler and P.~Schleper,
  ``Determination of supersymmetric masses using kinematic fits at the LHC,''
  arXiv:0911.2607 [hep-ph].

\bibitem{Cho:2009ve}
  W.~S.~Cho, J.~E.~Kim and J.~H.~Kim,
  ``Shining on buried new particles,''
  arXiv:0912.2354 [hep-ph].

\bibitem{Profumo:2009tb}
  S.~Profumo, K.~Sigurdson and L.~Ubaldi,
  ``Can we discover multi-component WIMP dark matter?,''
  JCAP {\bf 0912}, 016 (2009)
  [arXiv:0907.4374 [hep-ph]].

\bibitem{Boehm:2003ha}
  C.~Boehm, P.~Fayet and J.~Silk,
  ``Light and heavy dark matter particles,''
  Phys.\ Rev.\  D {\bf 69}, 101302 (2004)
  [arXiv:hep-ph/0311143].

\bibitem{Ma:2006uv}
  E.~Ma,
  ``Supersymmetric Model of Radiative Seesaw Majorana Neutrino Masses,''
  Annales Fond.\ Broglie {\bf 31}, 285 (2006)
  [arXiv:hep-ph/0607142].

\bibitem{Hur:2007ur}
  T.~Hur, H.~S.~Lee and S.~Nasri,
  ``A Supersymmetric U(1) -prime model with multiple dark matters,''
  Phys.\ Rev.\  D {\bf 77}, 015008 (2008)
  [arXiv:0710.2653 [hep-ph]].

\bibitem{Cao:2007fy}
  Q.~H.~Cao, E.~Ma, J.~Wudka and C.~P.~Yuan,
  ``Multipartite Dark Matter,''
  arXiv:0711.3881 [hep-ph].

\bibitem{Lee:2008pc}
  H.~S.~Lee,
  ``Lightest U-parity Particle (LUP) dark matter,''
  Phys.\ Lett.\  B {\bf 663}, 255 (2008)
  [arXiv:0802.0506 [hep-ph]].

\bibitem{Feng:2008ya}
  J.~L.~Feng and J.~Kumar,
  ``The Wimpless Miracle: Dark-Matter Particles Without Weak-Scale Masses Or
  Weak Interactions,''
  Phys.\ Rev.\ Lett.\  {\bf 101}, 231301 (2008)
  [arXiv:0803.4196 [hep-ph]].

\bibitem{SungCheon:2008ts}
  H.~Sung Cheon, S.~K.~Kang and C.~S.~Kim,
  ``Doubly Coexisting Dark Matter Candidates in an Extended Seesaw Model,''
  Phys.\ Lett.\  B {\bf 675}, 203 (2009)
  [arXiv:0807.0981 [hep-ph]].

\bibitem{Fairbairn:2008fb}
  M.~Fairbairn and J.~Zupan,
  ``Two component dark matter,''
  JCAP {\bf 0907}, 001 (2009)
  [arXiv:0810.4147 [hep-ph]].

\bibitem{Zurek:2008qg}
  K.~M.~Zurek,
  ``Multi-Component Dark Matter,''
  Phys.\ Rev.\  D {\bf 79}, 115002 (2009)
  [arXiv:0811.4429 [hep-ph]].

\bibitem{Arvanitaki:2009hb}
  A.~Arvanitaki, N.~Craig, S.~Dimopoulos, S.~Dubovsky and J.~March-Russell,
  ``String Photini at the LHC,''
  arXiv:0909.5440 [hep-ph].

\bibitem{Lester_code}
Stransverse mass library: http://www.hep.phy.cam.ac.uk/~lester/mt2/index.html

\bibitem{Davis_code}
Calculating MT2 by Bisection: http://daneel.physics.ucdavis.edu/~zhenyuhan/mt2.html



\end{thebibliography}
\end{document}